\documentclass[twocolumn]{aastex63}

\usepackage{csvsimple}
\usepackage{booktabs}

\received{}
\revised{}
\accepted{\today}

\submitjournal{ApJ}

\shorttitle{Stellar populations and SFH in ETGs}
\shortauthors{Ali et al.}
\graphicspath{{./}}

\begin{document}

\title{Probing the stellar populations and star formation history of early-type galaxies at $0 < z < 1.1$ in the rest-frame ultraviolet}

\correspondingauthor{Sadman S. Ali}
\email{sali@naoj.org}

\author[0000-0003-3883-6500]{Sadman S. Ali}
\affiliation{
Subaru Telescope,  National Astronomical
Observatory of Japan, 650 North Aohoku Place, Hilo, HI, 96720, USA}

\author[0000-0003-1455-7339]{Roberto De Propris}
\affiliation{Finnish Centre for Astronomy with ESO, University of Turku, Vesilinnantie 5, Turku, Finland}
\affiliation{Department of Physics and Astronomy, Botswana International University of Science and Technology, Private Bag 16, Palapye, Botswana}

\author[0000-0001-6812-4542]{Chul Chung}
\affiliation{Center for Galaxy Evolution Research, Department of Astronomy, Yonsei University, Seoul 03722, Republic of Korea}

\author{Steven Phillipps}
\affiliation{H. H. Wills Physics Laboratory,
University of Bristol, Tyndall Avenue, Bristol, BS1 TL8, United Kingdom}

\author{Malcolm N. Bremer}
\affiliation{H. H. Wills Physics Laboratory,
University of Bristol, Tyndall Avenue, Bristol, BS1 TL8, United Kingdom}

\author{Masato Onodera}
\affiliation{
Subaru Telescope,  National Astronomical
Observatory of Japan, 650 North Aohoku Place, Hilo, HI, 96720, USA}

\author{Marcin Sawicki}
\affiliation{Department of Astronomy and Astrophysics and Institute for Computational Astrophysics, Saint Mary’s University, 923 Robie Street, Halifax, NS B3H 3C3, Canada}

\author{Guillaume Desprez}
\affiliation{Department of Astronomy and Astrophysics and Institute for Computational Astrophysics, Saint Mary’s University, 923 Robie Street, Halifax, NS B3H 3C3, Canada}

\author{Stephen Gwyn}
\affiliation{NRC Herzberg Astronomy and Astrophysics, 5071 West Saanich Road, Victoria, BC V9E2E7, Canada}

\begin{abstract}

We measure the evolution of the rest-frame $NUV-V$ colors for early-type galaxies in clusters at $0<z<1.1$ using data from the Hyper Suprime-Cam Subaru Strategic Program (HSC-SSP), CFHT Large Area U-band Deep Survey (CLAUDS) and local SDSS clusters observed with GALEX. Our results show that there is an excess in the ultraviolet spectrum in most quiescent galaxies (compared to the expectations from models fitting their optical/infrared colors and spectra) below $z\sim0.6$, beyond which the excess UV emission fades rapidly. This evolution of the UV color is only consistent with the presence of a highly evolved, hot horizontal branch sub-population in these galaxies (amongst the majority cool and optically bright stars), comprising on average 10\% of the total stellar mass and forming at $z>3$. The blue UV colors of early-type galaxies at low-intermediate redshifts are likely driven by this sub-population being enriched in helium up to $\sim44\%$. At $z>0.8$ (when the extra UV component has not yet appeared) the data allows us to constrain the star formation histories of galaxies by fitting models to the evolution of their UV colors: we find that the epoch at which the stellar populations formed ranges between $3<z_{form}<10$ (corresponding to $0.5-2.2$ Gyrs after the Big Bang) with a star-formation e-folding timescale of $\tau=0.35-0.7$ Gyr, suggesting that these galaxies formed the majority of stars at very high redshift, with a brief yet intense burst of star-formation activity. The star formation history and chemical evolution of early-type galaxies resemble those of globular clusters, albeit on much larger scales.

\end{abstract}


\section{Introduction} \label{sec:intro}

Early-type galaxies (ETGs) consist of old and metal-rich stellar populations, implying that most of their stellar mass was formed (and likely assembled) at very high redshift, $\gtrsim 10\,\mathrm{Gyr}$ ago \citep{Kauffmann2003,Thomas2005,Thomas2010,Gallazzi2005,Treu2005}. 
Some galaxies are observed to be already massive and passively evolving at $z \gtrsim 2$ \citep{Cimatti2004,Daddi2005,vanDokkum2008,Kriek2009,Belli2014,Straatman2014,Marsan2015,Glazebrook2017,Marsan2017,Schreiber2018,Valentino2020,Forrest2020a,Forrest2020b,Forrest2023}.  Star formation in these galaxies must have been extremely intense, and quenched rapidly over timescales shorter than $\sim 1$ Gyr. This is consistent with the `quasi-monolithic' scenario of galaxy formation in which most massive galaxies assembled the majority of their stellar content rapidly and at high redshift  \citep{Pipino2004,PerezGonzalez2008}.

ETGs may also show a degree of residual star formation (RSF) from young stellar populations \citep{Vazdekis2016,Atlee2009,Rusinol2019,DePropris2022} especially for low mass galaxies \citep{Ree2012} and lenticulars \citep{Salim2012}. Young stars (age $\lesssim 1$ Gyr) contribute strongly to the rest-frame near ultraviolet. However, RSF and the short-wavelength tail of thermal emission from old stellar populations are not the only contributors to the UV spectra of ETGs in the local universe. It has long been known that ETGs exhibit an UV `excess' (or `upturn' - \citealt{code1979,oconnell1999}) and that this is most likely produced by blue or extreme horizontal branch (HB) stars \citep{Yi1997,Yi1999}, that emit efficiently in the far and near-UV spectral range. Note that blue or extreme HB stars may also affect age determinations from spectroscopic indices, via their broad Balmer lines and high luminosities \citep{DePropris2000,Lee2000,Percival2011}. This requires sampling a redshift range where blue HB stars have not  yet appeared or modelling the contribution to the UV light from old stellar populations if we wish to constrain star formation histories and the presence of young stars in ETGs.

In this paper we derive rest-frame $NUV$ photometry for galaxies in clusters at $0 < z < 1.1$. As shown in our previous work \citep{Ali2022} we can use $NUV$ photometry to constrain the strength and evolution of the UV excess component and confirm the mechanisms that create a blue HB feature in otherwise old and metal rich stellar systems (where such stars should not exist). At redshifts beyond $z=0.8$ where there is no doubt, from our data, that the blue HB stars have not yet appeared in the stellar population, we can use our rest-frame $NUV$ photometry to set stringent limits to the age, star formation history and RSF of cluster ETGs. Furthermore, the rest-frame $NUV$ is much less affected by the age-metallicity degeneracy than the optical \citep{Worthey1994} and multi-wavelength datasets that include the $NUV$ maintain their sensitivity to age and $\tau$ (the e-folding time of the star formation histories) across a wide range of masses and redshifts. The lack of evolved UV-emitting stellar populations at $z > 0.8$ therefore makes our conclusions more robust than can be achieved by similar studies in the local universe (e.g. \citealt{Kaviraj2007}).

Data from this paper come from GALEX \citep{Morrissey2007} $NUV$ imaging of nearby clusters of galaxies selected in \cite{DePropris2021}, $U$ band imaging of clusters at intermediate redshifts ($z=0.4-0.7$) from the CLAUDS survey \citep{Sawicki2019} and $g$ band imaging of higher redshift clusters ($z=0.85-1.1$) from the HyperSuprimeCam Subaru Strategic Program \citep{Aihara2018,Aihara2022}. These are described in section 2 of this paper. The analysis and results are presented in section 3 and discussed in section 4 and 5. Conclusions are summarized in section 6. Cosmological parameters are assumed to be H$_0 = 67$ km s$^{-1}$ Mpc$^{-1}$, $\Omega_m=0.3$ and $\Omega_{\Lambda}=0.7$ (\citealt{Planck2020}). All magnitudes quoted are in the AB system.

\section{Data and Photometry}

\begin{table*}
\centering
\begin{filecontents*}{Galaxies.csv}
Redshift,UV (Observed),UV (Rest),,U (Observed),U (Rest),HSC z (Rest),HSC y (Rest)
0.375-0.425,Megacam u,2527,,HSC g,3393,6429,7143
0.425-0.475,Megacam u,2440,,HSC g,3276,6207,6897
0.475-0.525,Megacam u,2359,,HSC g,3167,6000,6667
0.525-0.575,Megacam u,2283,,HSC g,3065,5806,6452
0.575-0.625,Megacam u*,2339,,HSC g,2969,5625,6250
0.625-0.675,Megacam u*,2268,,HSC g,2879,5455,6061
0.825-0.875,HSC g,2567,,HSC r,3378,4865,5405
0.875-0.925,HSC g,2500,,HSC r,3289,4737,5263
0.925-0.975,HSC g,2436,,HSC r,3205,4615,5128
0.975-1.025,HSC g,2375,,HSC r,3125,4500,5000
1.025-1.075,HSC g,2317,,HSC r,3049,4390,4878
1.075-1.125,HSC g,2262,,HSC r,2976,4286,4762
\end{filecontents*}
\csvautobooktabular{Galaxies.csv}
\caption{Table giving details of the observed bands and rest-frame central wavelengths of HSC clusters, divided into redshift bins of $0.05$.}
\label{table_1}
\end{table*}

\begin{figure}
{\includegraphics[width=0.45\textwidth]{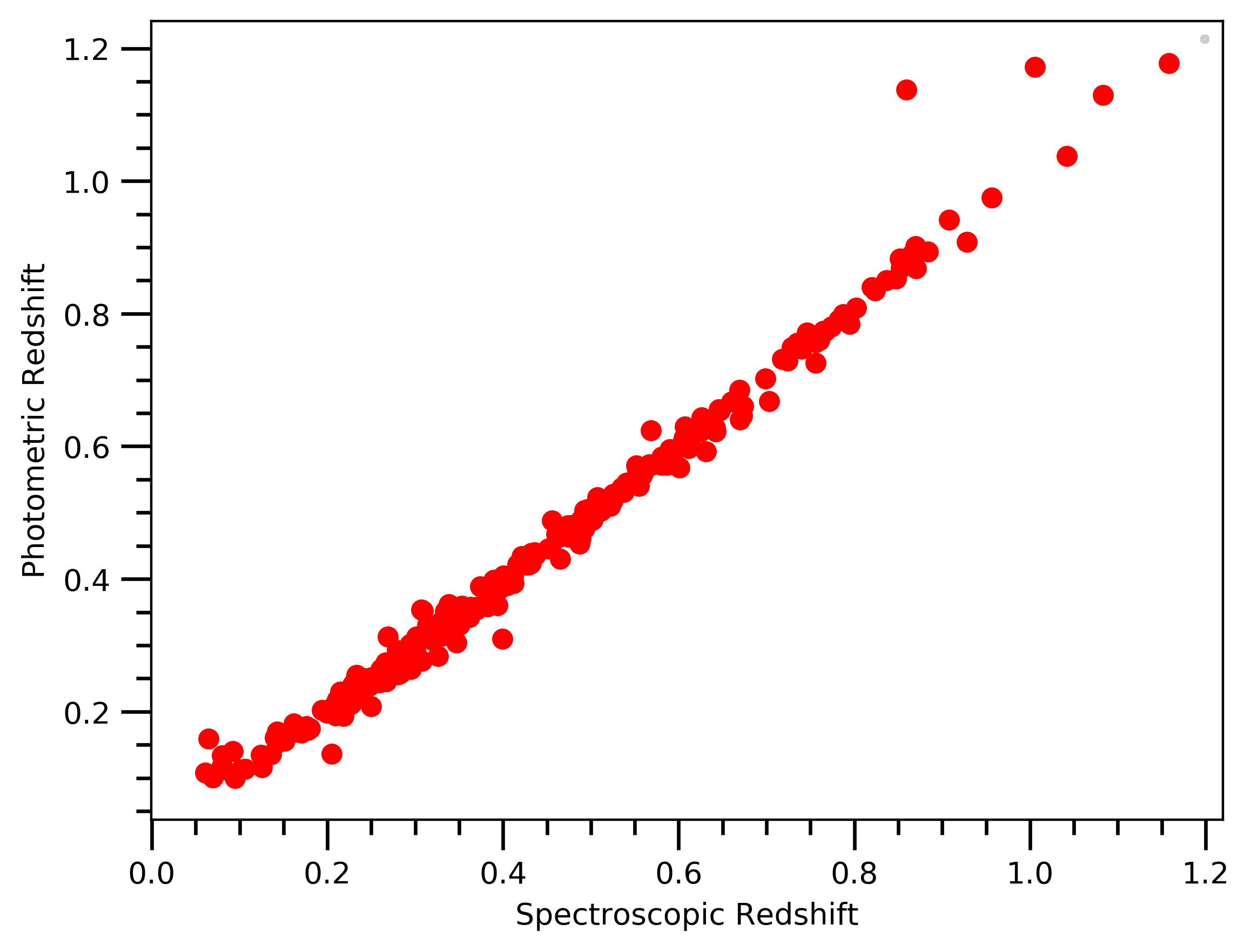}}
\caption{Spectrosopic vs photometric redshift of cluster galaxies for which spectroscopic data is readily available in the HSC SSP catalog.}
\label{fig:zcl}
\end{figure}

In order to analyse the evolution of the UV colors of galaxies, we made use of data from a number of surveys - primarily the Subaru Hyper Suprime-Cam Subaru Strategic Program (HSC SSP) and the CFHT Large Area U-band Deep Survey (CLAUDS\footnote{www.clauds.net}). The HSC SSP is a $grizy$ (plus several narrow-band) imaging survey consisting of three layers - wide, deep and ultra-deep, covering 1400, 26 and 3.5 deg$^2$ and reaching r-band $5\sigma$ point source depths of roughly 26, 27 and 28 respectively. The wide layer targets the North field, the Spring Equatorial field and the Fall Equatorial field. The deep layer targets the XMM-LSS, E-COSMOS, ELAIS-N1 and DEEP2-3 fields each with four pointings (apart from XMM-LSS which has 3), while the ultra-deep layer adds an extra pointing to the XMM-LSS field and a fifth overlapping pointing to E-COSMOS, dubbed `SXDS' and `COSMOS' respectively. For the purposes of this paper, we made use of the data from the deep fields, which already have a slew of existing multi-wavelength imaging and spectroscopy datasets in various archives. CLAUDS is a partner survey to the HSC-SSP, adding $u$-band data using the CFHT Megacam $u$ and $u^*$ filters (with central wavelengths at 3538\AA\ and 3743\AA\ respectively). Similar to the HSC-SSP, CLAUDS also consists of a deep and ultra-deep layer that target the exact same fields as the aforementioned survey. Though CLAUDS coverage largely overlaps with HSC-SSP, the deep and ultra-deep fields span somewhat smaller areas of 18.6 and 1.36 deg$^2$, reaching $u$-band 5$\sigma$ (2'' aperture) depths of 27.1 and 27.7 respectively. The ELAIS N1 and DEEP2-3 fields were imaged with the Megacam $u$ filter, while the XMM-LSS was imaged with the $u^*$ filter, and observations of the E-COSMOS field include both filters, which were combined in separate $u$ and $u^*$ stacks.

The clusters used in the analysis and their proposed member galaxies are drawn from the CAMIRA cluster catalog, which uses the CAMIRA red sequence finding algorithm (see \citealt{oguri2014,oguri2018} for details). The most up to date catalogs for the deep/ultra-deep (DUD hereafter) layers\footnote{The latest publicly available catalog is available at $https://github.com/oguri/cluster\_catalogs/tree/main/hsc\_s20a\_camira$} are derived from Data Release 4, which uses a similar setup to the cluster catalogs from Data Release 1 \citep{oguri2018}. The catalogs contain the photometric redshift of each cluster and when available the spectroscopic redshift of the brightest cluster galaxy (BCG). Generally for clusters with confirmed spec-z, the photo-z is within $\pm 0.1$ of the spec-z (Fig. \ref{fig:zcl}) and as such we use the photo-z for the rest of the clusters when necessary for our analysis given that we are only dealing with broadband data and do not require highly precise redshift measurements. Crucially, the catalogs also provide a list of proposed member galaxies for each cluster, which are the main targets of our study.

Photometric data was taken from the internal Data Release 4 (Data Release 3 is currently publicly available). The $u$-band (both MegaCam $u$ and $u^*$) data was taken from the HSC-joint catalog that is part of Data Release 2. For each cluster galaxy in the DUD catalog we extract the $ugrizy$ cmodel magnitudes (a measure of the total flux), magnitudes in fixed 1.0/2.0/3.0/4.0/5.7" diameter apertures and their associated errors.

\begin{figure*}
\centering
{\includegraphics[width=0.325\textwidth]{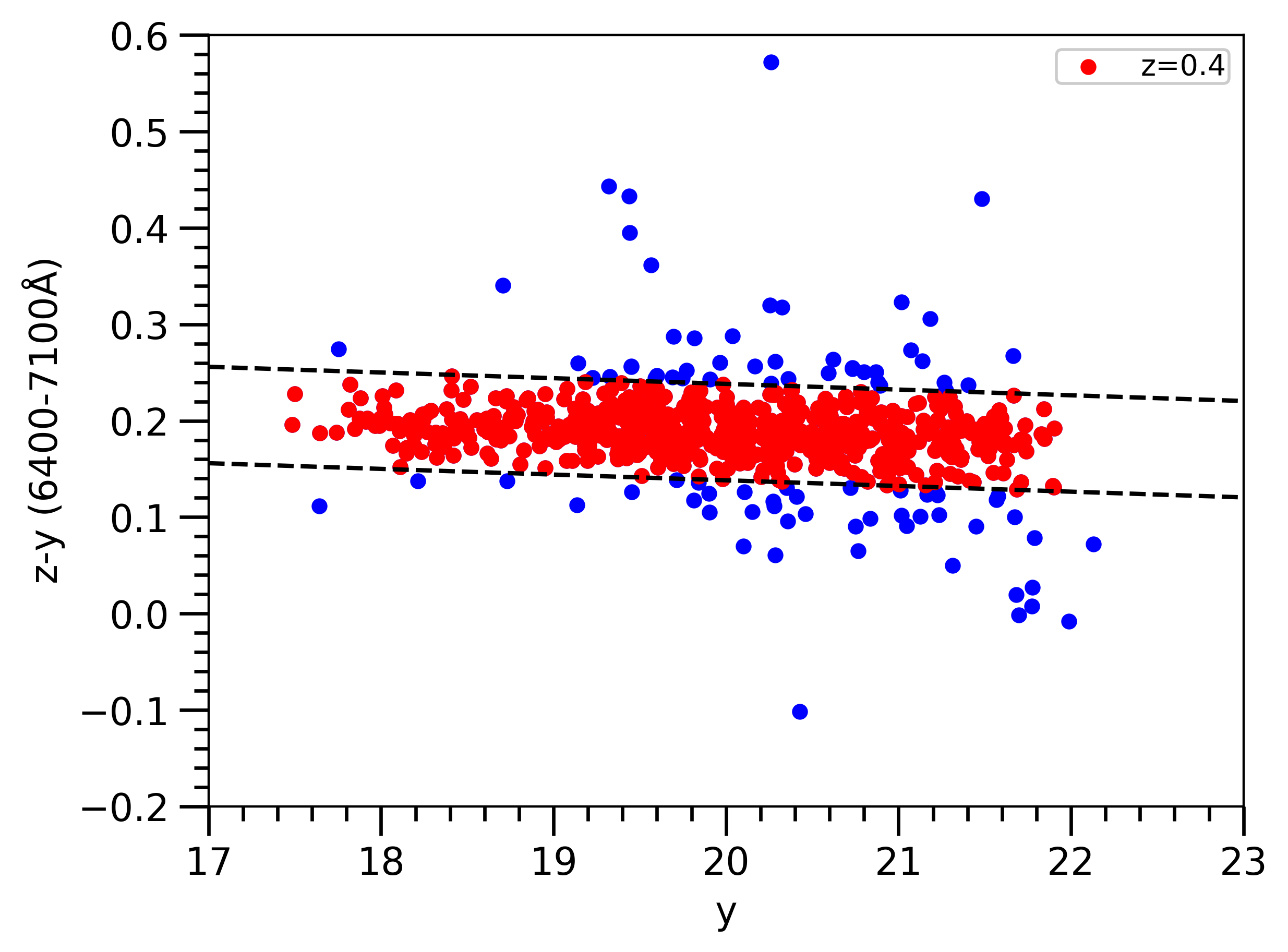}}
{\includegraphics[width=0.325\textwidth]{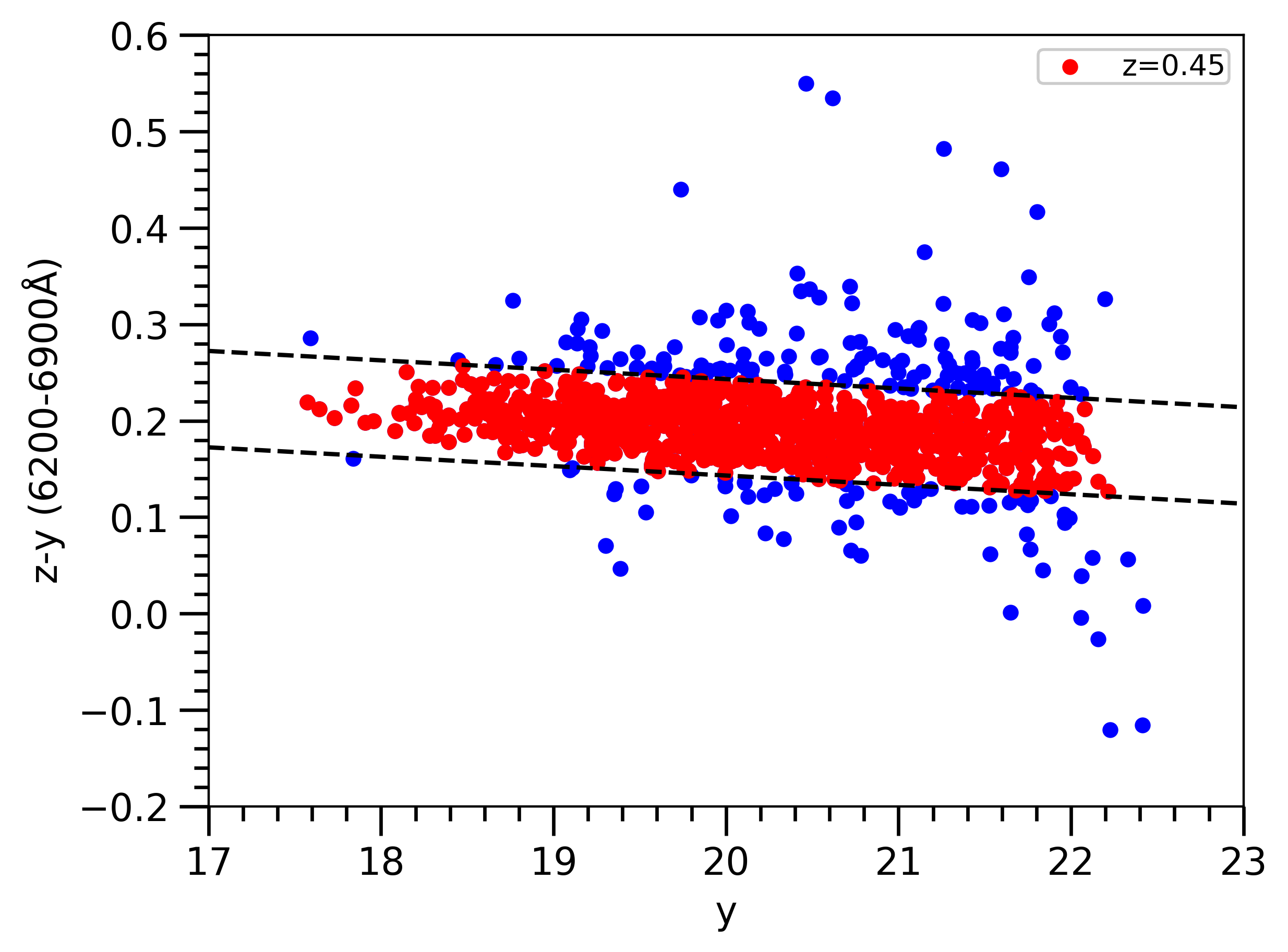}}
{\includegraphics[width=0.325\textwidth]{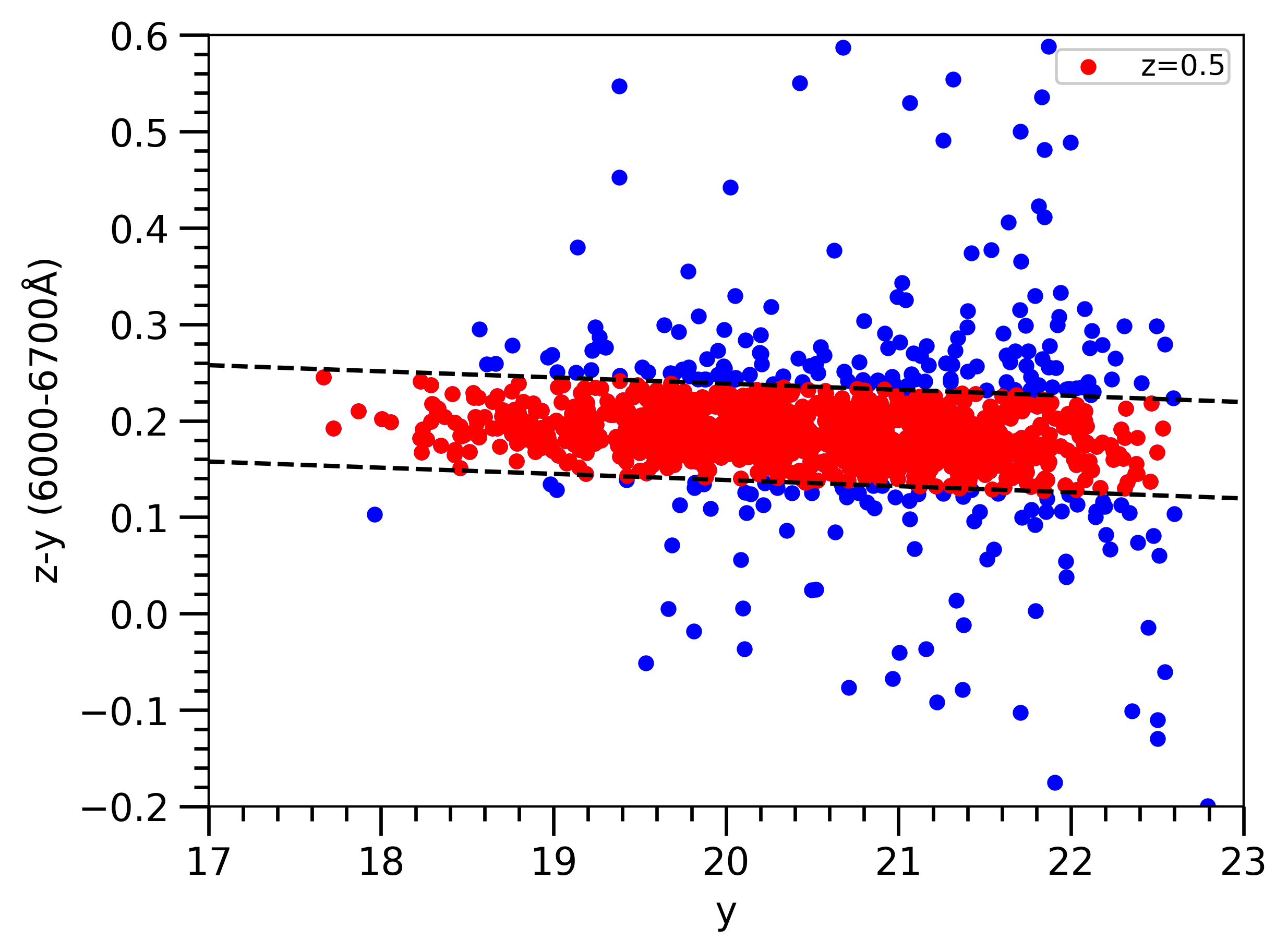}}
{\includegraphics[width=0.325\textwidth]{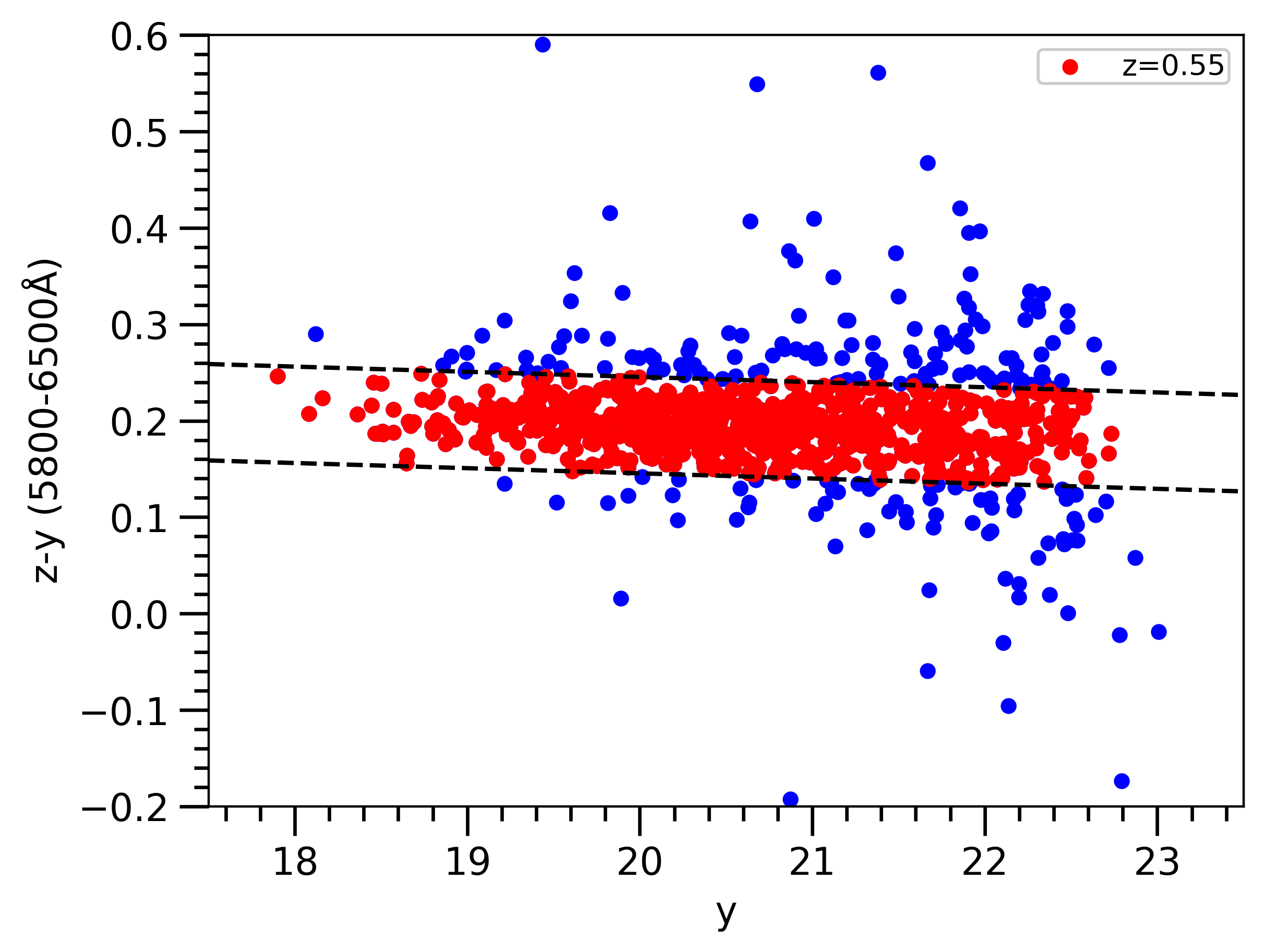}}
{\includegraphics[width=0.325\textwidth]{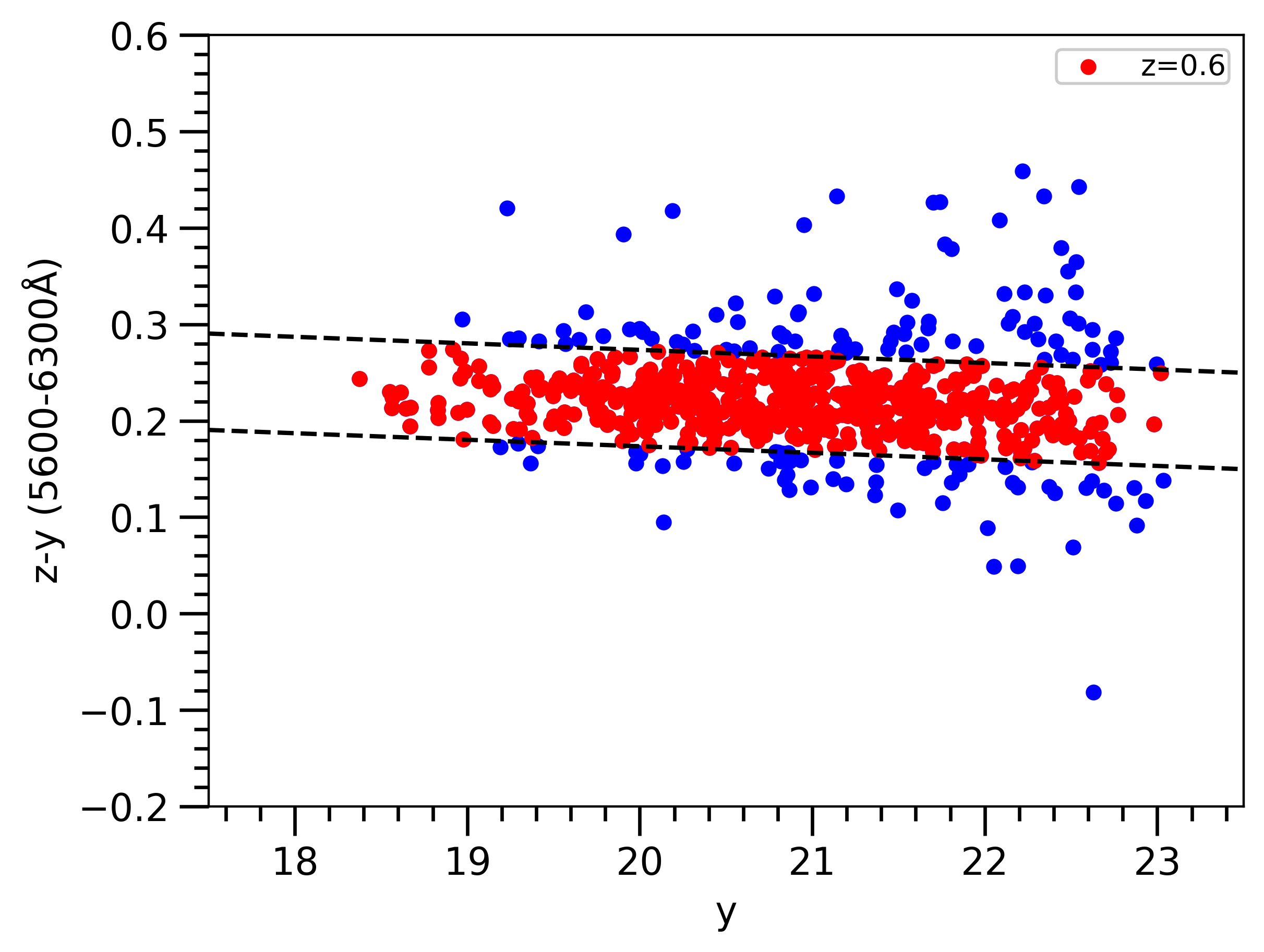}}
{\includegraphics[width=0.325\textwidth]{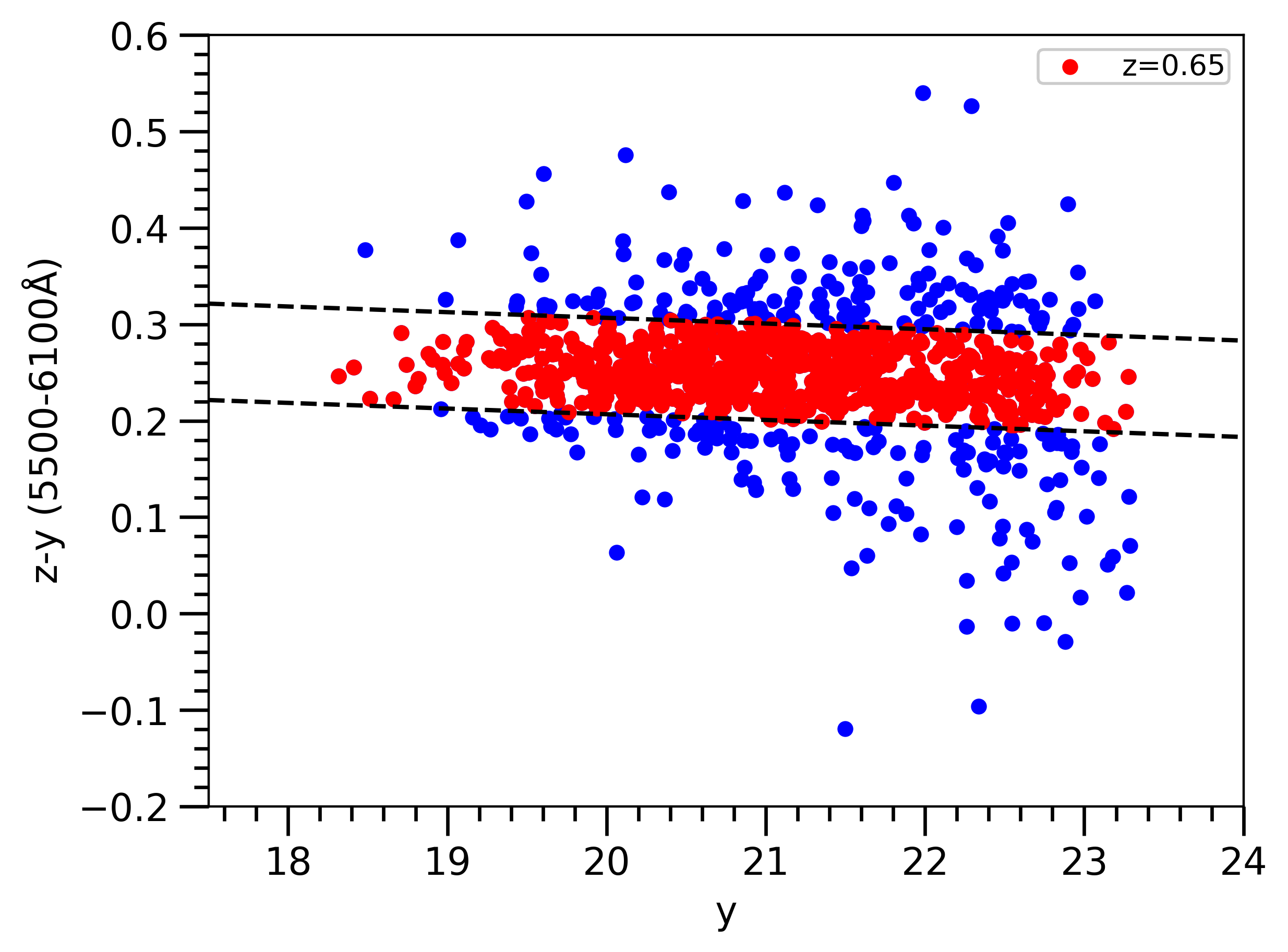}}
{\includegraphics[width=0.325\textwidth]{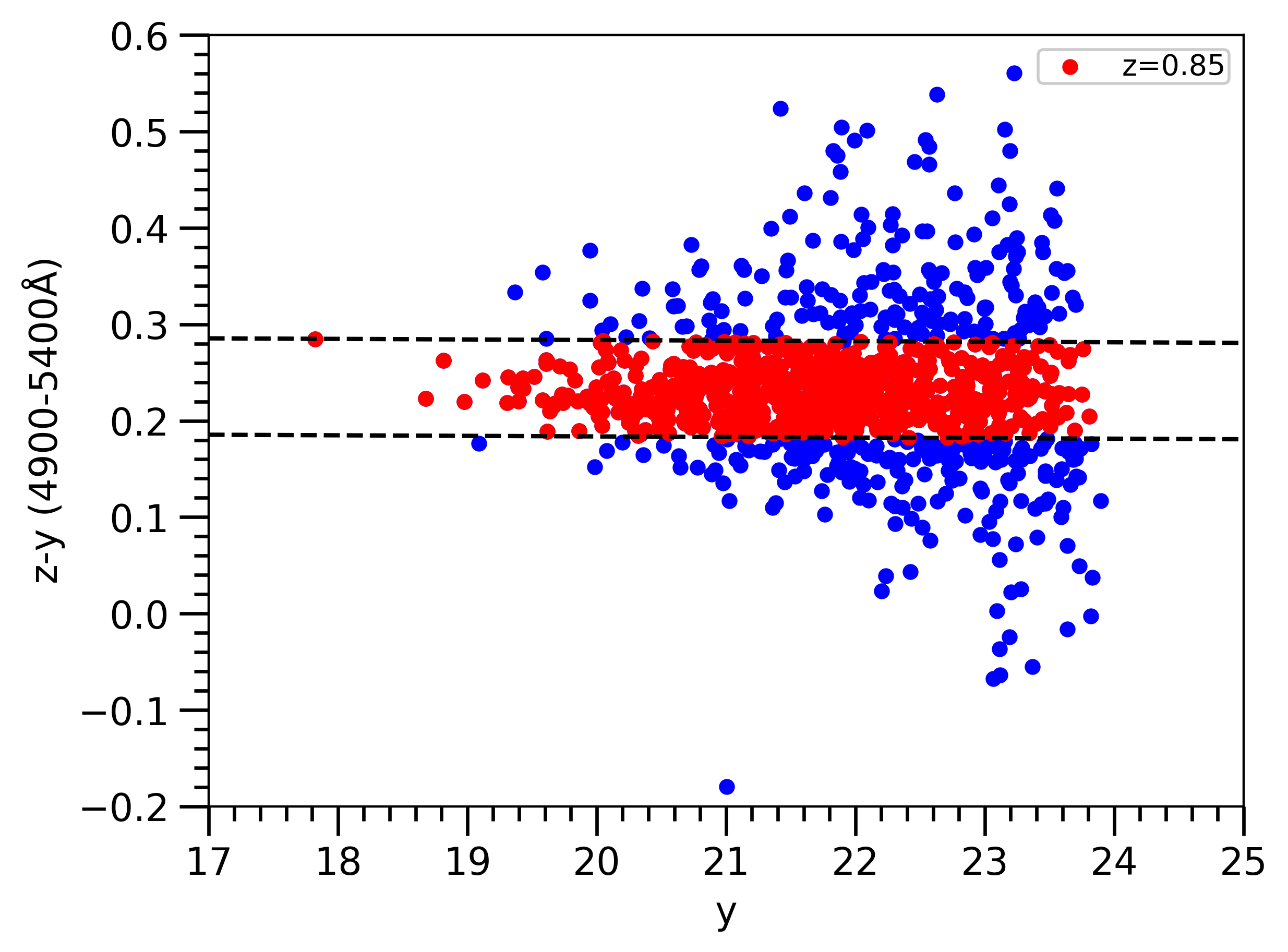}}
{\includegraphics[width=0.325\textwidth]{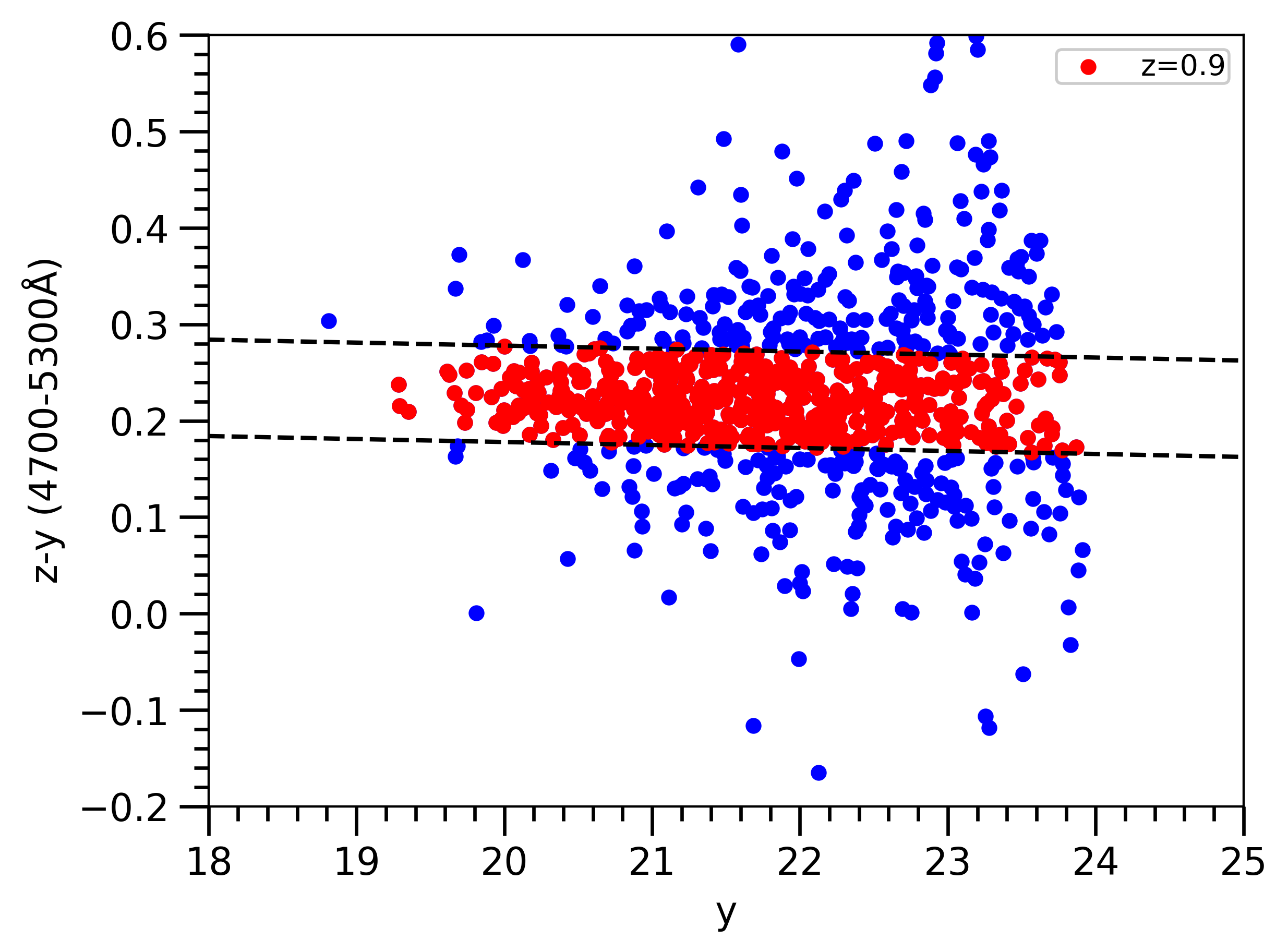}}
{\includegraphics[width=0.325\textwidth]{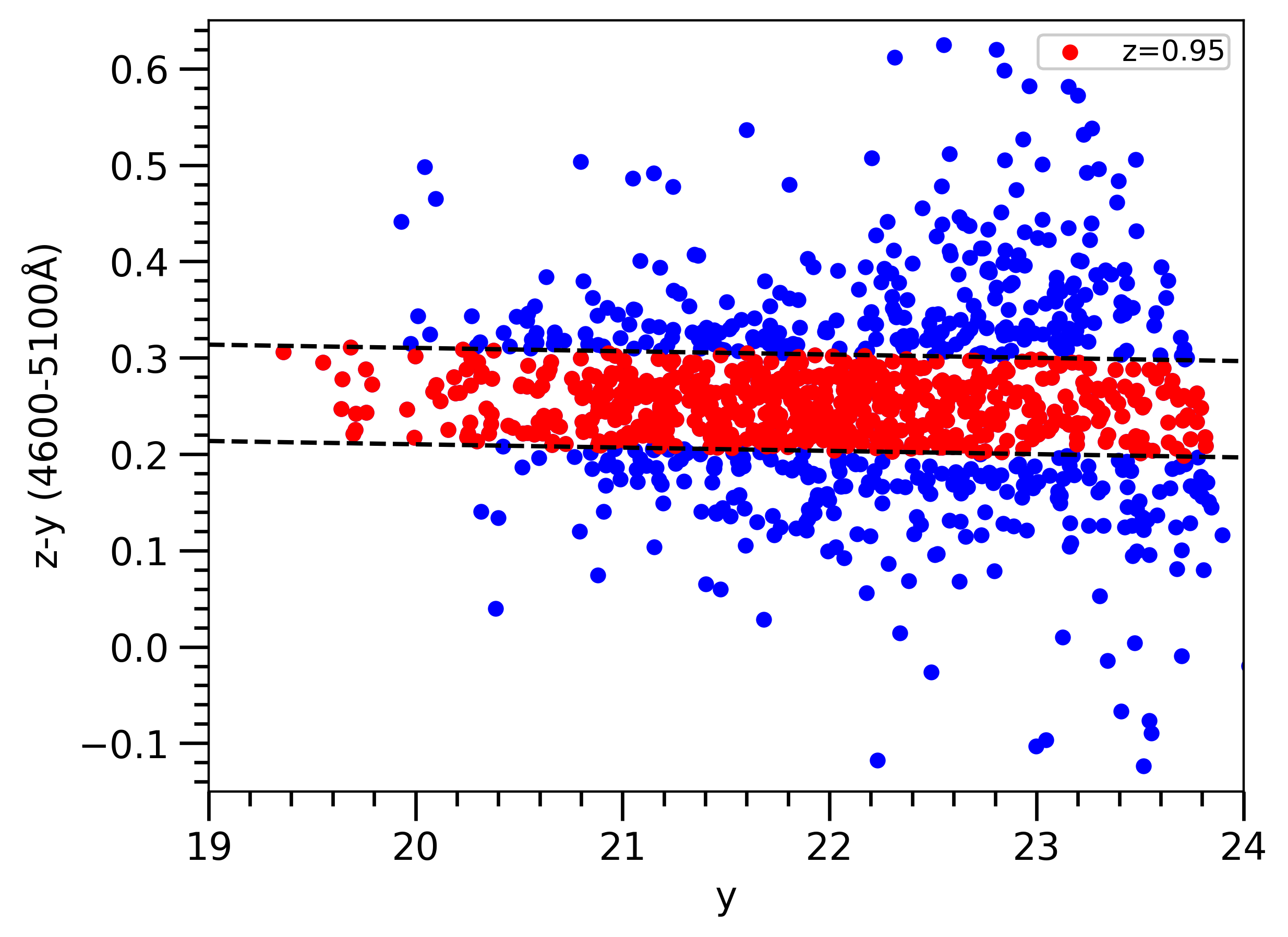}}
{\includegraphics[width=0.325\textwidth]{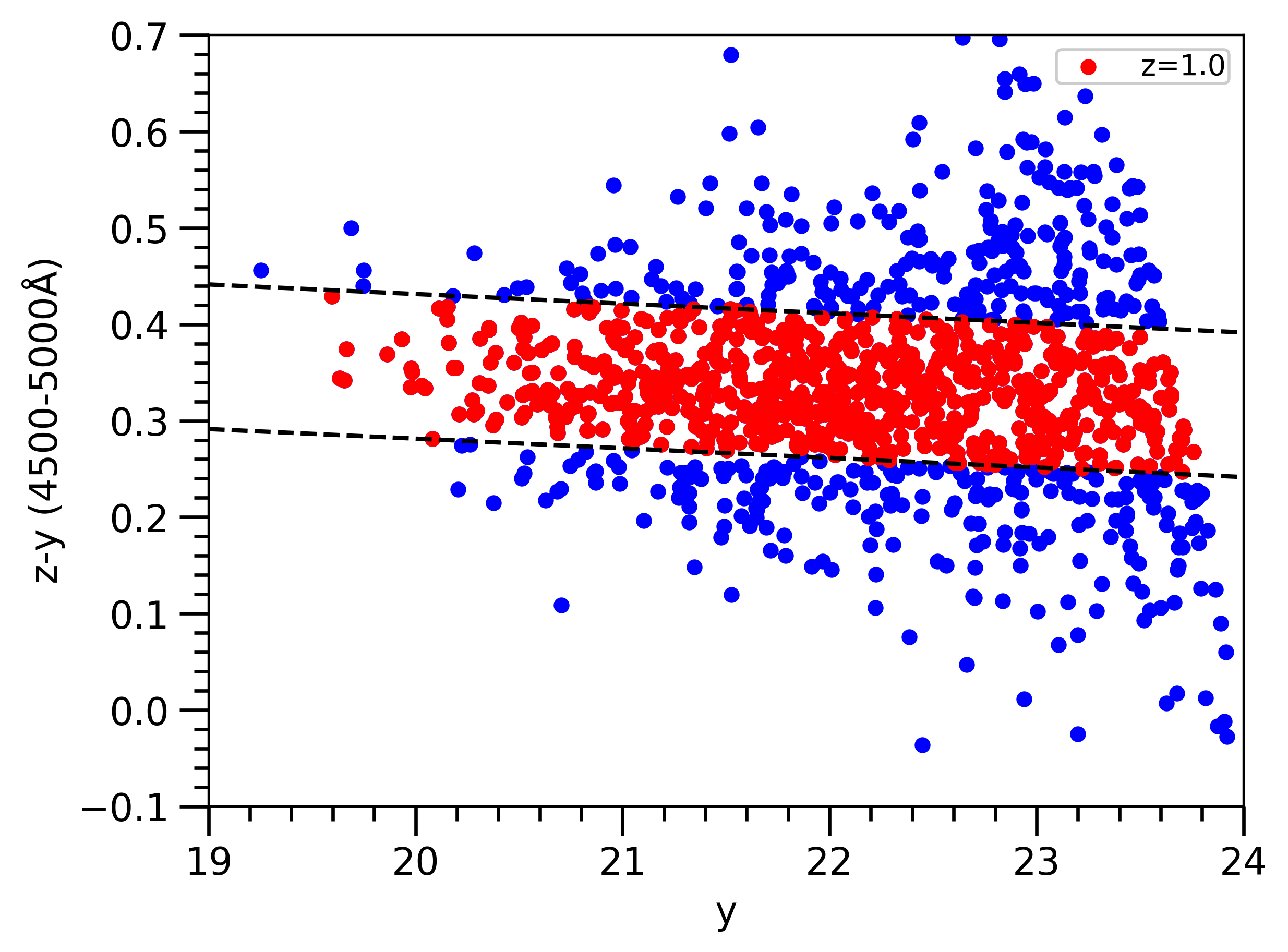}}
{\includegraphics[width=0.325\textwidth]{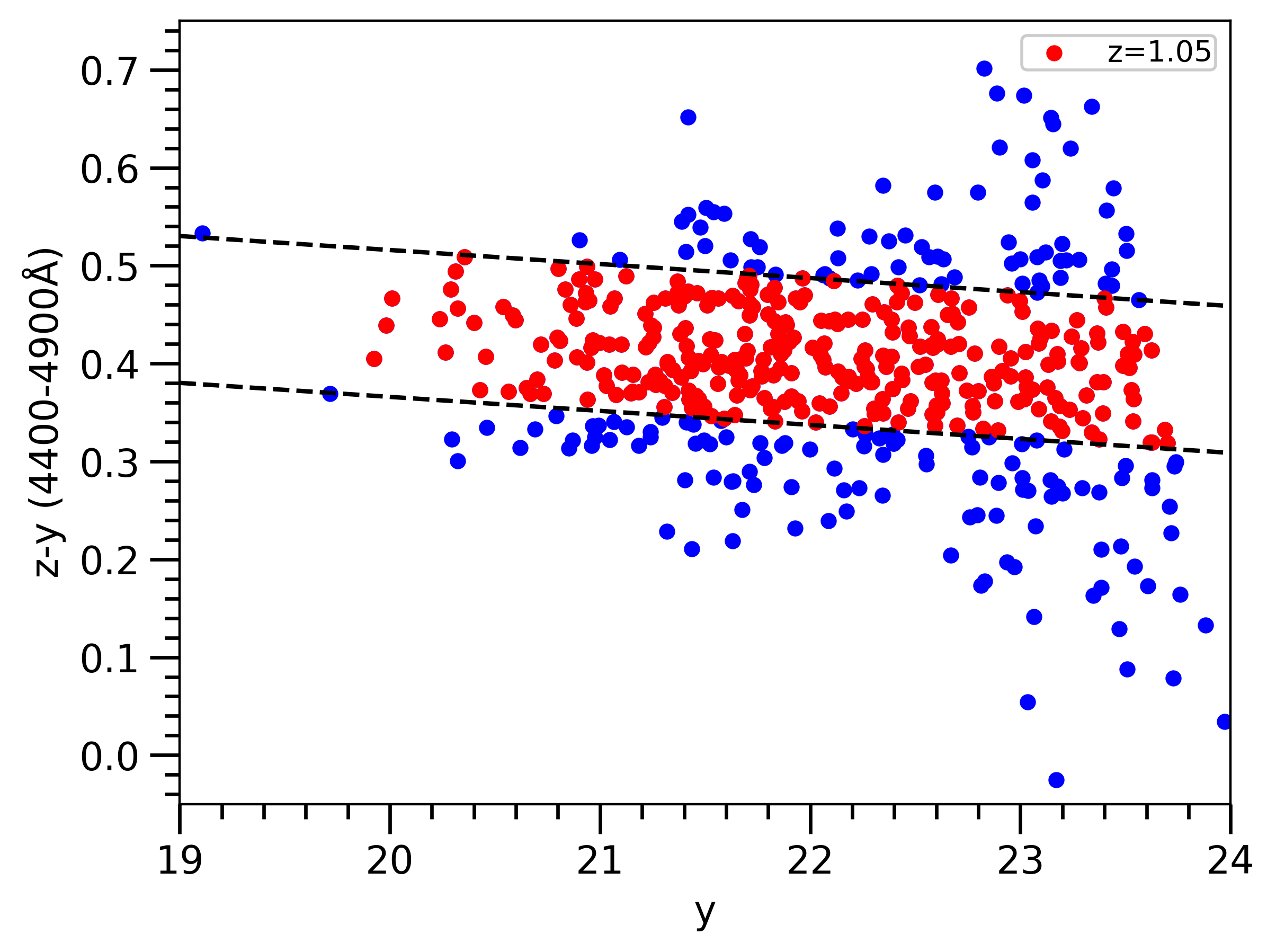}}
{\includegraphics[width=0.325\textwidth]{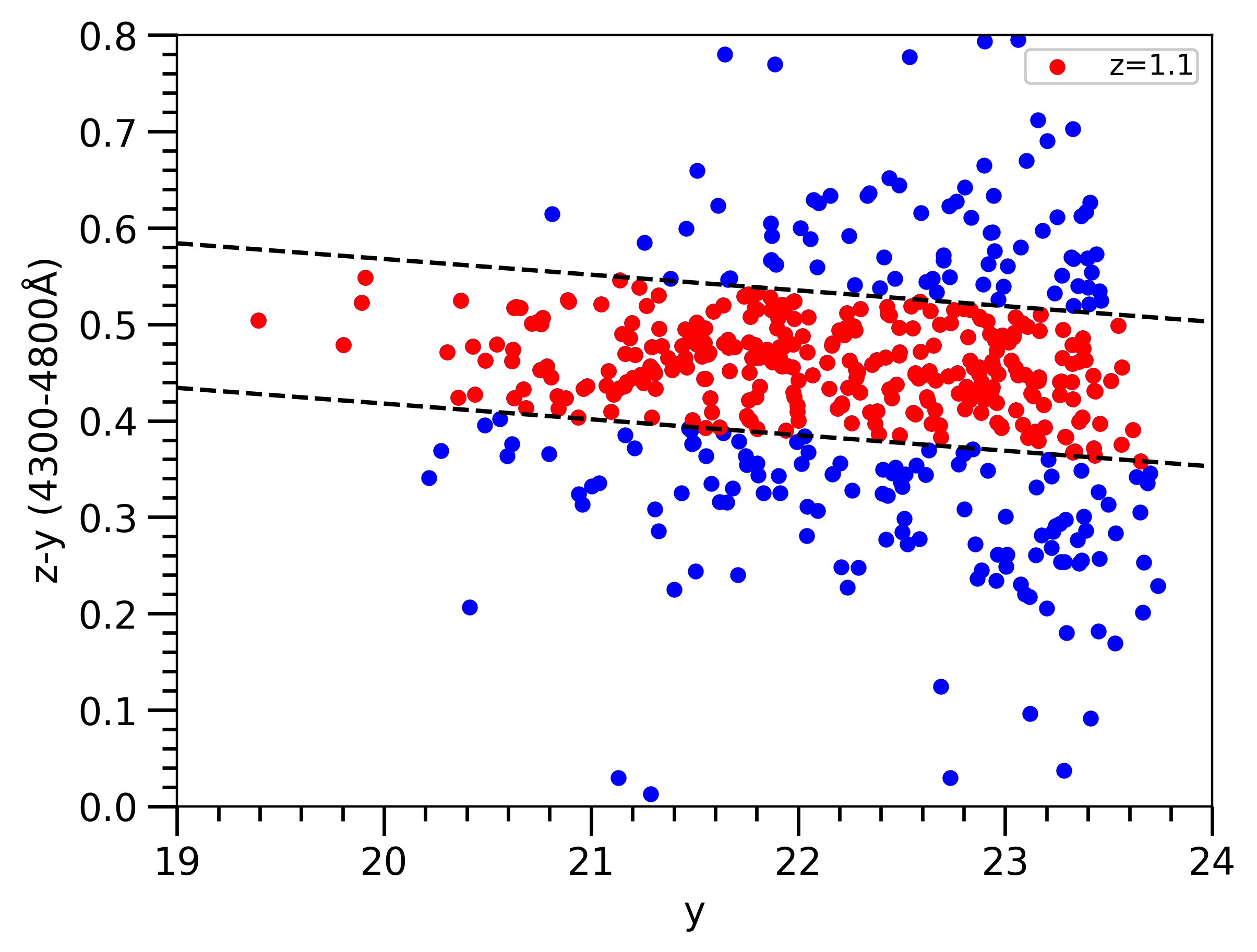}}
\caption{Observed $z-y$ (rest-frame optical as noted in brackets) color-magnitude diagrams of cluster galaxies between $z=0.4-1.1$ from the HSC SSP survey, separated into bins of $0.05$ in redshift. The red sequence galaxies are denoted with the red filled circles within the dashed lines and have photometric uncertainties of $<0.05$ magnitudes in their optical colors. The dashed lines show the selection region for quiescent cluster ETGs (as described in the text, roughly within $\pm 0.1$ mag. of the mean ridge line for the red sequence).}
\label{fig:op}
\end{figure*}

\begin{figure*}
\centering
{\includegraphics[width=0.325\textwidth]{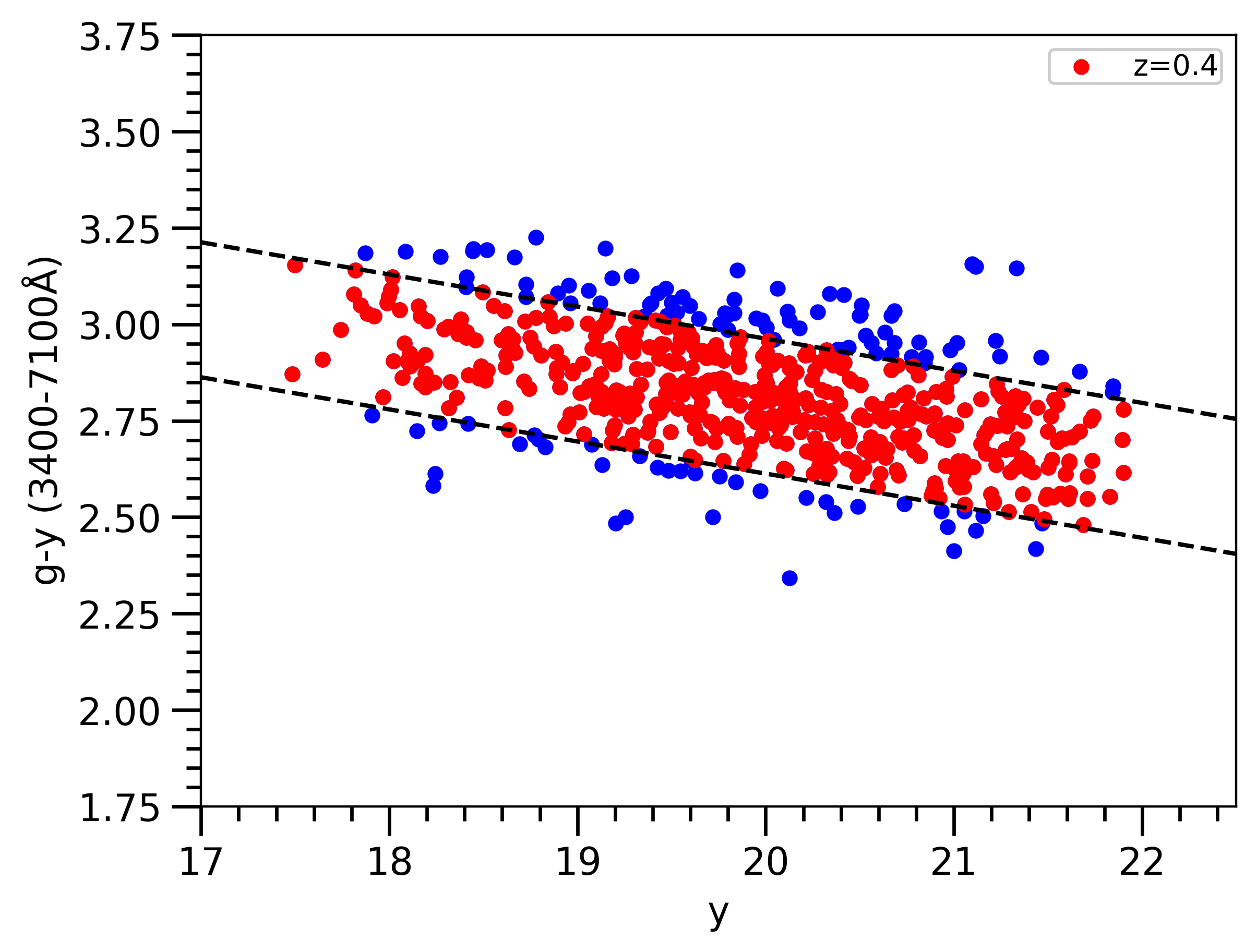}}
{\includegraphics[width=0.325\textwidth]{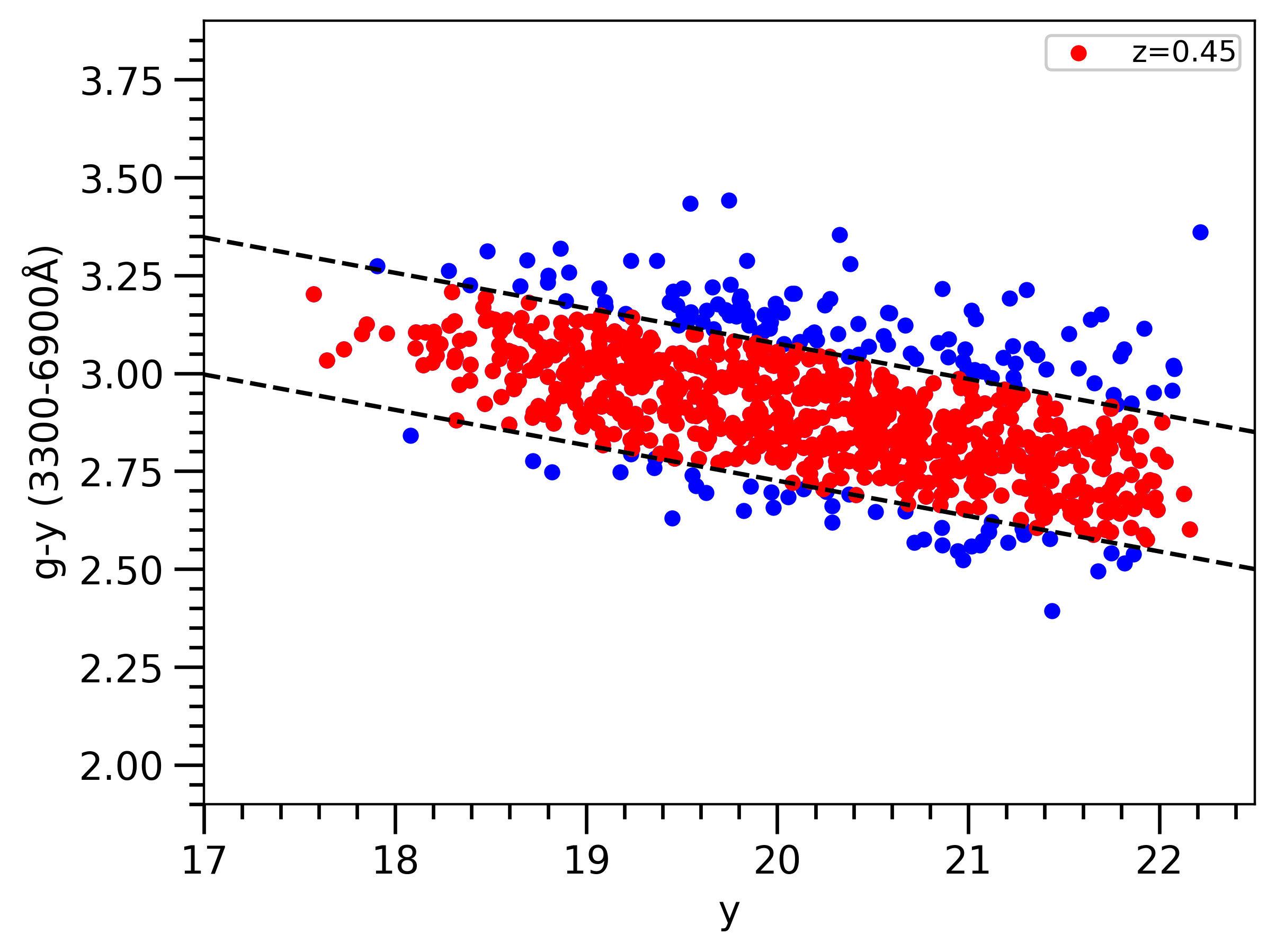}}
{\includegraphics[width=0.325\textwidth]{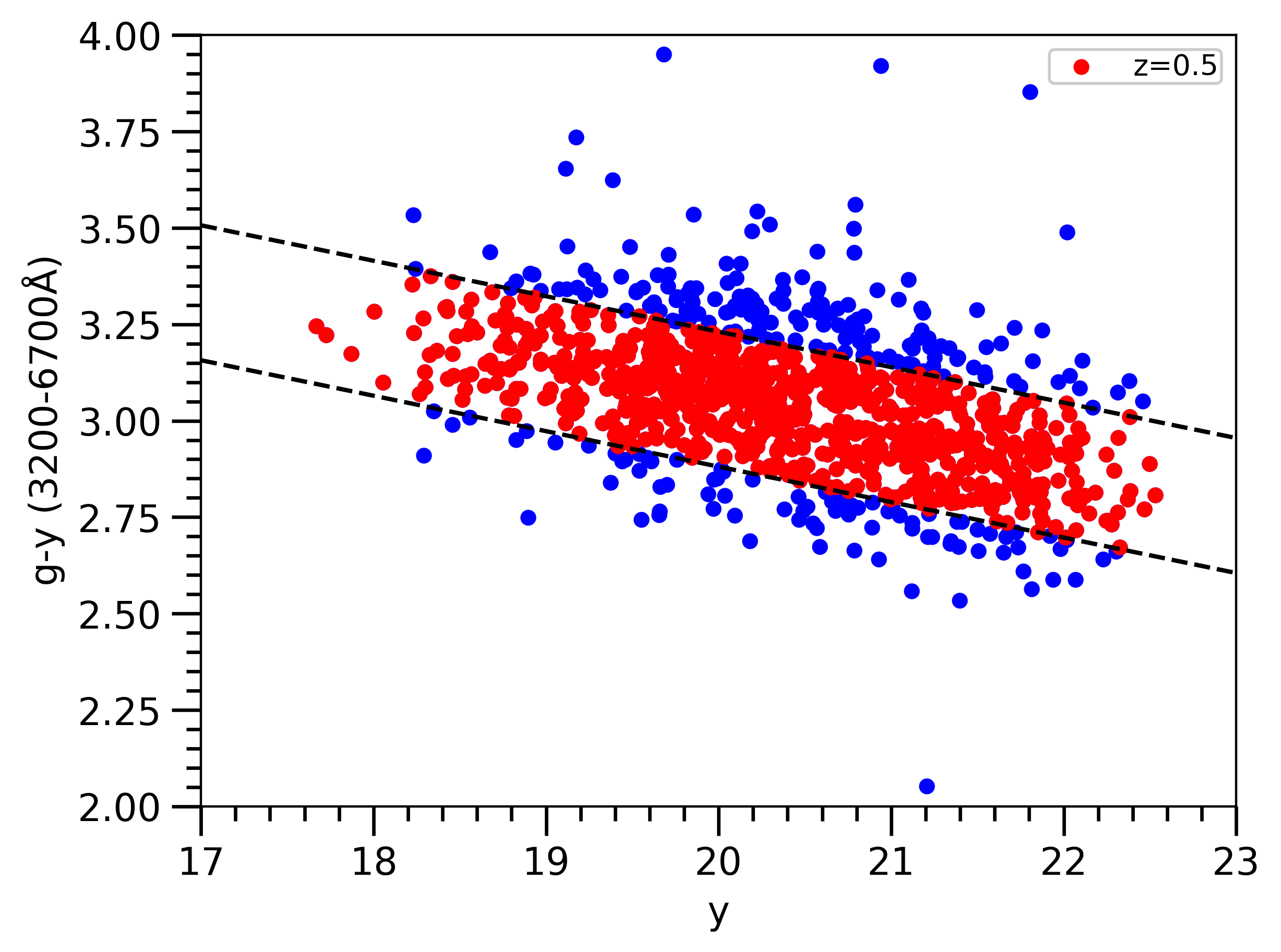}}
{\includegraphics[width=0.325\textwidth]{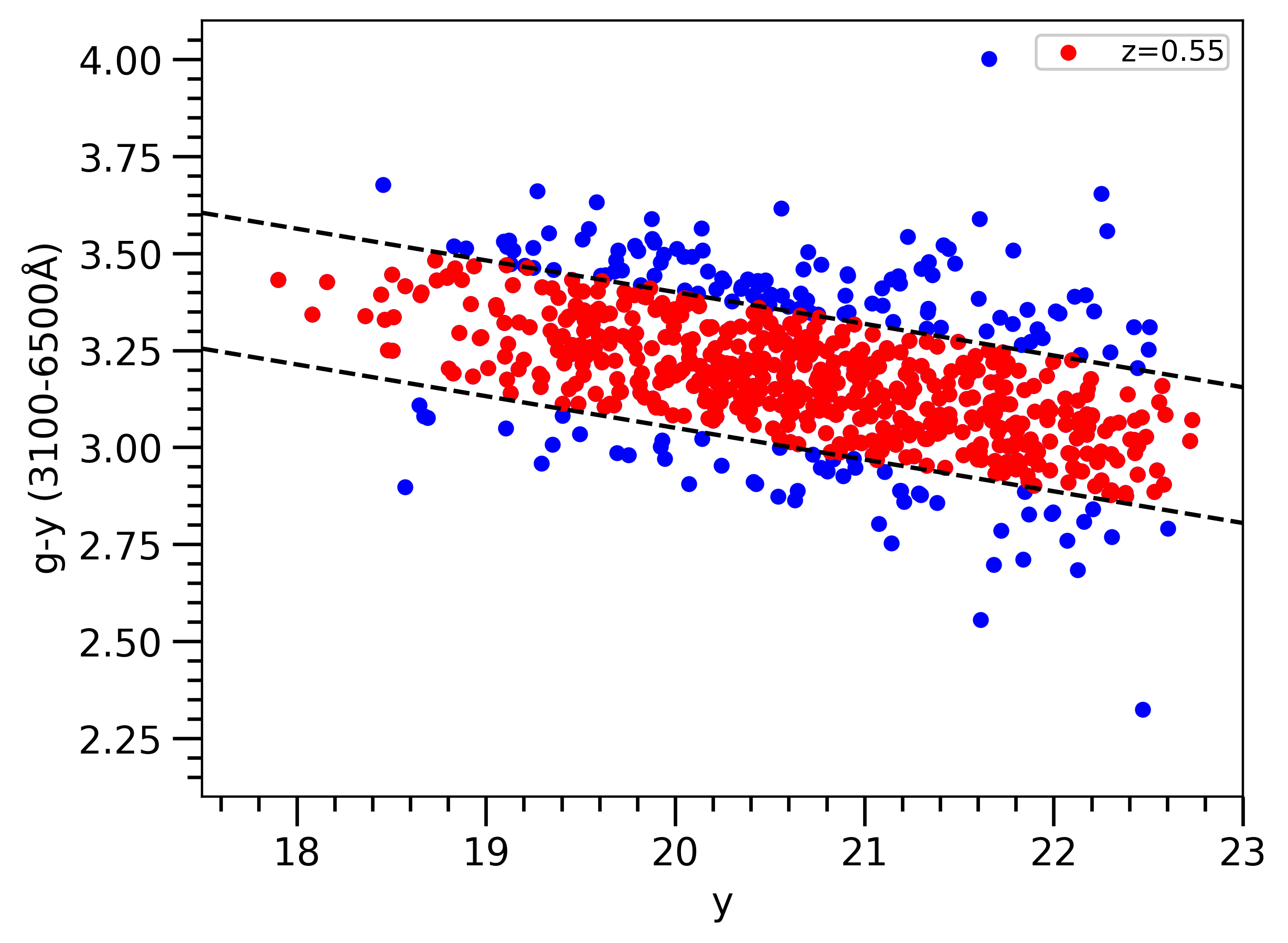}}
{\includegraphics[width=0.325\textwidth]{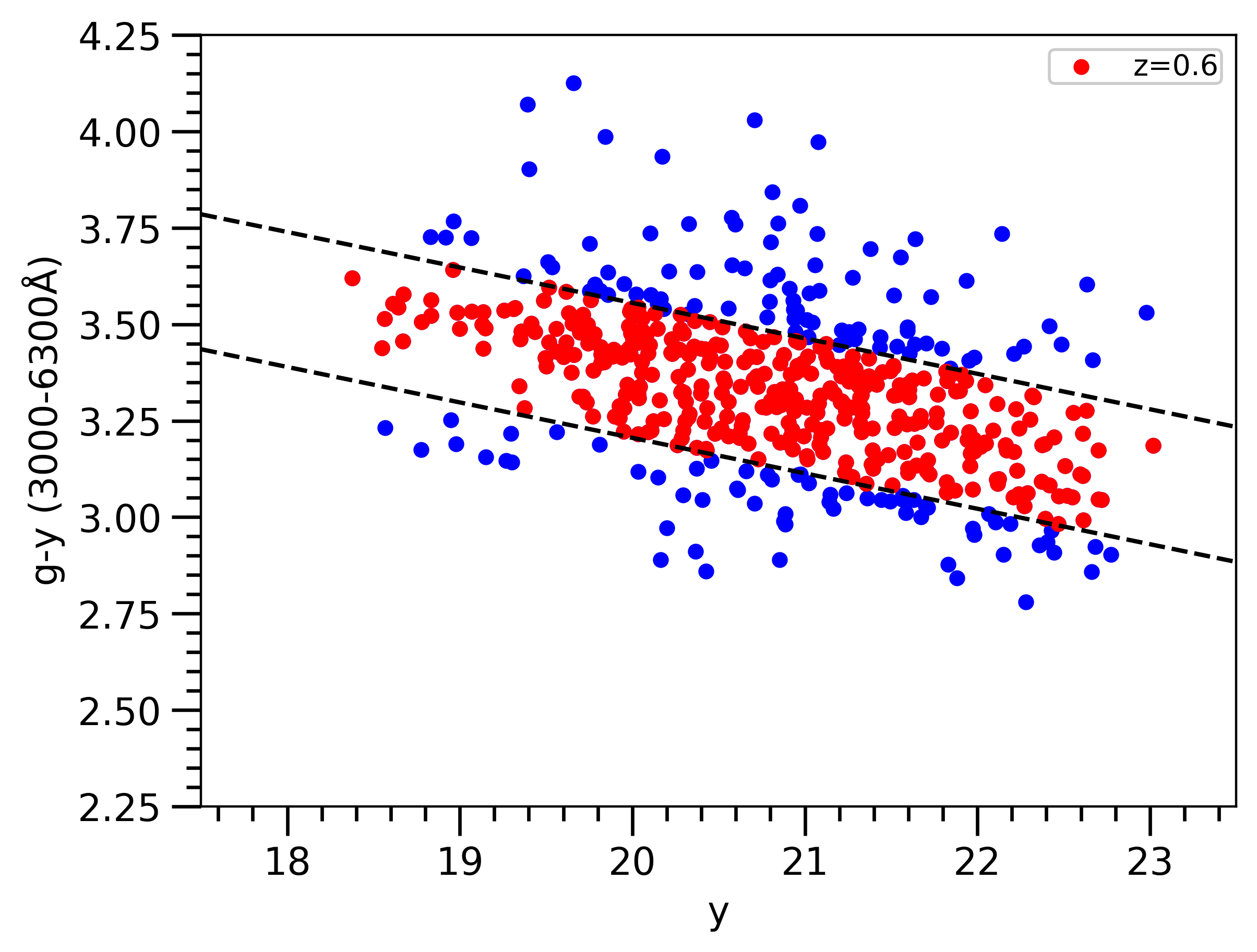}}
{\includegraphics[width=0.325\textwidth]{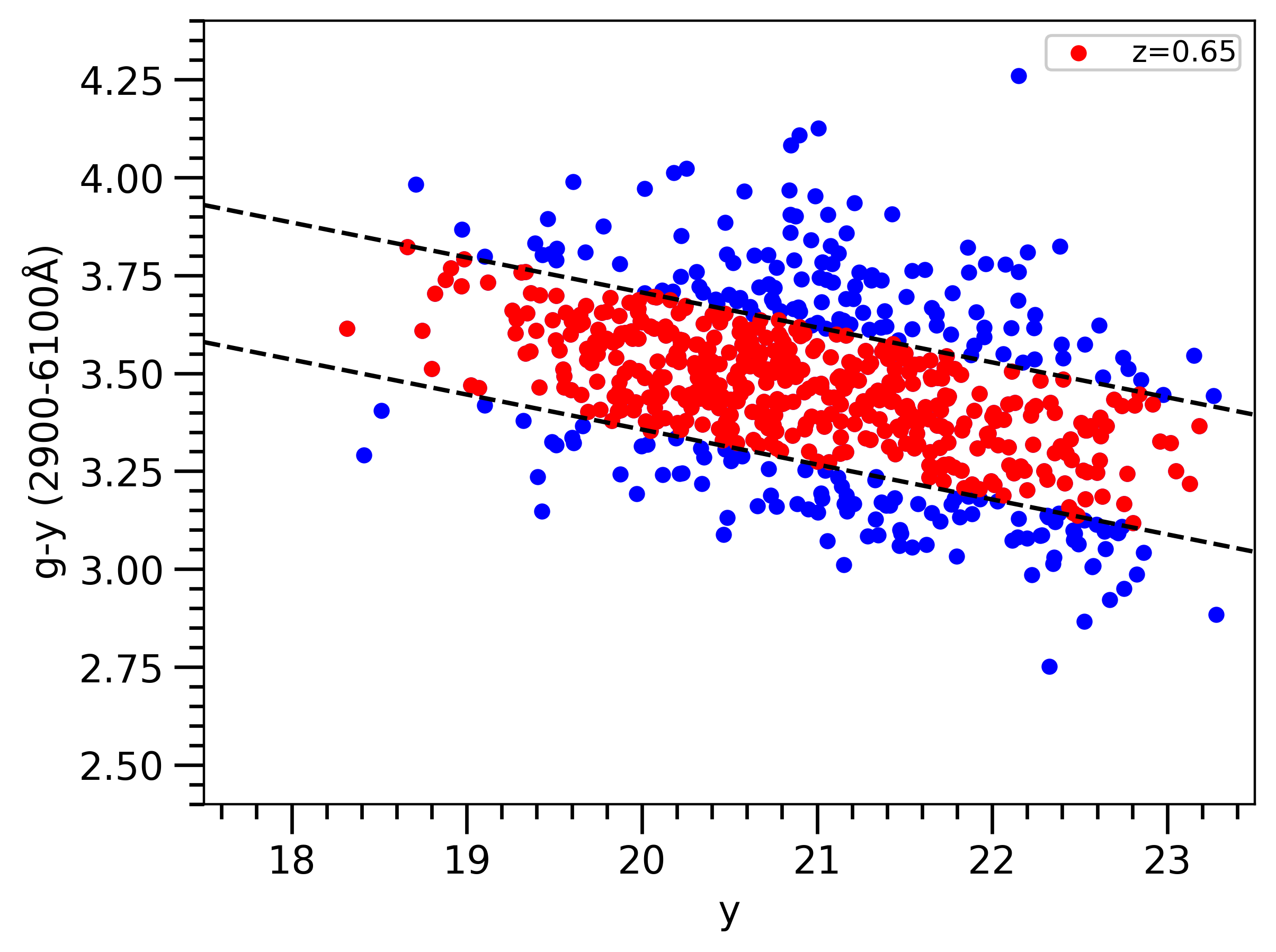}}
{\includegraphics[width=0.325\textwidth]{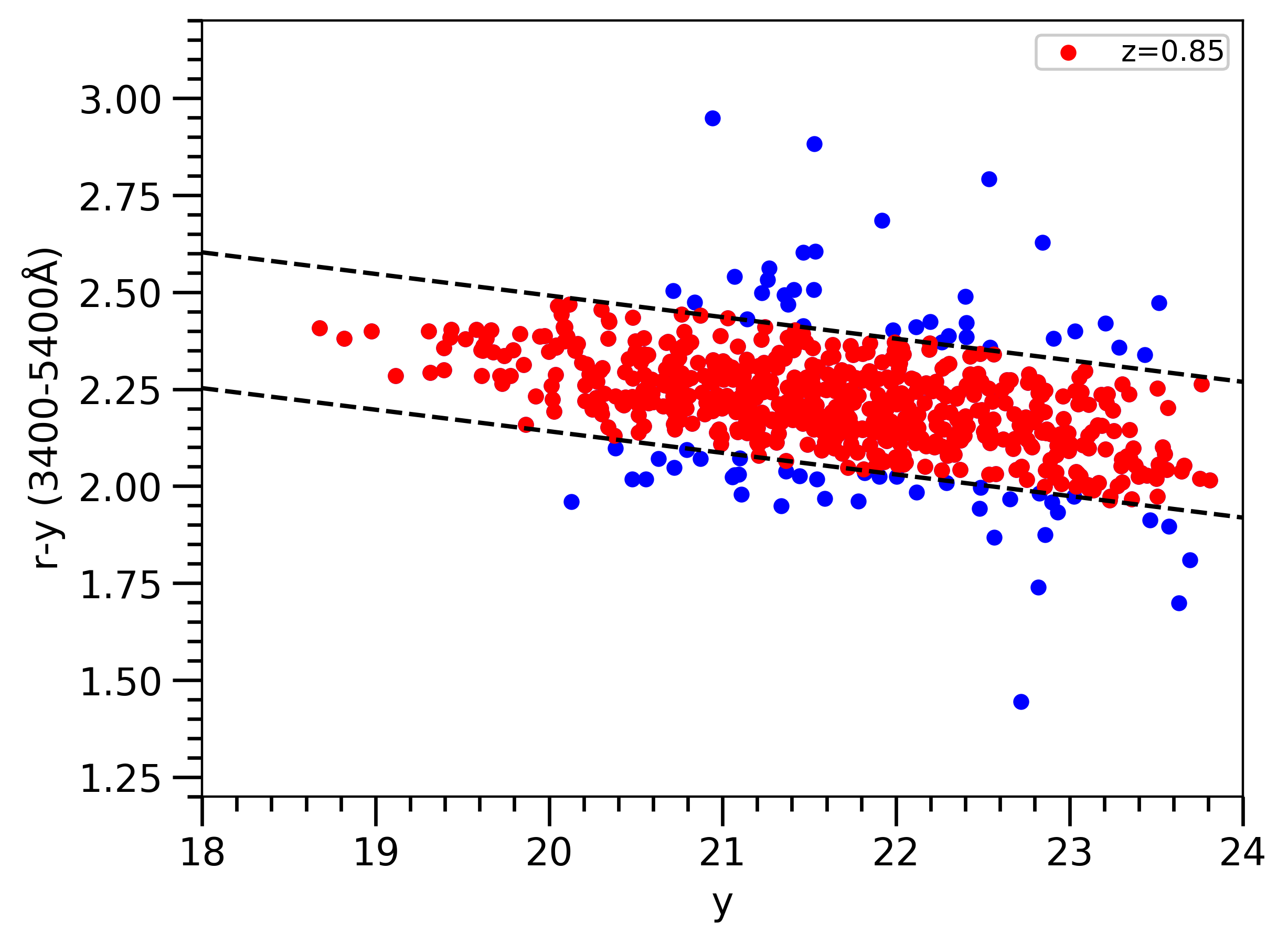}}
{\includegraphics[width=0.325\textwidth]{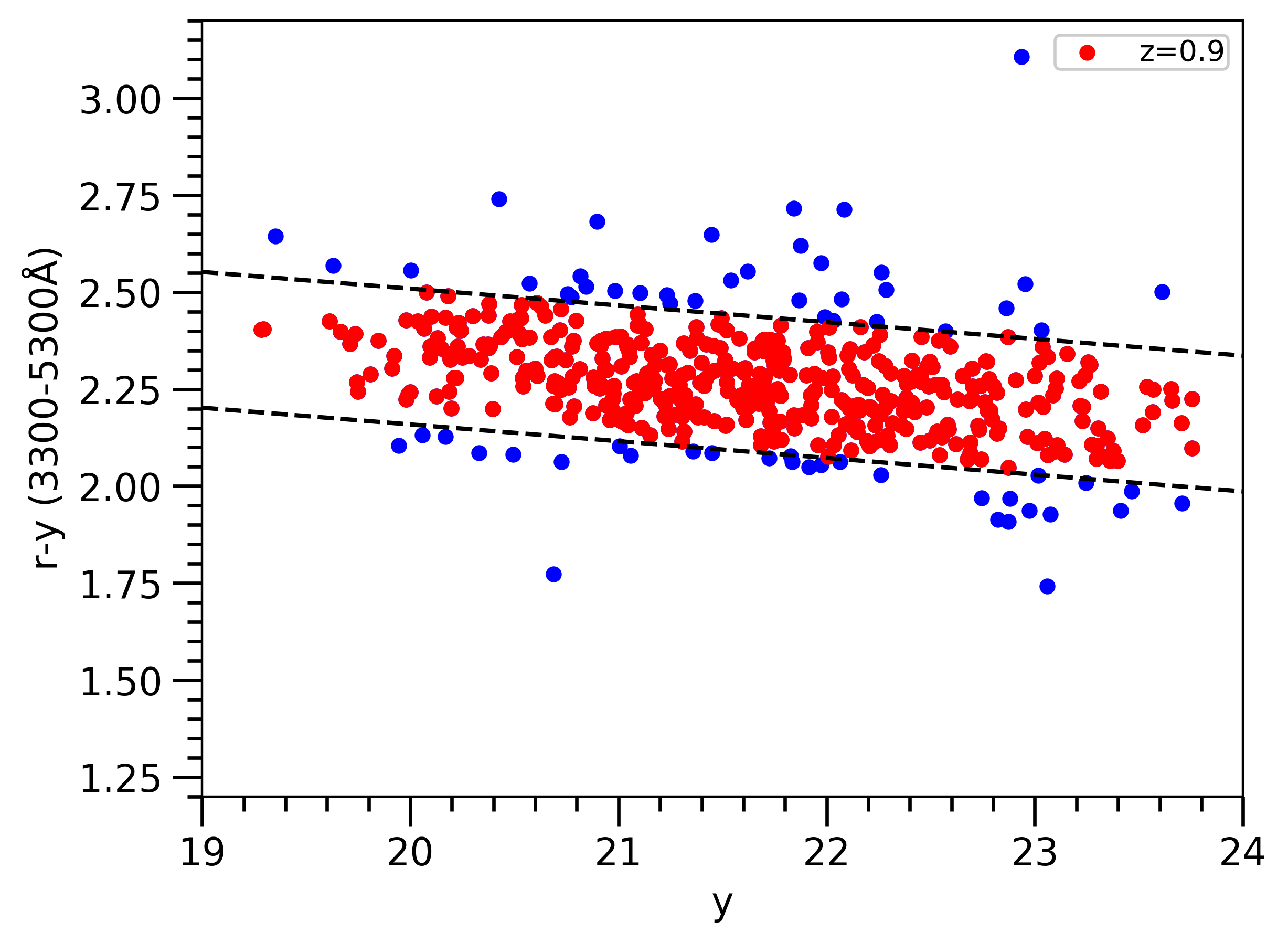}}
{\includegraphics[width=0.325\textwidth]{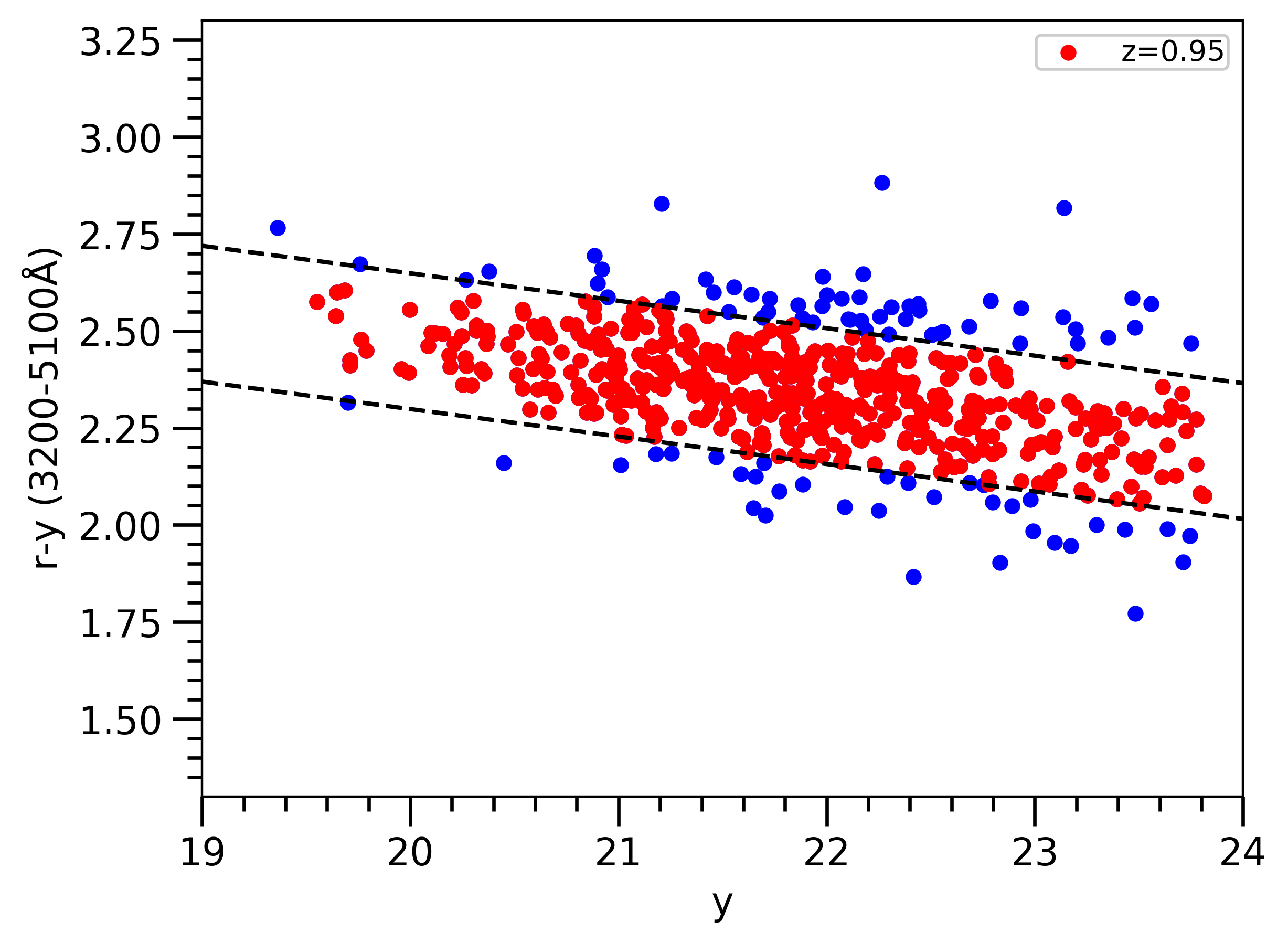}}
{\includegraphics[width=0.325\textwidth]{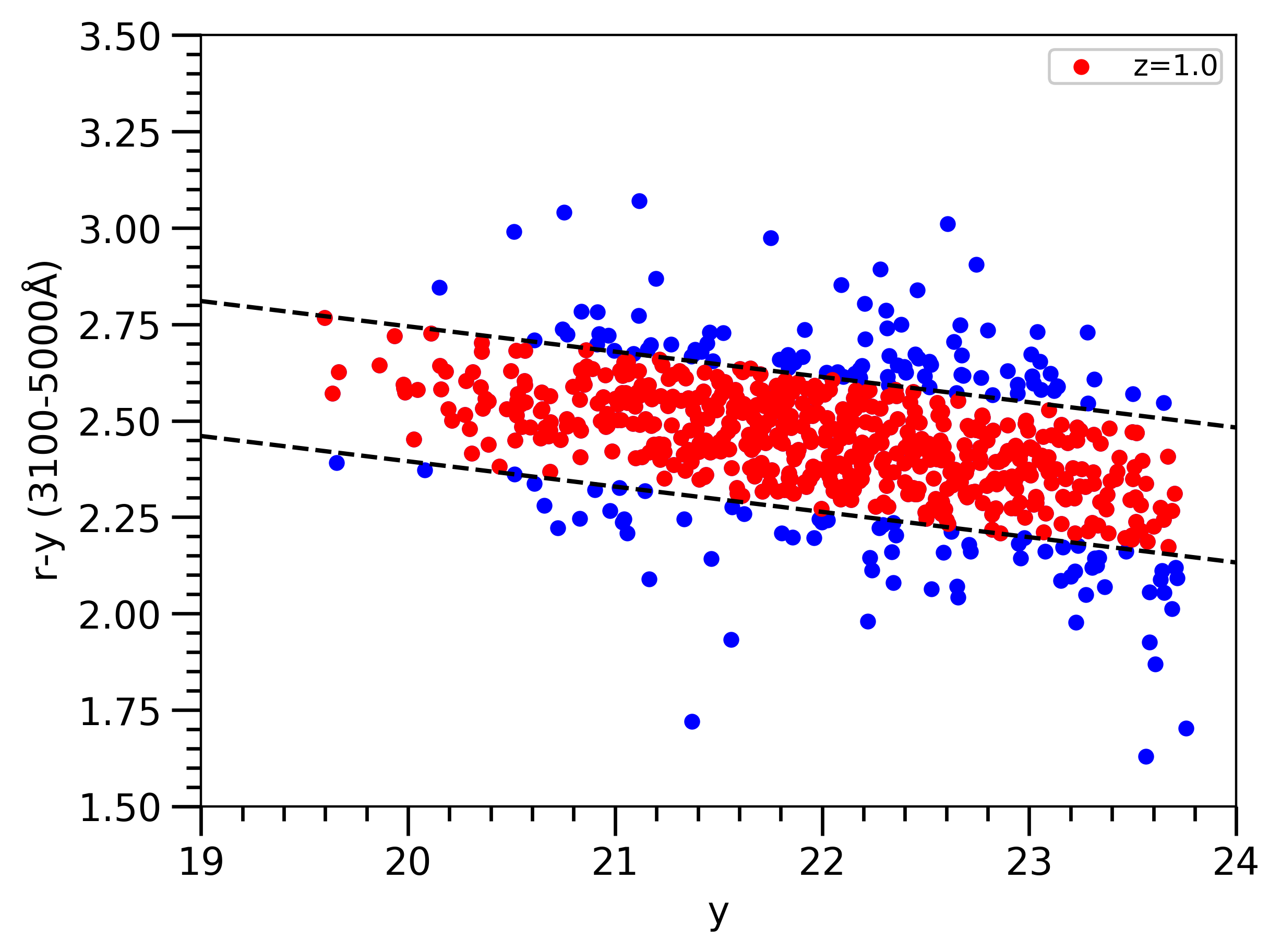}}
{\includegraphics[width=0.325\textwidth]{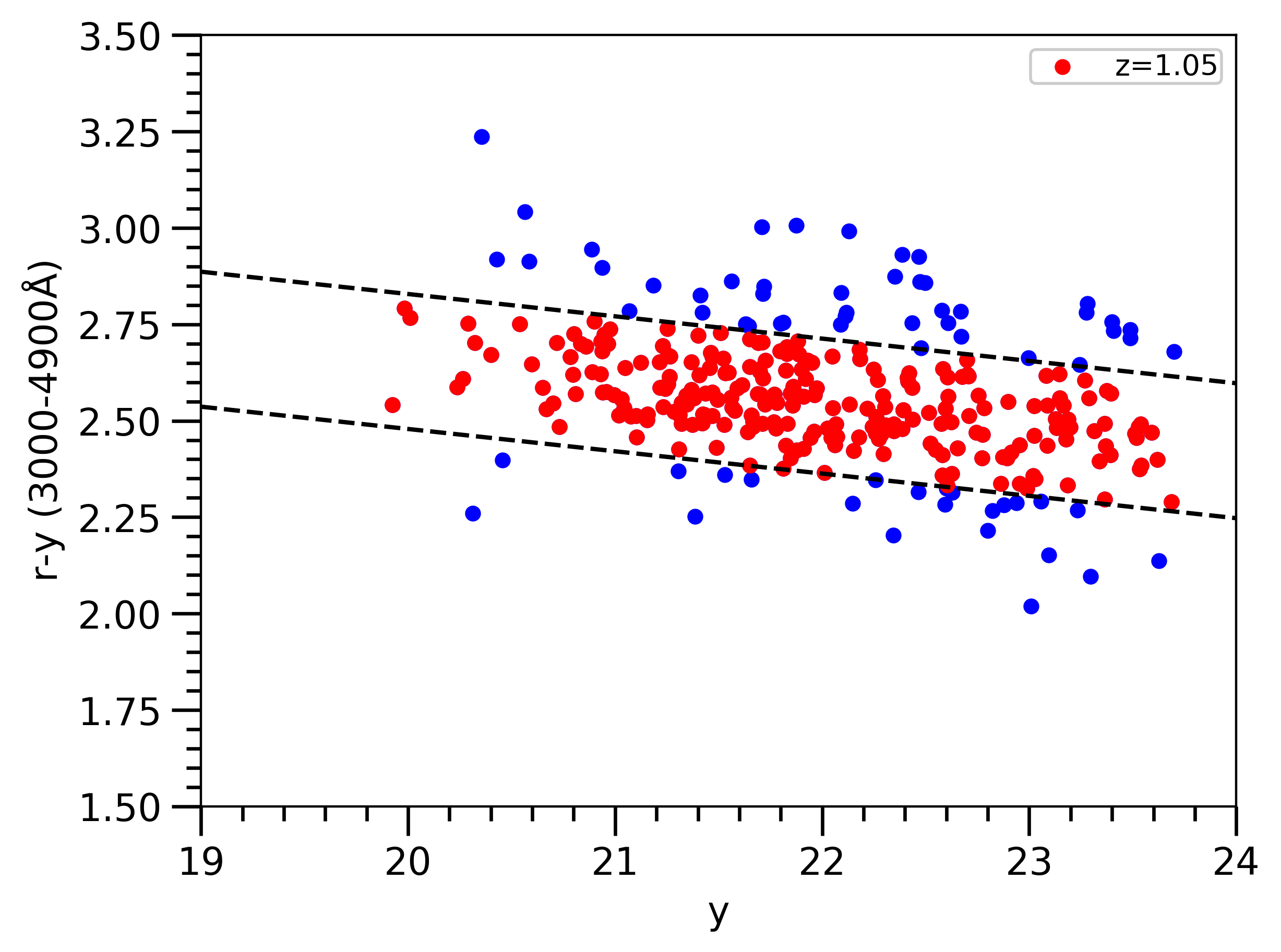}}
{\includegraphics[width=0.325\textwidth]{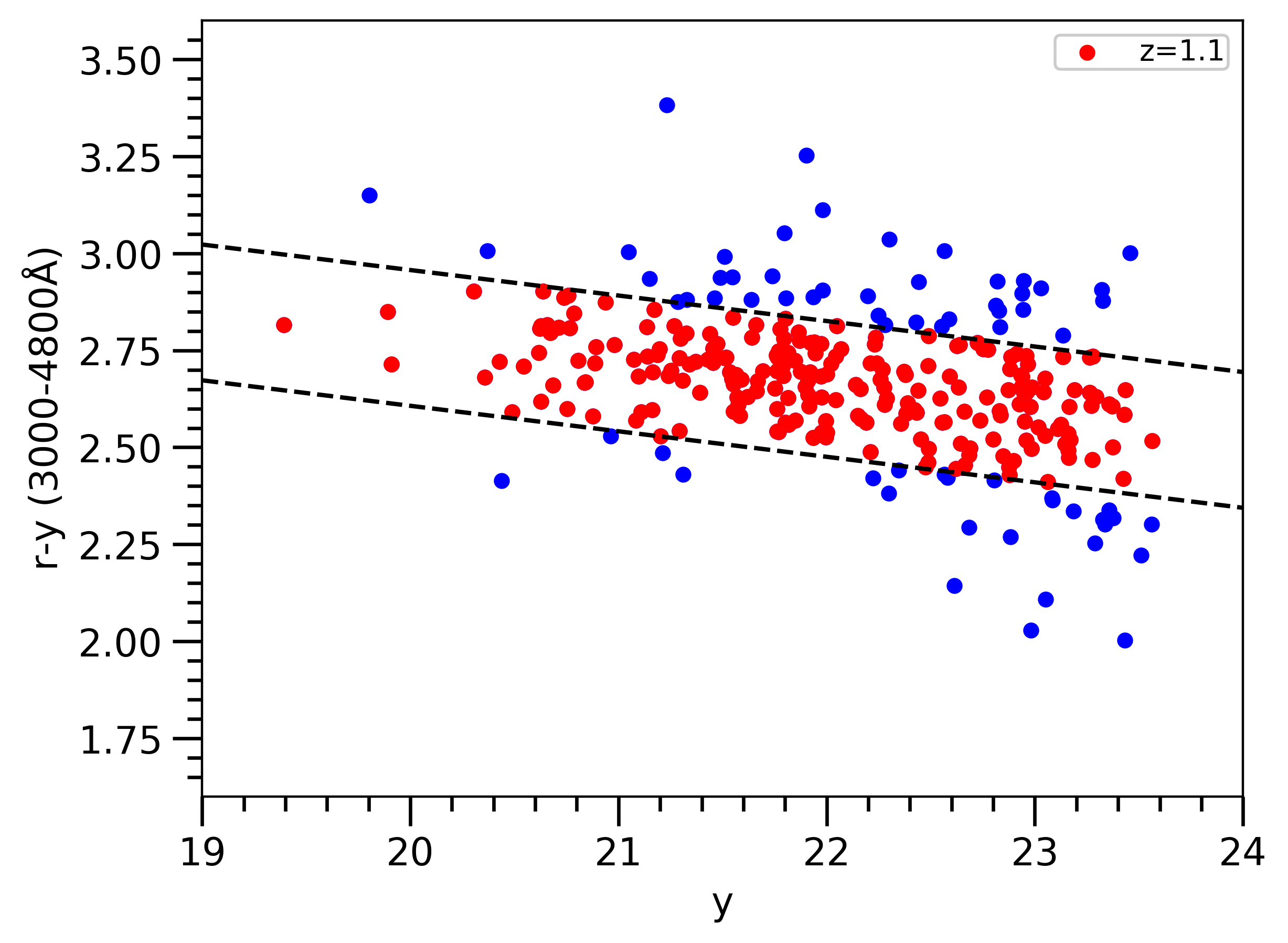}}
\caption{Observed $g-y$ and $r-y$ (rest-frame $u$-band as noted in brackets) color-magnitude diagrams of cluster galaxies between $z=0.4-1.1$ from the HSC SSP survey, separated into bins of $0.05$ in redshift. The red sequence galaxies are denoted with the red filled circles within the dashed lines and have photometric uncertainties of $<0.1$ magnitudes in their optical colors. The dashed lines show the selection region for quiescent cluster ETGs (as described in the text, roughly within $\pm 0.15$ mag. of the mean ridge line for the red sequence).}
\label{fig:u}
\end{figure*}

For our analysis we have chosen all clusters and their associated member galaxies between $z\sim0.4-0.65$ and $z\sim0.85-1.1$ (i.e., where the CLAUDS $u$/$u^*$ bands and the HSC $g$ band match the rest-frame $NUV$ in each redshift interval respectively). In total we have used data of 378 galaxy clusters, consisting of approximately 8000 potential member ETGs. We split the clusters up into small redshift bins of width $z=0.05$ and as in \cite{Ali2022} first determine the red sequence in a rest-frame optical color - in this case observed $z-y$.
%which in rest-frame roughly corresponds to the optical $r-i$/$V-r$ at $z=0.4-0.5$ and $g-V$/$B-g$ at $z=0.85-1.1$.
Then from the $z-y$ red sequence galaxies, we select a further red sequence in a rest-frame $u$-band color - which are observed $g-y$ at $z=0.4-0.65$ and observed $r-y$ at $z=0.85-1.1$. 
%corresponding roughly to rest-frame $u-i$/$u-r$ at $z=0.4-0.5$ and $u-V$/$u-g$ at $z=0.85-1.1$.
The red sequence in $z-y$ generally has a width of $\pm0.1$ and the $g-y$/$r-y$ $\pm0.15$, though this changes slightly with redshift, as shown in Figs \ref{fig:op} and \ref{fig:u}. This double red sequence selection method allows us to select almost entirely a passive sample of ETGs, rejecting even most residual star-forming galaxies. Furthermore, the dual red sequence selection helps to retain the most likely cluster members in the sample given that red sequence members are found to have a 80\% likelihood of being cluster members \citep{Annunziatella2014,Annunziatella2016,Rozo2015,DePropris2016}. Finally, for the passive sample that remained after the color cuts, we compute the rest-frame $NUV-optical$ color using the appropriate observed band ($u/u^*/g$) that corresponds to the rest-frame NUV and the observed $y$-band data. The specific bands that were used for each redshift bin and their rest-frame wavelengths are shown in Table \ref{table_1}.

For nearby ($z < 0.15$) clusters we used galaxies in the sample of \cite{DePropris2021} that are confirmed spectroscopic members of each cluster (using public redshifts from the NASA Extragalactic Database) within $r_{200}$ of the brightest cluster galaxy. We have collated photometry from SDSS \citep{Alam2015} with data from GALEX \citep{Morrissey2007} by matching positions within $3"$. The galaxies were split into two redshift bins - $z=0-0.05$ and $0.05-0.10$, then a red sequence was selected for each bin in $g-r$, followed by $u-r$ as with the HSC data described above in order to select as purely a quiescent sample as possible (\citealt{Meng2023}). The GALEX NUV data was then used to calculate the $NUV-r$ color. We also carried out redvisual classification on all galaxies and rejected any non-ETGs. This sample will be the subject of a future paper analyzing environmental effects and RSF in nearby clusters. 

\section{Results}

\begin{figure*}
\centering
{\includegraphics[width=0.325\textwidth]{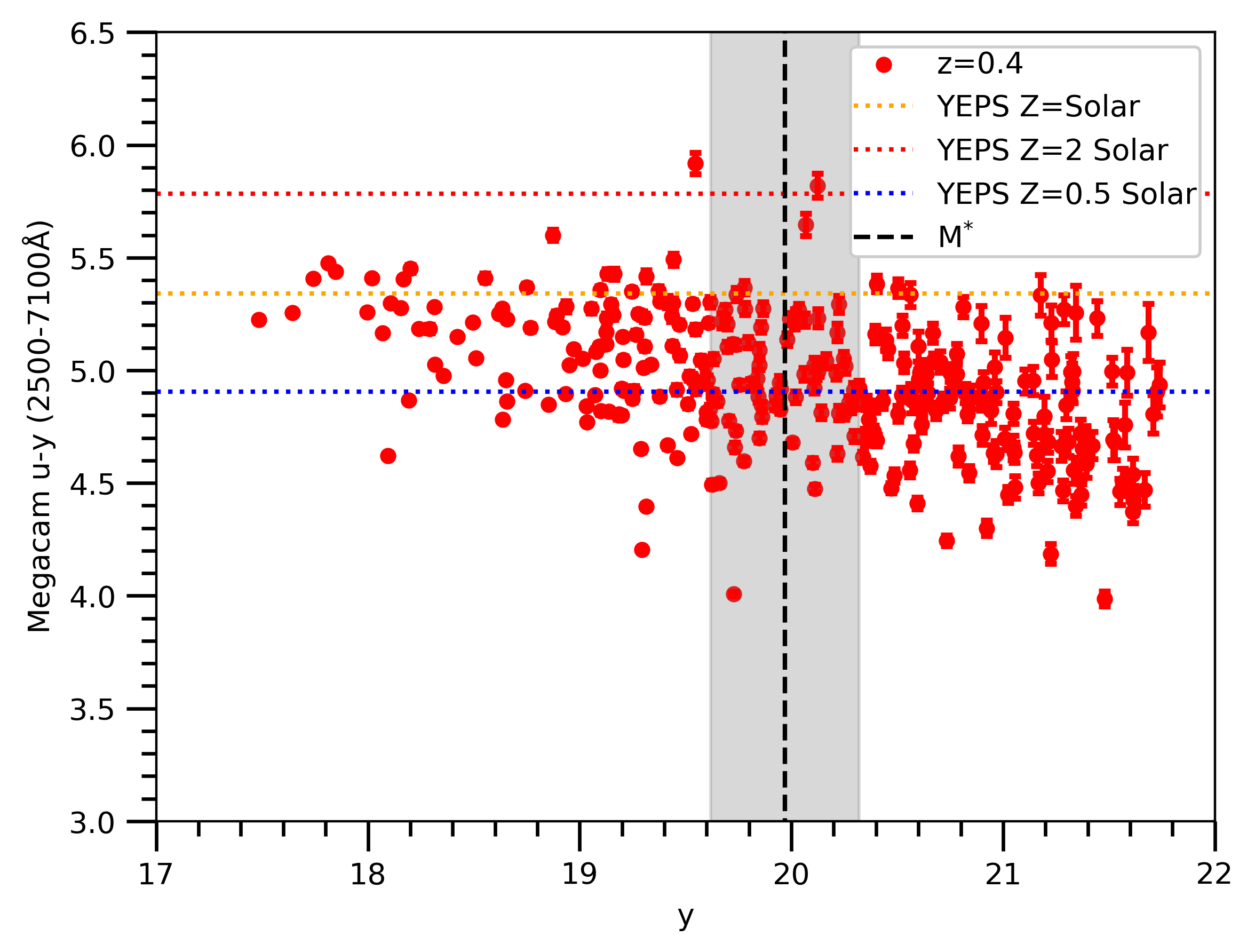}}
{\includegraphics[width=0.325\textwidth]{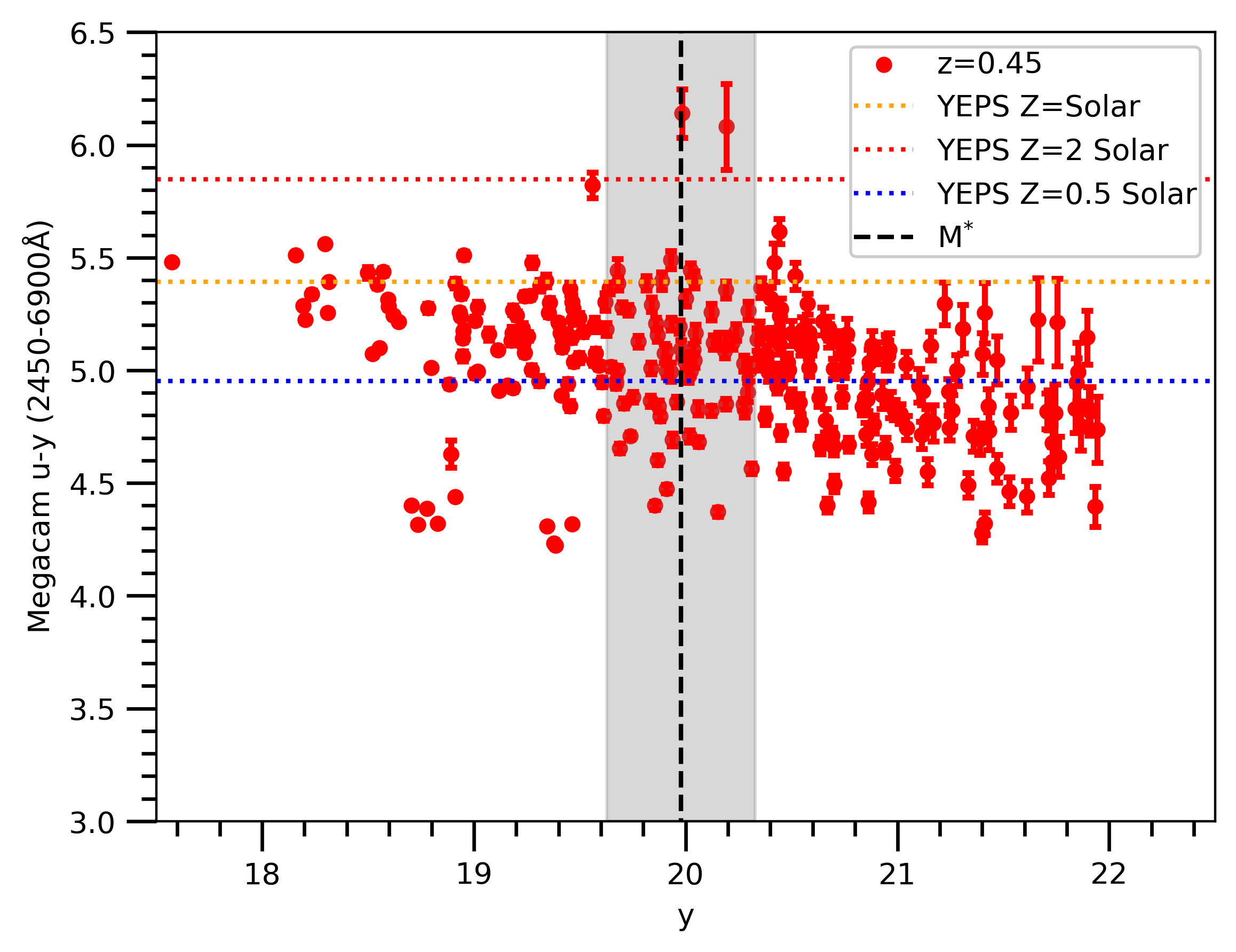}}
{\includegraphics[width=0.325\textwidth]{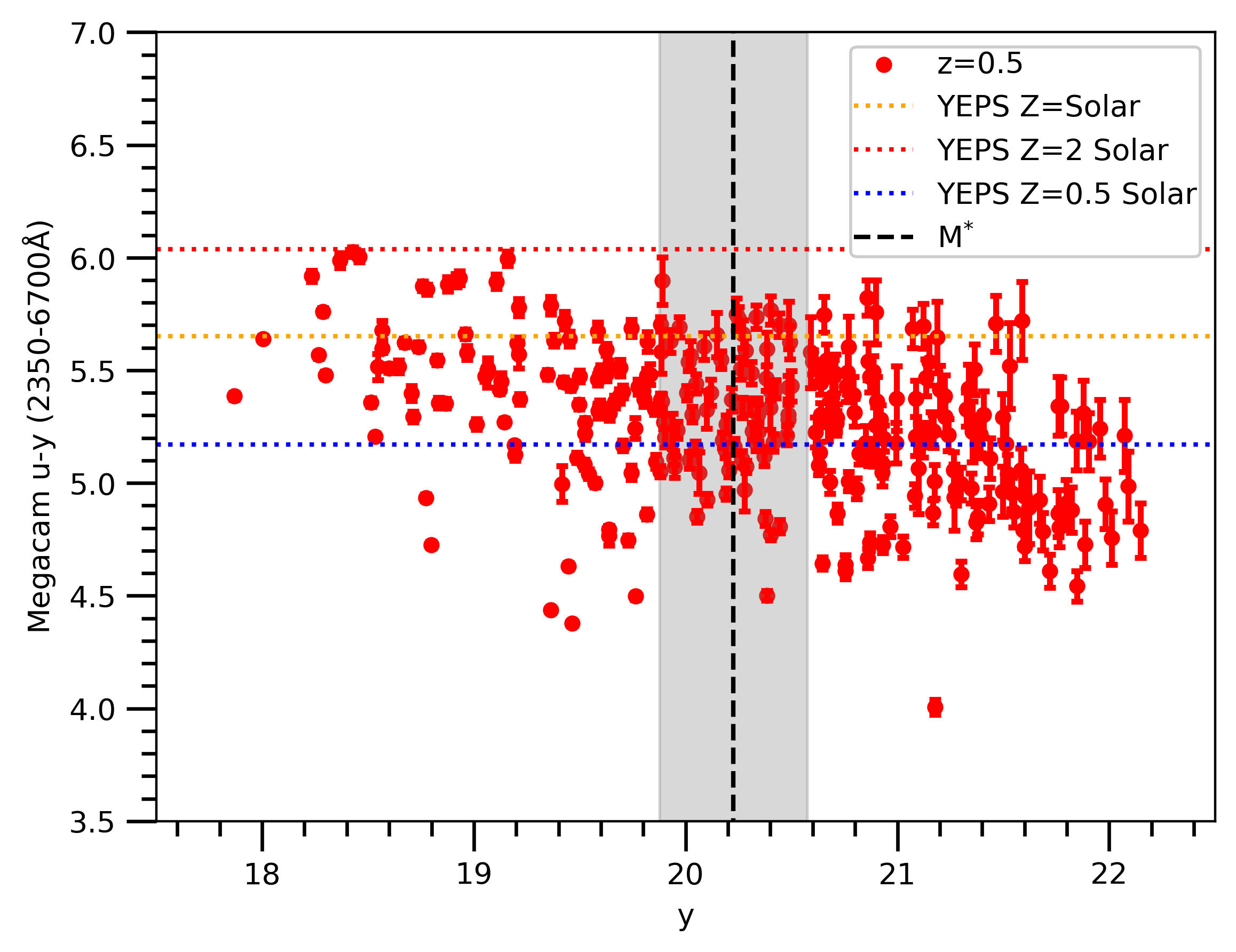}}
{\includegraphics[width=0.325\textwidth]{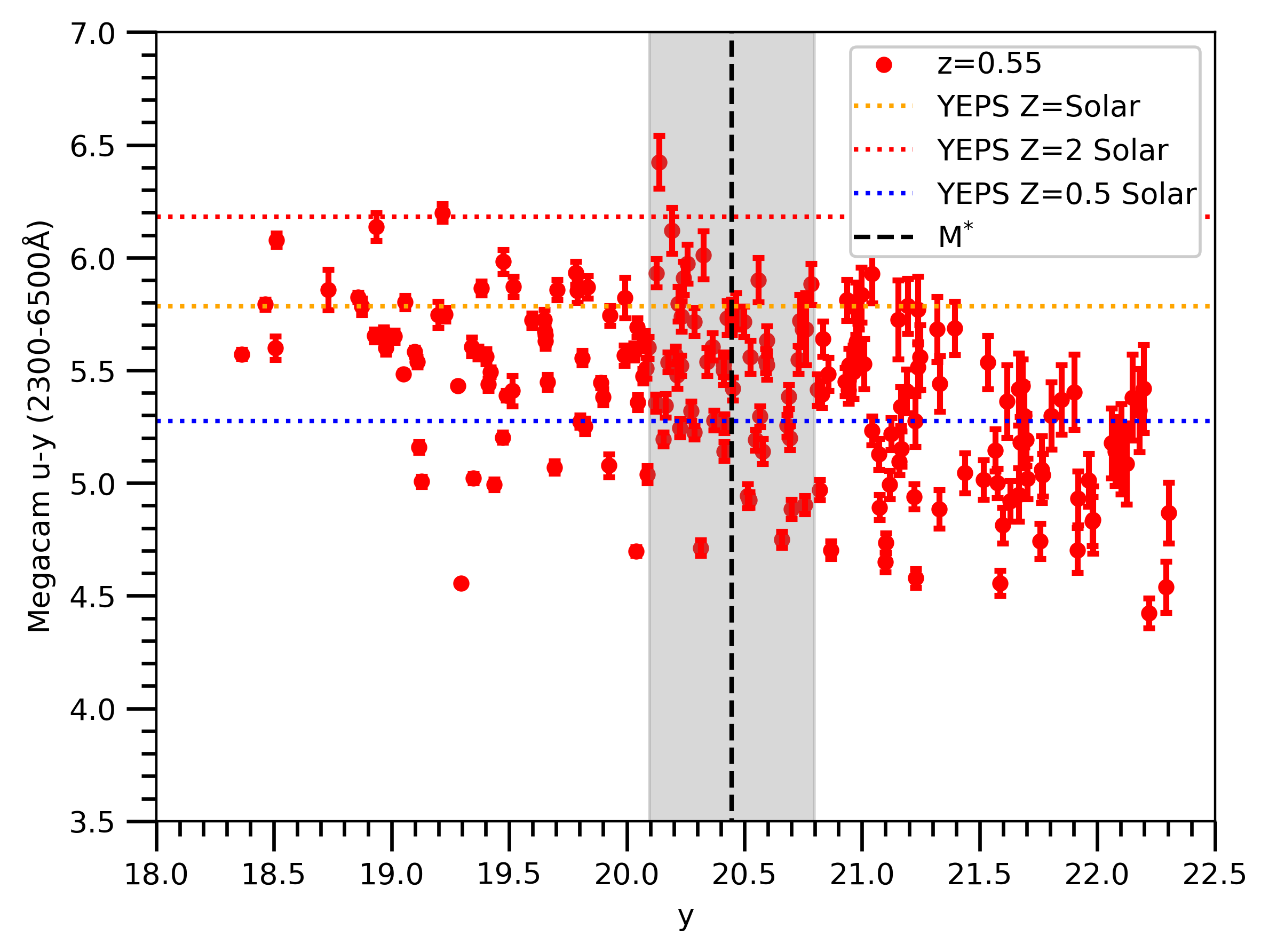}}
{\includegraphics[width=0.325\textwidth]{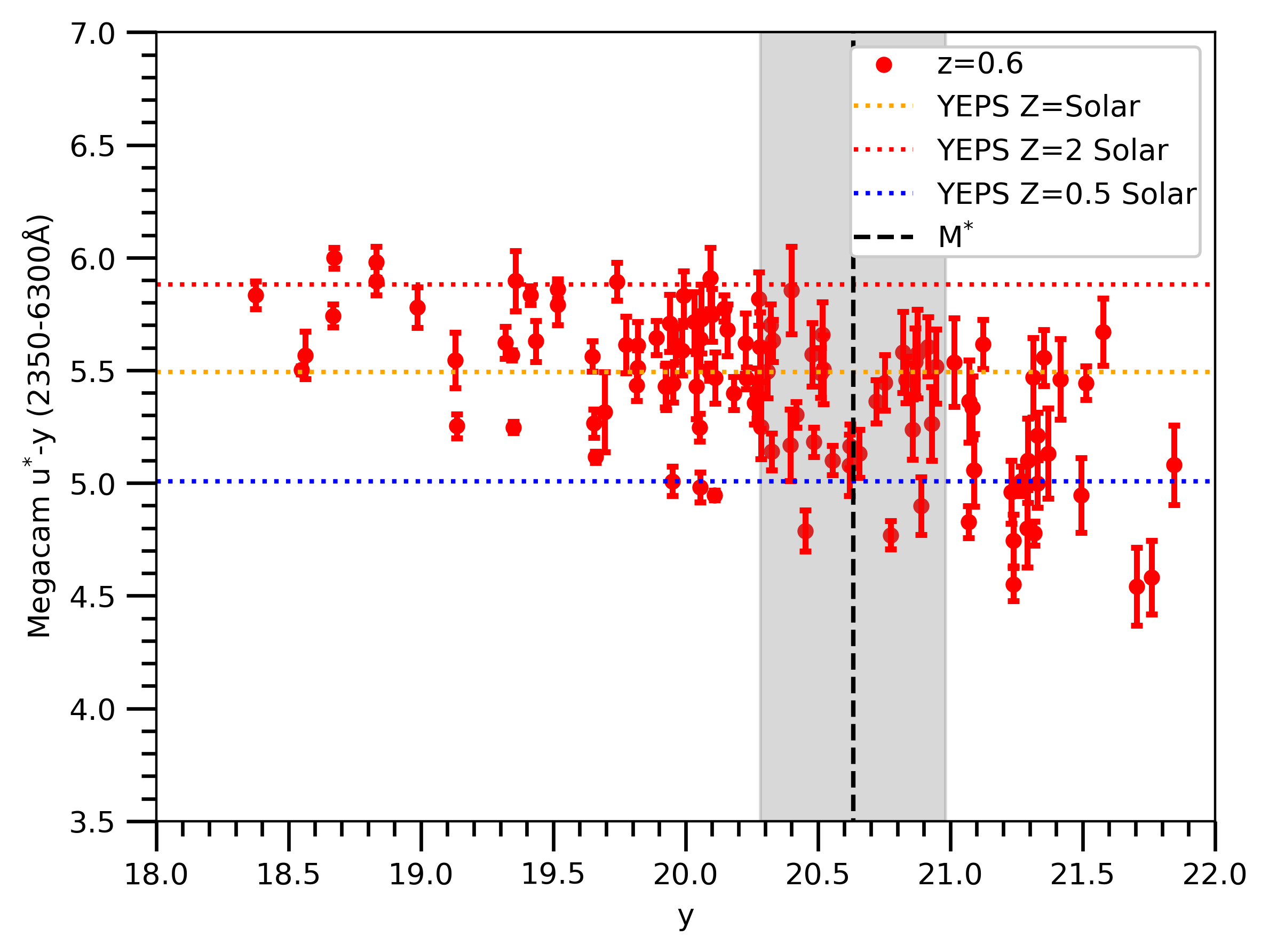}}
{\includegraphics[width=0.325\textwidth]{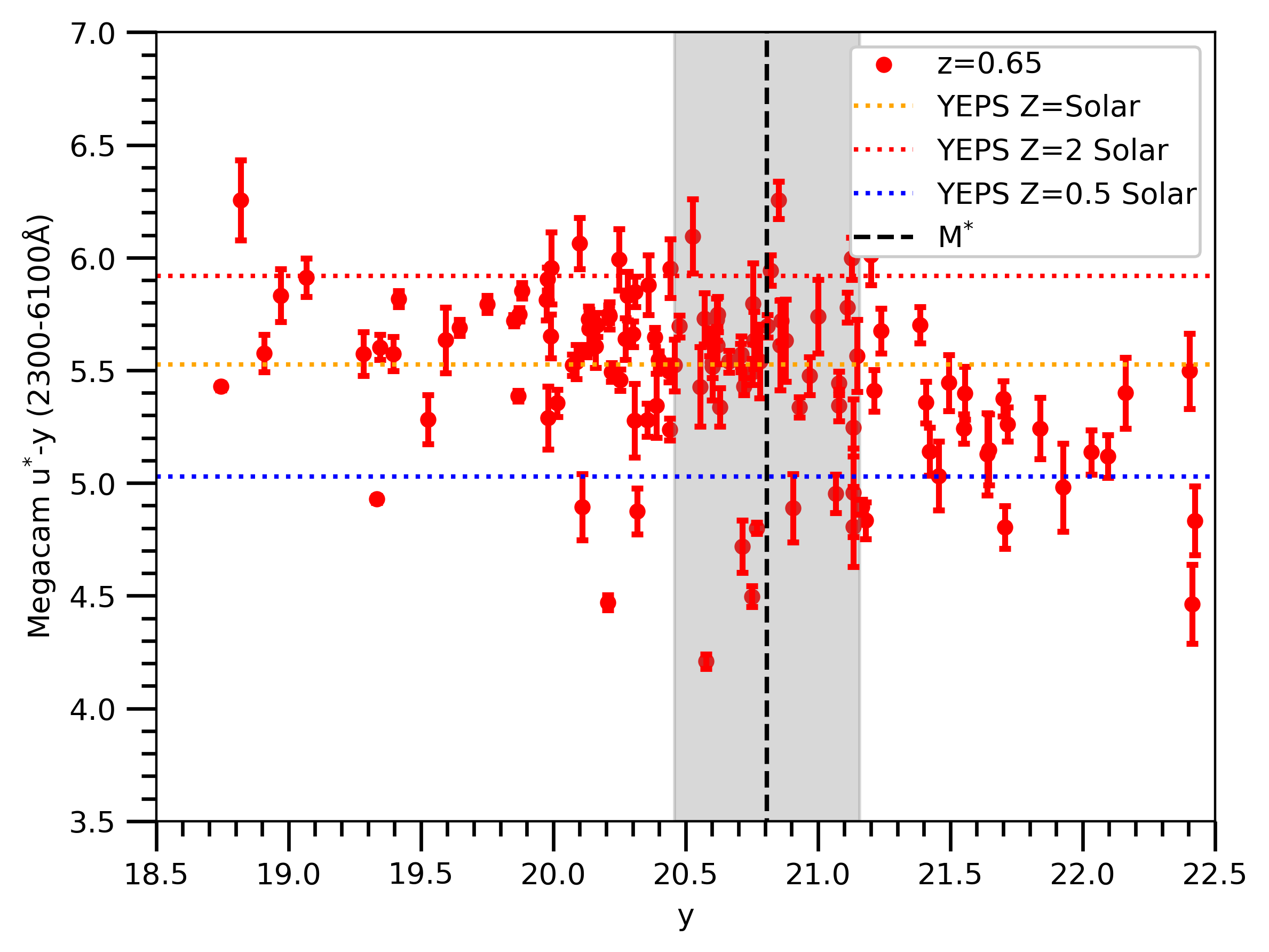}}
{\includegraphics[width=0.325\textwidth]{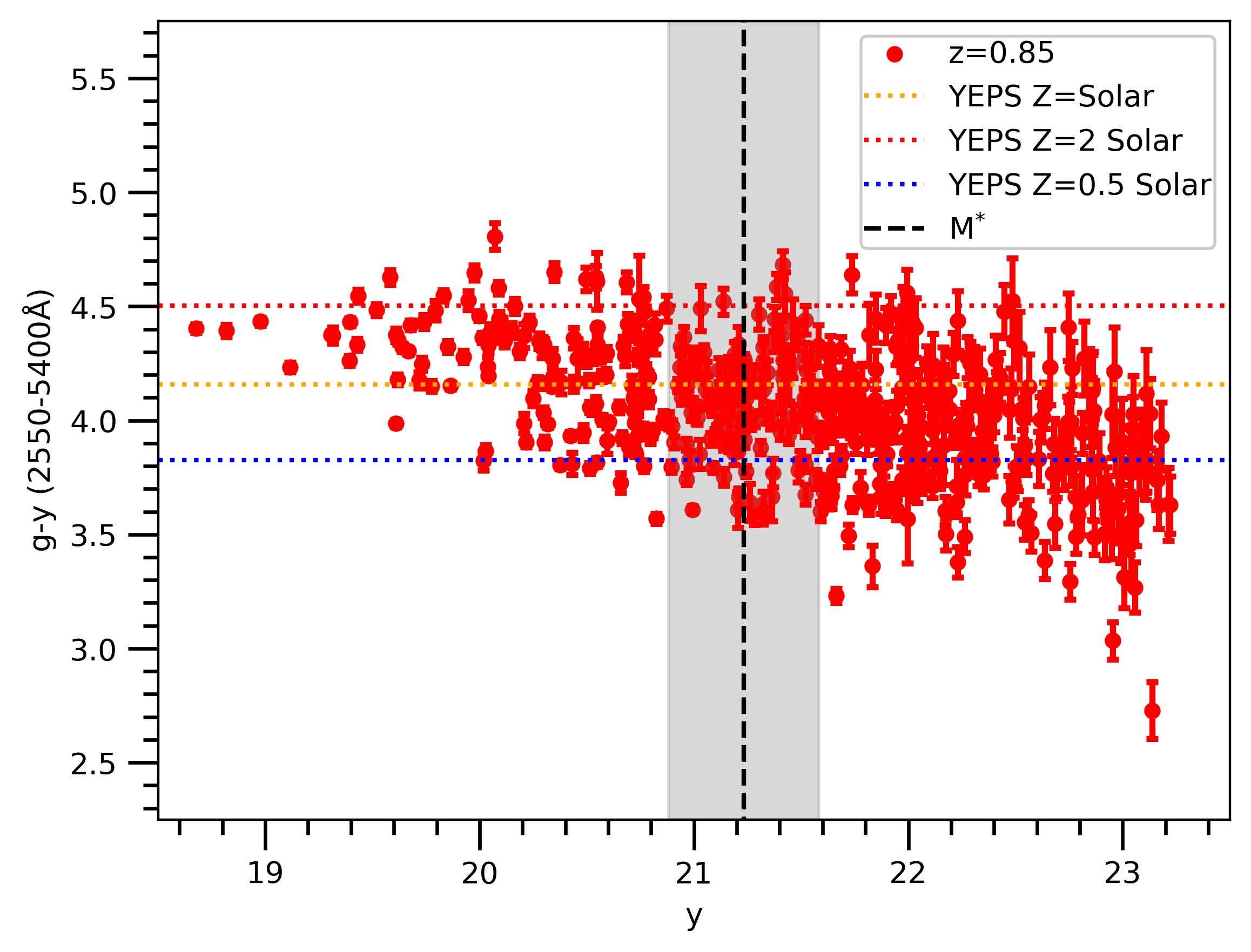}}
{\includegraphics[width=0.325\textwidth]{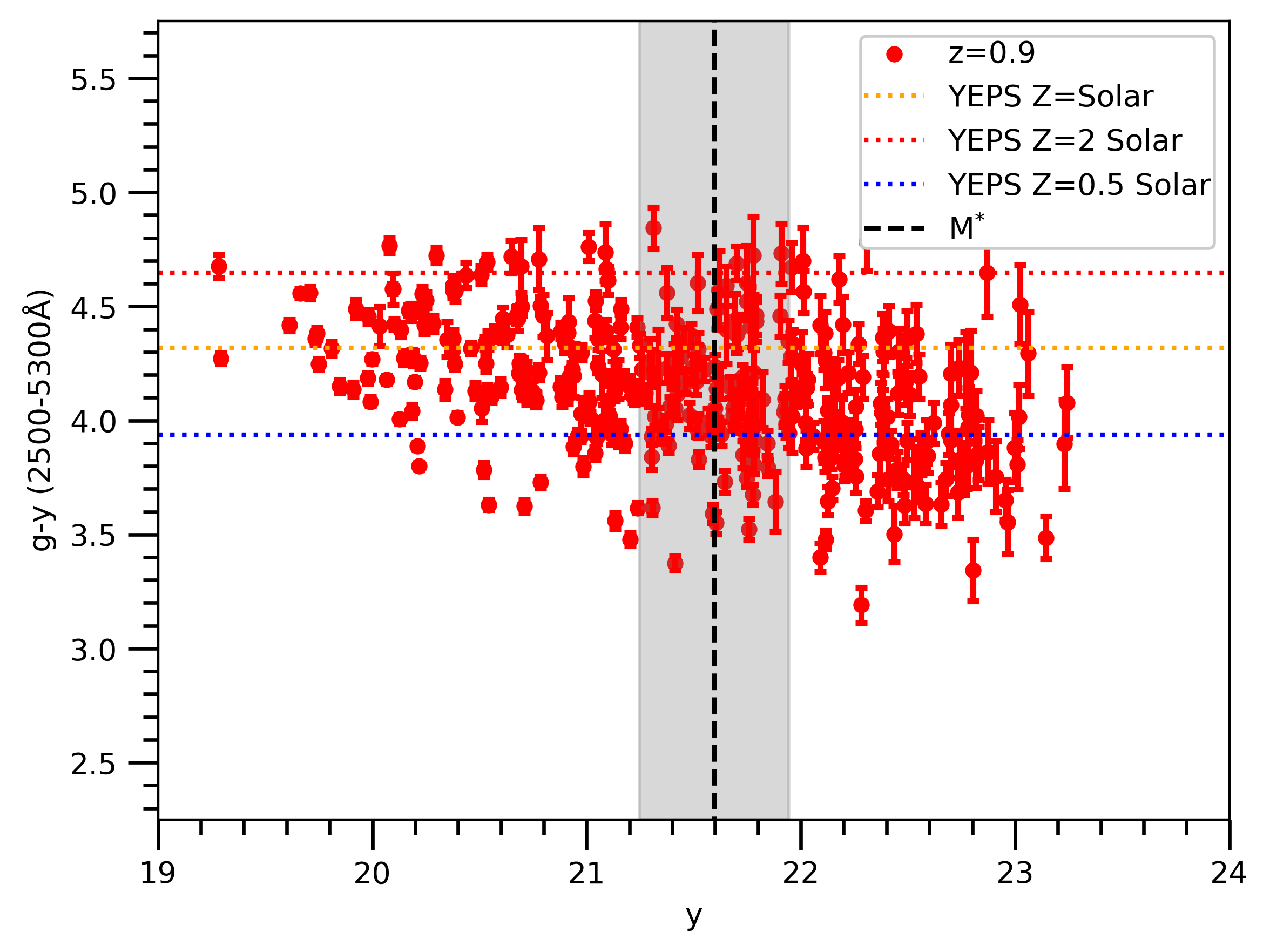}}
{\includegraphics[width=0.325\textwidth]{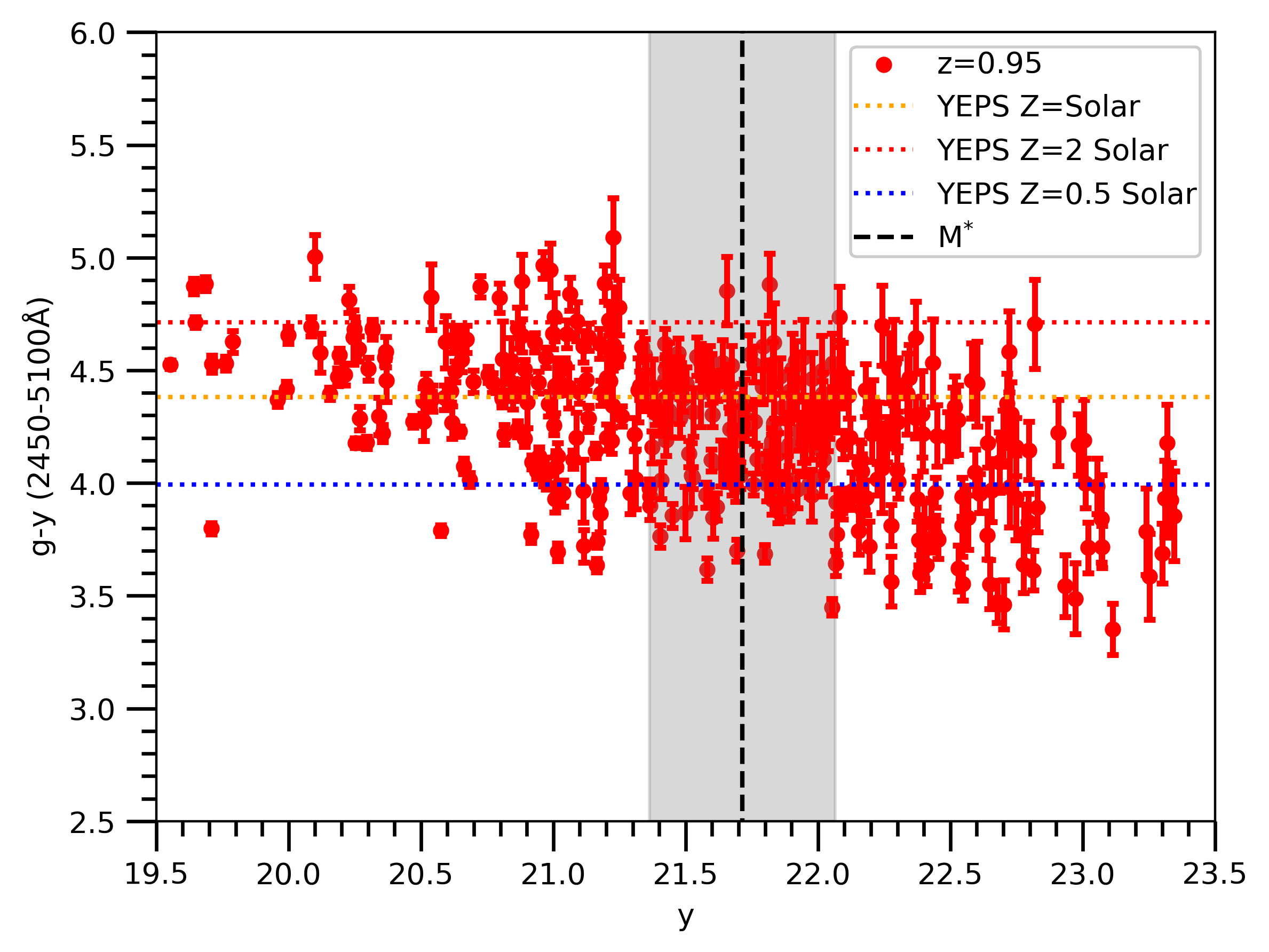}}
{\includegraphics[width=0.325\textwidth]{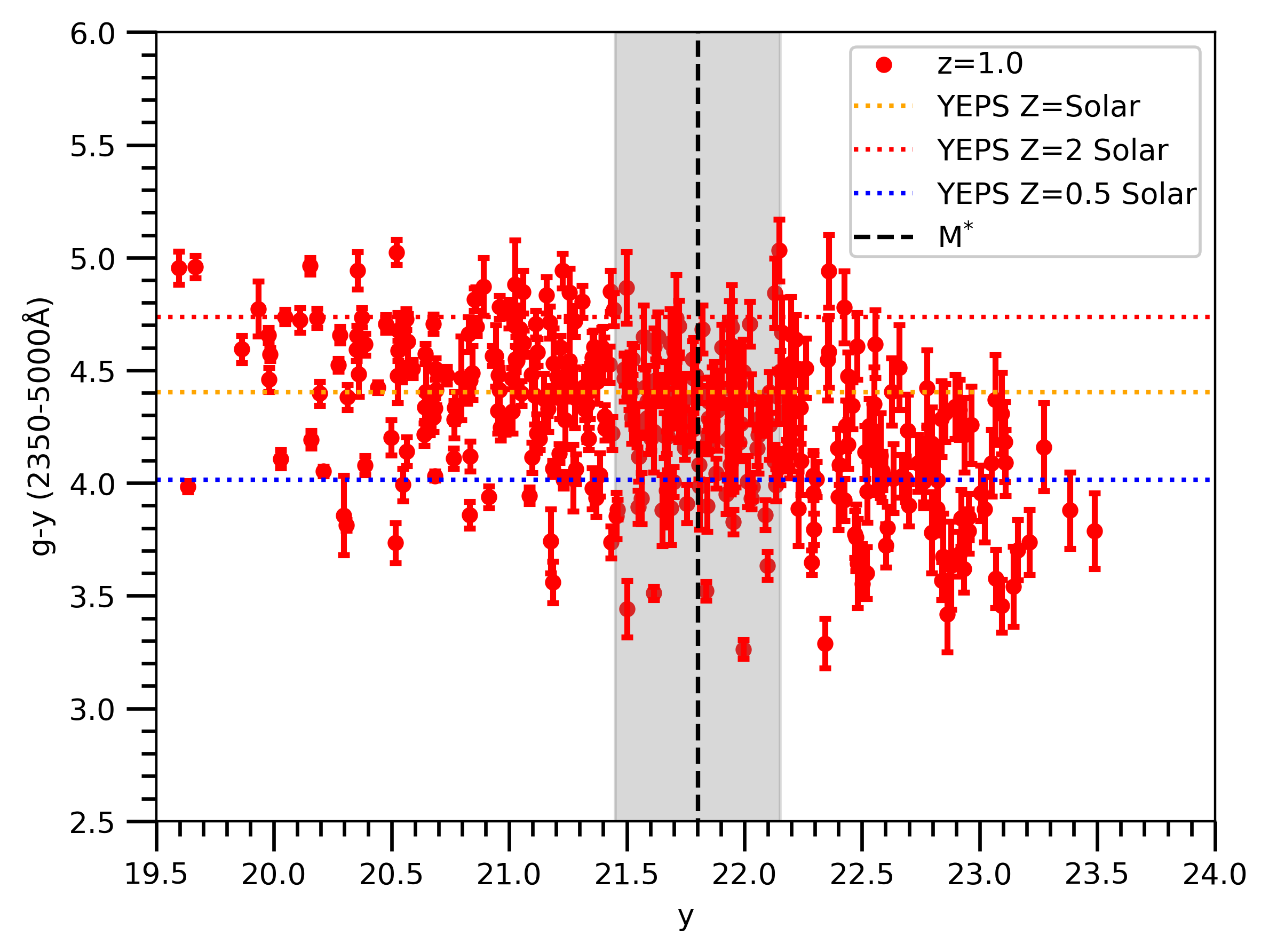}}
{\includegraphics[width=0.325\textwidth]{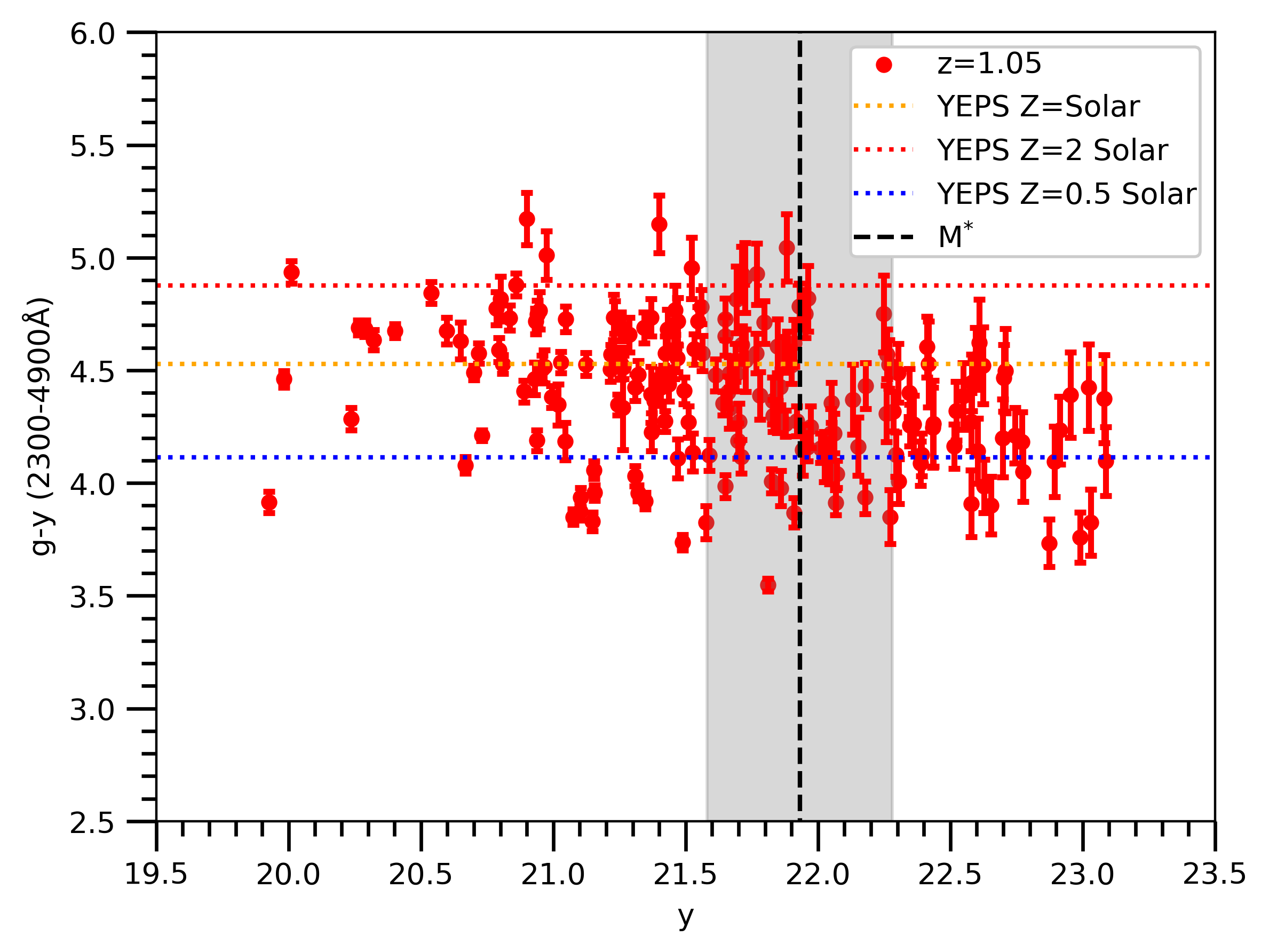}}
{\includegraphics[width=0.325\textwidth]{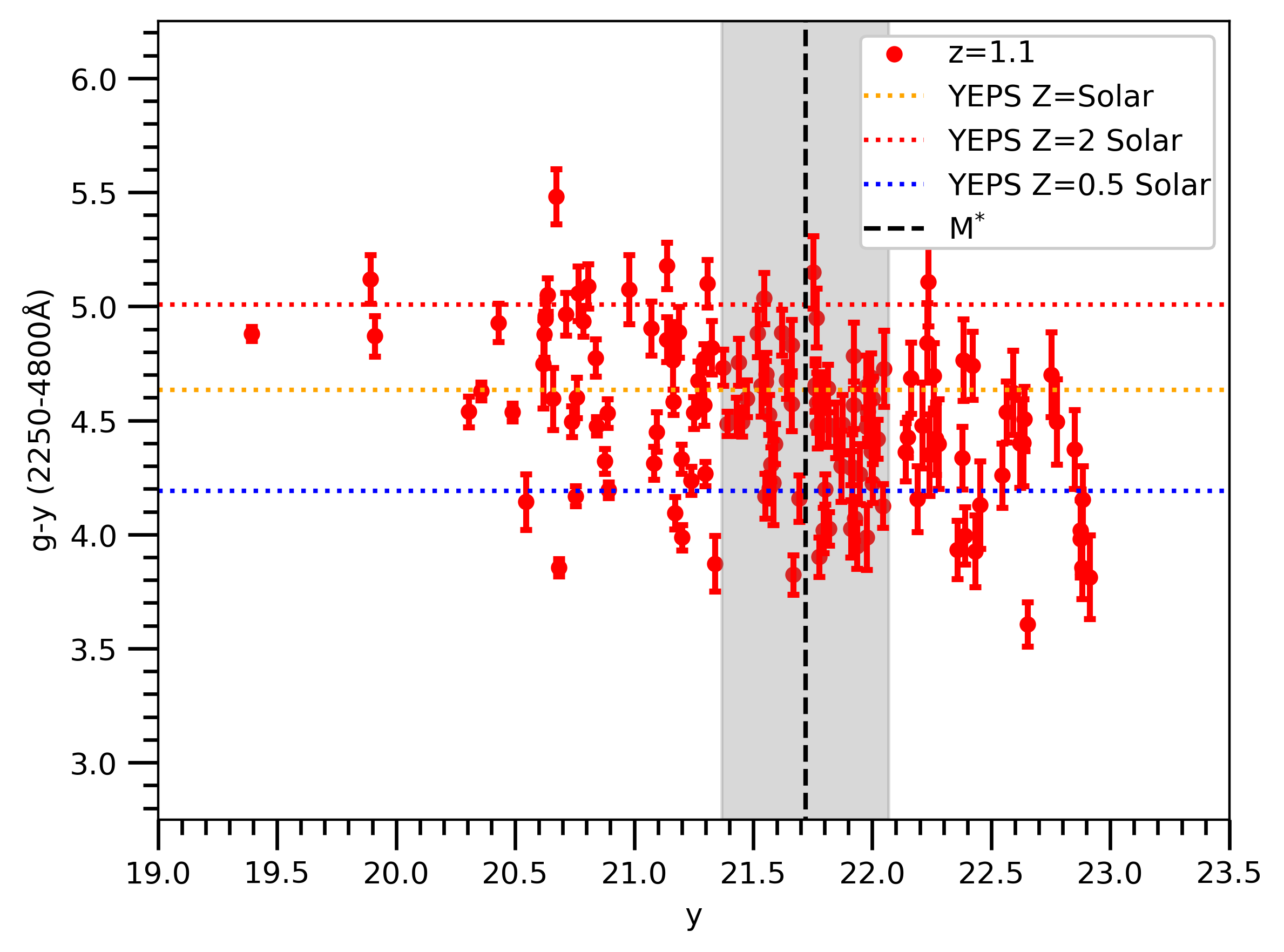}}
\caption{Observed Megacam $u/u^*-y$ and $g-y$ (rest-frame $NUV-optical$ as noted in brackets) color-magnitude diagrams of cluster galaxies between $z=0.4-1.1$, separated into bins of $0.05$ in redshift. The red sequence galaxies are those selected from the optical colors in Fig. \ref{fig:op} and \ref{fig:u}. Photometric uncertainties are $<0.15$ magnitudes.}
\label{fig:uv}
\end{figure*}

For the red sequence galaxies selected from the optical and $u$-band colors, we calculate the $NUV-optical$ colors at each redshift bin using the rest-frame UV band as shown in Table \ref{table_1} and the HSC $y$-band as the optical band (the exact rest-frame wavelengths are given in the table). We use fixed aperture magnitudes in all bands to calculate the colors - 2.0" diameter between $z=0.4-0.6$ redshift bins, 1.5" between $z=0.6-0.7$ and 1.0" between $z=0.85-1.15$. These aperture sizes correspond roughly to a fixed physical diameter of 10 kpcs across redshift. Fixed apertures are particularly useful as the UV emission is often centrally concentrated in these systems and does not follow the same surface brightness profile as the optical, due to different stellar populations being responsible for the output at each wavelength region \citep{Carter2011,Jeong2012}. As such fixing the aperture to a diameter that covers most of the UV emission and using this aperture in all other bands allows for the calculation of a more accurate color that represents the sub-population responsible for the upturn in these galaxies.

The UV excess has classically and ideally been measured using the GALEX $FUV$ band, previous studies have shown shown that the $NUV$ is also a suitable proxy and is largely sensitive to the phenomenon (\citealt{schombert2016,Ali2022}). In this paper we have selected galaxies which have been observed in the rest-frame $NUV$ centred between $2250-2500$\AA. As seen in the analysis of \cite{chavez2011}, the emission at $<2500$\AA\ in ETGs is largely dominated by the stellar sub-population responsible for the upturn, with the output from the standard population of stars that makes up the bulk of ETGs and dominates the optical wavelengths $>3000$\AA\ being nearly negligible in comparison (see also detailed explanation in \citealt{Ali2022}). In \cite{ali2018a} we showed that the observed SEDs of Coma ETGs were best fit by 2-component models - one component being a standard SSP model which represents the majority main sequence/red giant branch stars and a second blackbody component superimposed on top, representing the highly evolved blue horizontal branch stars that are believed to be the driver of the UV output. It could clearly be seen that the first component had little effect below $3000$\AA, while the same was true for the second component above $3000$\AA. This suggests that the UV-bright stellar population in ETGs is entirely distinct from the optically-bright stellar population, as is postulated in previous studies \citep{ali2018a,ali2018b}. The $NUV$ region is also much more sensitive to age than the optical or $NIR$. As such, in the absence of an upturn population, the $NUV$ can be used to measure the age and star-formation history of the majority stellar population in ETGs at large, more efficiently than any other wavelength regime \citep{Kaviraj2007}. We will take advantage of this key feature of the $NUV$ to explore the star-formation history of ETGs in our analysis of higher redshift galaxies.

In Fig. \ref{fig:uv} we show the $NUV-optical$ color for each individual redshift bin from $z=0.4-1.1$. At $z<0.6$ we detect objects in the UV at or above 5$\sigma$ up to $M^*$+2, but at $z>0.8$ we reach only up to about $M^*$+1.5, where $M^*$ is the characteristic magnitude in the Schechter luminosity function of red sequence galaxies in the observed $y$-band. Virtually all red sequence galaxies in the rest-frame optical and $u$ bands are detected to a 5$\sigma$ level or higher. Of these galaxies, $>80\%$ (the percentage being higher at lower redshift bins) also have 5$\sigma$ or higher detections in the rest-frame UV, down to $\sim M^*$+1.5. As such the vast majority of our optically selected galaxies also have strong UV detections. In the case of non-detections, which are mostly low luminosity ($>M^*$) UV-red galaxies - if we extend the detection limit down to $3\sigma$, this includes most such galaxies and their inclusion does not change in any way the overall statistics and trends discussed henceforth. However, we exclude them from our analyses for consistency.

For each redshift bin we also include the color in the observed bands from the YEPS composite stellar population (CSP) model \citep{2013ApJS..204....3C} using an infall prescription and standard cosmological He abundance (see section 4 for further details). The models are for $Z$=\(Z_\odot\), $Z$=2\(Z_\odot\) and $Z$=0.5\(Z_\odot\), with a redshift of formation ($z_{form}$) of 3 and a delta burst. Given that our cluster data roughly encompasses magnitudes of $M^*\pm2$, the metallicity range is derived from the Coma cluster for the same magnitude range (\citealt{Thomas2005,Thomas2010}). For instance, a typical $M^*$ galaxy in Coma has an average metallicity of about $Z$=1.15\(Z_\odot\), while $M^*+2$ and $M^*-2$ roughly correspond to $Z$=1.75\(Z_\odot\) and $Z$=0.65\(Z_\odot\) respectively. We show a slightly broader metallicity range of $0.5-2$\(Z_\odot\) in order to take into account the error in the $M_{r}^*$ calculation ($\sim0.15$ mags - \citealt{Beijersbergen2002}) and the spread in the mass-metallicity relation. As such, the models shown in Fig. \ref{fig:uv} have a metallicity range and a formation history typical of ETGs in our luminosity range and provide a reference point of what the (upper and lower limits of) UV colors should be without any upturn component, i.e. the colors are driven purely by the age and metallicity of the overall stellar population in the galaxy. Below $z\sim0.6$ there is a large excess of galaxies that have colors bluer than even the $Z$=0.5\(Z_\odot\) model, which for most ETGs in our magnitude range is already lower than their known metallicities from optical studies in the Coma cluster as noted above. Changing $z_{form}$ to 2 (a difference of about 1 Gyr from $z_{form}=3$) only changes the color by about $\sim0.1$ mags blueward, which still cannot account for most galaxies. Given that a large number of galaxies have colors bluer than even the bluest model suggests an external parameter at play that cannot simply be explained by an age/metallicity effect. Conversely, in all redshift bins above $z>0.8$, the vast majority of galaxies are reasonably well encapsulated by the 0.5\(Z_\odot\)$<Z<$2\(Z_\odot\) models, with most galaxies no longer exhibiting a strong UV excess. Hence the UV colors at these redshifts are mostly driven by age/metallicity. In essence there is a clear evolution in the UV color of ETGs at around $z=0.6-0.8$, which will be explored through the lens of several models in the next section in order to understand the fundamental physical properties of these galaxies.

\section{The Ultraviolet Upturn}

\begin{figure*}
\includegraphics[width=0.9\textwidth]{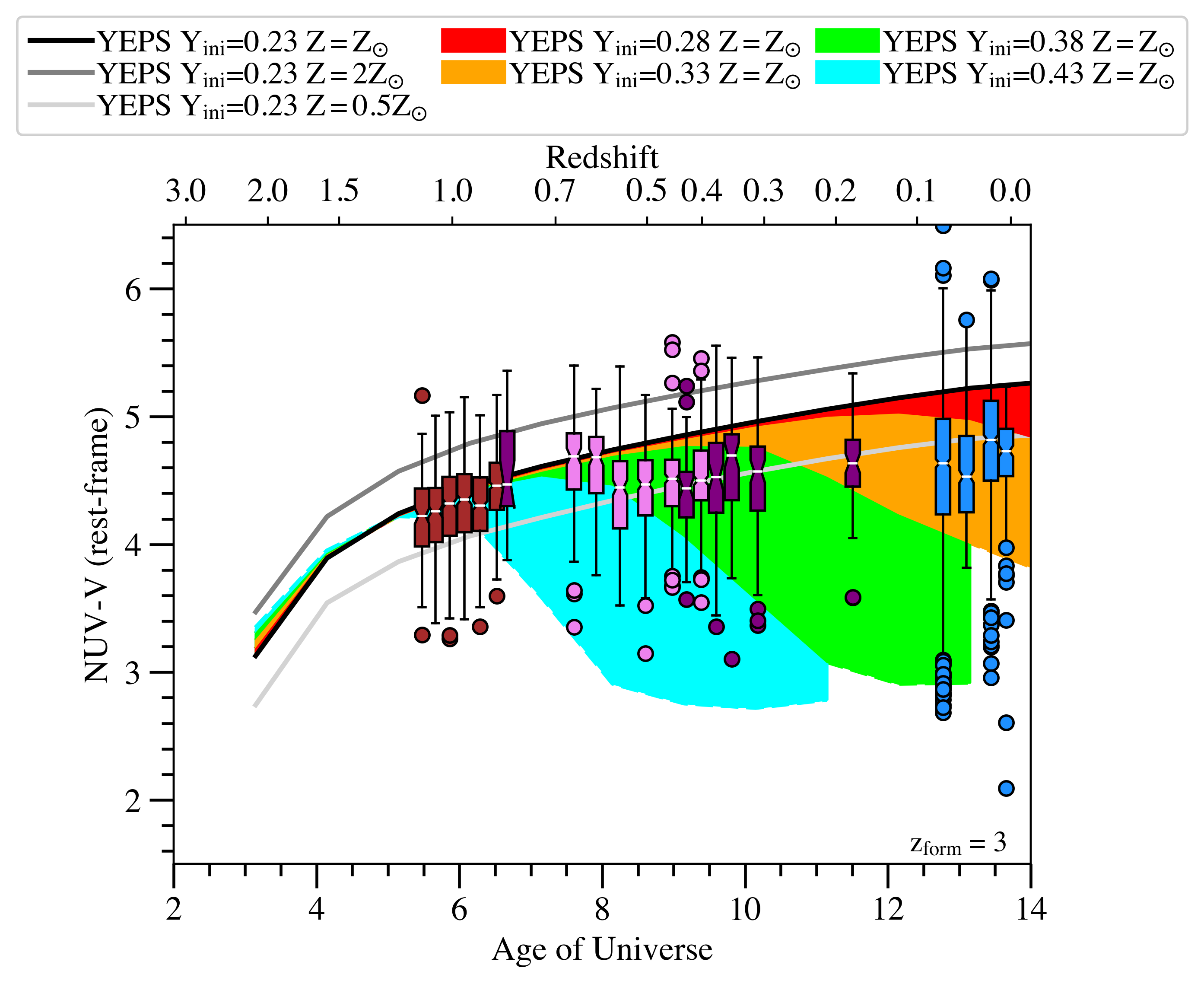}
\caption{Evolution of the rest-frame $NUV-V$ (observed $g-y$ at $z=1$) color over redshift/lookback time as given by the YEPS spectrophotometric (infall) models for a range of initial helium abundances - $Y_{ini}=0.28,0.33,0.38,0.43$ (shaded areas), with $z_{form}=3$ and metallicities as detailed in the figure legends. Also included is the evolution of the same color for infall models with $Y_{ini}=0.23$ (i.e. standard cosmological He abundance with no upturn) for $Z$=\(Z_\odot\), 0.5\(Z_\odot\) {\rm and} 2\(Z_\odot\) (solid lines). Plotted on top are box plots which show the rest-frame $NUV-V$ colors of cluster galaxies in bins between $z=0-1.1$, with different colors indicating the datasets from which the galaxies are derived, as labelled in the figure. Photometric uncertainties in color are $<0.2$ magnitudes.}
\label{fig:evo3}
\end{figure*}

\begin{figure}
\includegraphics[width=0.45\textwidth]{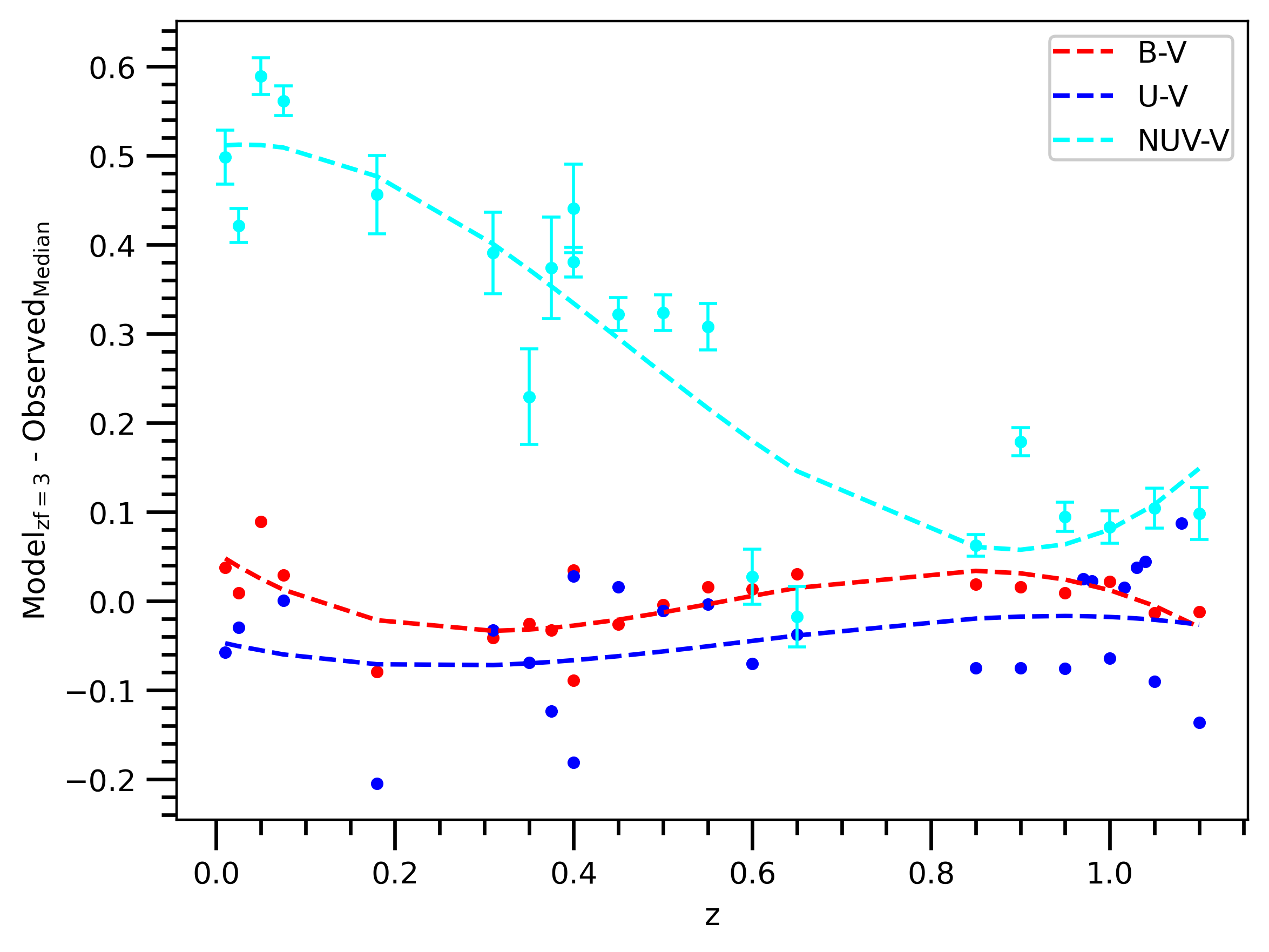}
\caption{The difference between the median $NUV-V$ color of each redshift bin (white notches in each box plot in the Fig. \ref{fig:evo3}) subtracted from the $NUV-V$ color of the $Y_{ini}=0.23$ solar metallicity infall model formed at $z_{form}=3$ (black line in Fig. \ref{fig:evo3}, i.e. standard passively evolving model with no upturn). The plot shows that the difference between the observed and model color is large at low redshift ($\sim0.5$ mags) but become nearly zero at $z>0.8$, as would be expected when the upturn has faded. This evolution is not seen in the rest-frame $B-V$ or $U-V$ colors, where a simple passively evolving model fits the observed data reasonably well at all redshifts sampled by our data, albeit with some stochasticity.}
\label{fig:evo4}
\end{figure}

As introduced above, ETGs are known to exhibit a significant excess of UV flux compared to the extrapolation of the spectral energy distribution of their optical and infrared stellar populations (e.g., see \citealt{oconnell1999,yi2010} for a review). Recent RSF may provide some of the UV flux in a few cases \citep{Yi2005,Boselli2005}, but this explanation does not account for the majority of the galaxies where UV upturns have been observed \citep{Ferguson1991,Brown1997,Brown2000,Boselli2005,Han2007}. The broad consensus is that hot, or extreme, horizontal branch (EHB) stars must be the key contributor to the UV fluxes of ETGs \citep{dorman1993,dorman1995,Han2007,Lisker2008}. Blue HB stars are directly observed in the bulge of M32 \citep{tbrown2004} and identified as the source of the UV light, while \cite{Rosenfield2012} argue that such stars must also provide the bulk of the unresolved UV emission in the bulge of M31.

These blue HB stars may originate from a metal-poor stellar population \citep{park1997}, through normal stellar evolution channels, a metal-rich stellar population with extra mass loss \citep{Yi1997}, binaries stripped of their envelope during the first ascent of the red giant branch \citep{Han2007,Hernandez2014} or helium enriched stars \citep{tantalo1996,chung2011,chung2017}. The observed evolution of the UV color in \cite{ali2018c,Ali2022} shows that the blue HB stars appear in the population at $z \lesssim 0.7$ and this is only consistent with a model where they originate from a population of stars enriched in helium at up to 40\%, with a mass fraction of typically 10\%, consistent with the model of \cite{Buzzoni2008}. Similar stars likely produce the anomalously  blue HBs observed in globular clusters in the Galaxy \citep[e.g.,][]{2005ApJ...621L..57L, 2013MNRAS.430..459D, marino2014}.

In Fig. \ref{fig:evo3} we plot the evolution of the $NUV-V$ color with redshift of all red sequence galaxies up to $\sim M^*+1.5$, such that we consistently analyse cluster ETGs in the same mass range across redshift. The rest-frame NUV data at all redshifts is centred between $\sim$2300-2500\AA (as seen in Table \ref{table_1}), so $k-$corrections were not large for the UV data. In the optical even though we use the same observed $y$-band at all redshifts, in rest-frame the effective band changes significantly in the wavelength regime, spanning rest $g$ to $i$ from $z=0.4-1.1$. Hence larger corrections were required, though this could be achieved with reasonable accuracy using standard models given that ETGs have consistent optical colors that evolve passively over cosmic time. In order to account for the slight difference in bandpass shapes (in the UV) and rest-frame wavelengths, we followed a similar $k-$correction prescription from \cite{Ali2022} to correct the $UV-optical$ colors in each redshift bin to bring them in line with the rest-frame HSC $g-y$ at $z=1$ (rest-frame $NUV-V)$, such that the data across all redshifts could be directly compared. To do so we made use of a standard YEPS CSP model with $Z$=\(Z_\odot\), $z_{form}=3$, $Y_{ini}=0.23$ and age corresponding to the redshift of the galaxy bin being corrected. The model was then shifted to the redshift of the bin and $z=1$, from which the $k-$correction to the observed color was calculated. Note that we used a fixed age as we do not wish to apply an evolutionary correction (given that we look to measure it in the data) and only apply a $k-$correction (to mostly account for the bandpass differences in the observed bands at different redshifts). Ultimately once the corrections are applied, it is possible to observe the evolution of the $NUV-V$ color over redshift and fit models to the observations. Beside the main HSC SSP and CFHT CLAUDS data that is the focus of this paper, we have also included in the plot the data from our previous analyses of cluster galaxies from the HST Frontier Fields and CLASH surveys (\citealt{ali2018c, Ali2022}), Abell 1689 (\citealt{ali2018b}), the 2dF survey (\citealt{ali2019}) and the SDSS/GALEX clusters as described in Section 2 (\citealt{DePropris2021}). The datasets are labelled with different colors in the figure.

The $NUV-V$ is plotted in the form of box plots where the box represents the middle 50\% of the color distribution of all galaxies in the redshift bin and the notch within the box represents the median, which also roughly corresponds to the typical $M^*$ galaxy in each bin. The whiskers then show the upper and lower 25\% of the color distribution, with any outliers plotted as filled circles. The box plot has the benefit of allowing us to track both the overall evolution of galaxy colors with redshift (i.e. using the median), while also keeping an eye on the scatter in the color over redshift. Alongside the observed colors we have also included a number of YEPS He-enhanced spectrophotometric models \citep{chung2011,chung2017} for comparison. The metallicity distribution function derived from the infall chemical evolution model proposed by \citet{kodama1997} is employed in these composite stellar population (CSP) models for ETGs. %These are composite stellar population (CSP) models with an infall prescription from \citet{kodama1997}. 
The solid lines in black, dark gray and light gray represent models with the mean metallicities of $Z$=\(Z_\odot\), $Z$=2\(Z_\odot\) and $Z$=0.5\(Z_\odot\) - these act as the baseline model for the standard stellar population with no Helium enhancement as in Fig. \ref{fig:uv}. As previously noted, this metallicity range was chosen as it roughly aligns with that of Coma cluster galaxies in a similar magnitude range, though the sub-solar model shown here is slightly bluer than would be expected of the lower mass galaxies in each redshift bin on average, but can account for any outliers in the lower end of the luminosity function. Also plotted in shaded regions are the colors of increasingly He-enhanced models, with $Y_{ini}$\footnote{The helium abundance $Y$ of a stellar population is related to the initial helium abundance $Y_{ini}$ and the metallicity $Z$ through the following equation: $Y = \Delta Y/\Delta Z \times $ $Z$\, +\, $Y_{ini}$, where $\Delta Y/\Delta Z$ is the galactic helium enrichment parameter, assumed to be 2.0.}$=0.23,0.28,0.33,0.38,0.43$ (and solar metallicity). The blue envelope of the shaded region indicates the color of the He-enhanced model when 100\% of the population is He-enhanced (to the model's $Y_{ini}$) and the red envelope is simply the solar metallicity model with no He-enhancement, with the shaded regions representing between $0-100\%$ of the population being enhanced. All models have a $z_{form}$ of 3, though it should be noted that increasing $z_{form}$ to 4 (or a similar reasonable value) only moves the color redward by approximately 0.1 mags (as shown in \citealt{Ali2022}) and hence the results would remain almost the same, with the strength of the upturn appearing slightly stronger in comparison to the CSP. The He-enhanced models initially trace the standard solar metallicity model at high redshift when most stars in the population are still on the main sequence, as the He-enhanced population only become extremely hot in late stages of stellar evolution. However stars with increased He abundance progress through the MS and RGB phases at a faster rate than standard stars and reach the HB phase earlier. As such models with the highest He abundance start becoming bluer at higher redshifts as their blue HBs have already become populated, with those with the lowest He abundance only start becoming blue at much lower redshift.

In order to test whether the red sequence selection criterion (as detailed in section 2 and shown in Figs. 2 and 3) in the rest-frame optical and $u$ bands had any impact on the evolutionary trend observed in the UV, we both doubled and halved the width of the red sequence in the observed $z-y$ and $g-y$/$r-y$ bands to re-run all of our analyses using the alternative selection criteria and galaxy samples. We find that at every redshift bin altering the selection criteria only changed the median $UV-optical$ color of all galaxies within $\pm0.1$ magnitude and hence did not have any clear impact on the overall UV color evolution as seen in Figs. 5 and 6. Significantly broadening the red sequence width does introduce some outliers in the $UV-optical$ both in the red and blue end of the color distribution - likely dusty red galaxies in the former case and residual star-forming galaxies in the latter.

To visualise the evolution of the $NUV-V$ color of galaxies in comparison to conventional galaxy evolution models, in Fig. \ref{fig:evo4} we subtract median $NUV-V$ measured for each redshift bin (as seen in the box plots in Fig.~\ref{fig:evo3}) from the model $NUV-V$ color of a solar metallicity model with $z_{form}=3$ and standard He abundance (shown as the solid black line in the top figure), which is then plotted against redshift. As mentioned previously, this model represents the evolution of the $NUV-V$ of a standard $M^*$ ETG (with the models being calibrated in the optical/NIR), without any extreme HB stars. As such, if the evolution of the observed color matches this standard model, the difference in the model and observed colors should remain roughly constant with redshift. To demonstrate that this is indeed the case in the optical wavelengths (where these models are known to replicate the results well), we also plot the difference in the model and observed median colors for both rest-frame $U$-band and optical colors (the exact rest-frame wavelengths are shown in Table \ref{table_1}) for all redshift bins. At a glance it can be clearly seen that the difference between the model and $U$-band/optical colors remains roughly constant and nearly zero between $z=0-1.1$, suggesting that the standard solar metallicity CSP model with $z_{form}=3$ can reasonably replicate the median color of quiescent cluster ETGs at all redshift bins in our sample. This is of course to be expected as most galaxy evolution models are designed to primarily interpret the evolution of the optical and NIR data. However, when we look at the $NUV-V$ color, we find a difference of about $\sim0.5$ magnitude between the model and observed colors particularly at $z<0.6$, suggesting that the galaxies are far bluer in the UV than is expected from a passively evolving stellar population (with standard He abundance). At $z>0.6$ we start to see the difference become smaller gradually until eventually at $z>0.8$, the observed colors match reasonably closely the prediction from the model, as is the case in the $U$-band and optical colors, indicating that at higher redshift the $NUV-V$ color behaves as is expected from a quiescent, non-starforming stellar population. This unusual form of evolution in the UV color can be physically explained by invoking a sub-population of He-enhanced blue HB stars driving the UV output amidst the majority red MS/RGB stars that are traditionally known to occupy ETGs. An evolutionary pattern where the UV color remains consistently blue from $z=0\sim0.6$, then declines and eventually reaches its reddest by $z\sim0.8$ and beyond requires the hot HB sub-population to be He-enhanced to $Y=0.44$ or higher ($Y_{ini}=0.40$ as seen in Fig. \ref{fig:evo3}), formed at $z_{form}\sim3-4$ and making up roughly 10\% of the overall population. From a physical standpoint the strong UV emission is explained by the presence of the highly evolved HB stars at low redshift, which naturally starts to disappear at intermediate redshifts (starting at $z\sim0.6$ and fully by $z\sim0.8$) as there simply has not been enough time at that epoch for these stars to have evolved through the MS/RGB (with them becoming sufficiently hot only in the HB phase). Given that the UV emission is a near ubiquitous feature in the vast majority of old stellar populations at low redshift, this poses further questions on how a sub-population of stars can become He-enhanced to nearly twice the cosmological level by at least a redshift of 3, which suggests that the chemical evolution must have happened in earlier generations at even higher redshifts. 

The observed evolution in Fig. \ref{fig:evo3} is inconsistent with low metallicity HB stars \citep{park1997} as these would only appear at $z\sim 0.3$. Such stars, if present in the amounts needed to explain the UV flux, would also dramatically weaken the strong metal lines in the observed spectra of ETGs. The results are also inconsistent with high metallicity HB stars with extra mass loss \citep{Yi1997} as these would also appear at much later times than observed here - note also that there is no evidence for a metallicity-dependent mass loss rate in the data of \cite{Miglio2012} and \cite{McDonald2015}. Binary stars \citep{Han2007,Hernandez2014} would always be present in the population even at high redshift and therefore the match to the standard $NUV$ colors at $z\sim 1$ and the onset of the UV upturn at $z<0.8$ also excludes this hypothesis. In the case of binary stars, no significant evolution would have been expected in UV colors at low-intermediate redshifts after its initial onset at very high redshift, which is quite clearly not the case in our observations. These observations strongly imply that a sub-population of He-rich stars is present within most ETGs at early epochs (see also \citealt{ali2018c} for a direct comparison of observed UV colors of cluster ETGs with predictions from alternative models).

Helium-rich stars are believed to contribute to the anomalously blue horizontal branch morphology of several clusters in
our Galaxy (e.g., \citealt{Piotto2015}), the Magellanic Clouds \citep{Mucciarelli2009,Dalessandro2016}, the Sagittarius dwarf \citep{Carretta2014} and the Fornax dwarf \citep{Larsen2014}. A large number of these GCs are considerably metal-poor and hence have lower requisite degrees of He-enhancement (observed to be up to $Y\sim0.35$) in order to produce blue HB stars compared to more metal-rich systems. However in one of the most extreme cases, NGC 2808, more than half the population is He-rich, with around 10\% having He abundance of $\sim 0.43$, consistent with that inferred in ETGs. Metal-rich clusters in our Galaxy such as NGC 6388 and NGC 6441 \citep{2008ApJ...677.1080Y, tailo2020} have a large fraction of stars with $Y > 0.38$ and $FUV/NUV$ colors similar to those of ETGs. Metal-rich clusters in M31 \citep{Schiavon2013} and M87 (\citealt{Peacock2017} - the closest comparable system to our galaxies) also have $FUV/NUV$ colors typical of local ETGs. This led \cite{Goudfrooij2018} to propose that the UV upturn population in ETGs originates from the disruption of large numbers of metal rich clusters in the early universe.

Given their small remaining envelopes, a fraction of low mass HB stars may not undergo the traditional AGB phase, thus refereed to as `AGB-manqu\'e' and are the hot UV-bright progeny of the most He-enriched HB stars \citep{LaGioia2021,Carlos2022}. Abundances of light elements in Galactic globular clusters show evidence of enrichment in N and Na and depletion in C and O \citep{Milone2014}  suggestive of hot bottom burning at temperatures of 70-100 MK, consistent with deep dredge-up in massive AGB stars \citep{Ventura2001} or fast-rotating massive stars \citep{Decressin2007} whose Hydrogen envelope is removed. One of the AGB-manqu\'e stars in \cite{Carlos2022} is observed to be O-poor, a hallmark of the hot bottom burning process (see also \citealt{Gratton2010}). Such processes, at much higher metallicities, may produce the hot HB/EHB stars that contribute to the UV upturn in ETGs.

\section{The star formation history of ETGs}
\begin{figure*}
\centering
{\includegraphics[width=0.3\textwidth]{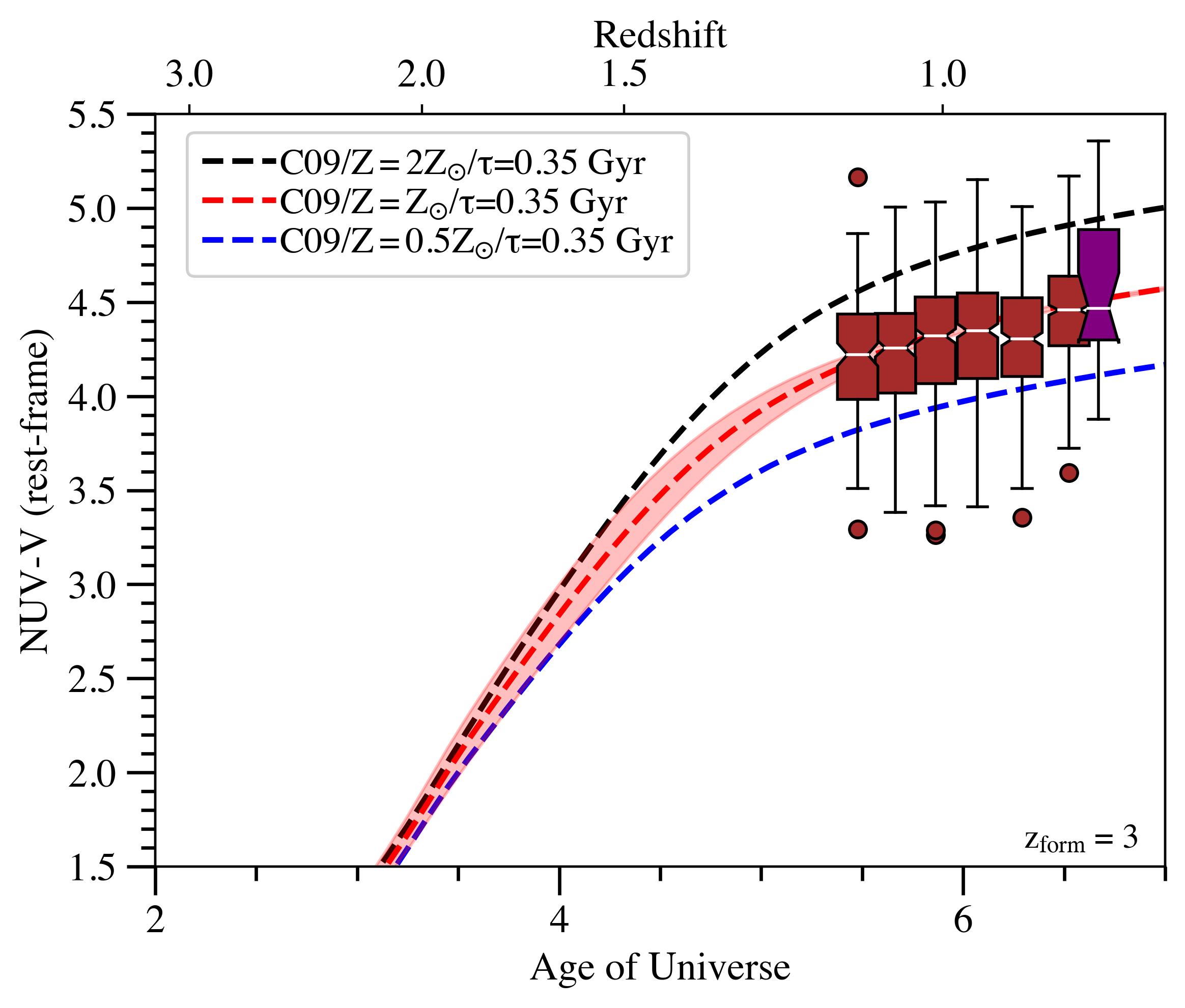}}
{\includegraphics[width=0.3\textwidth]{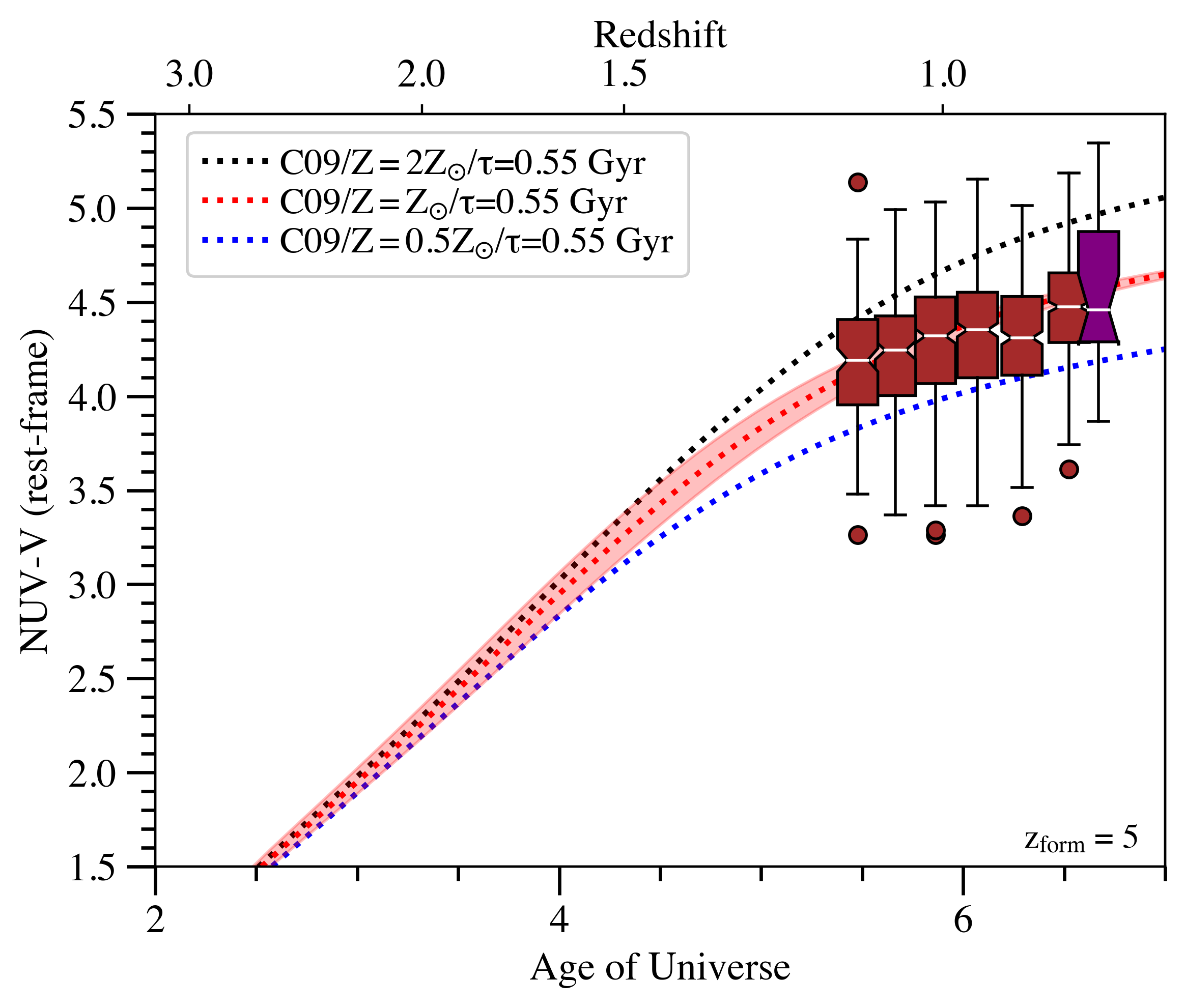}}
{\includegraphics[width=0.3\textwidth]{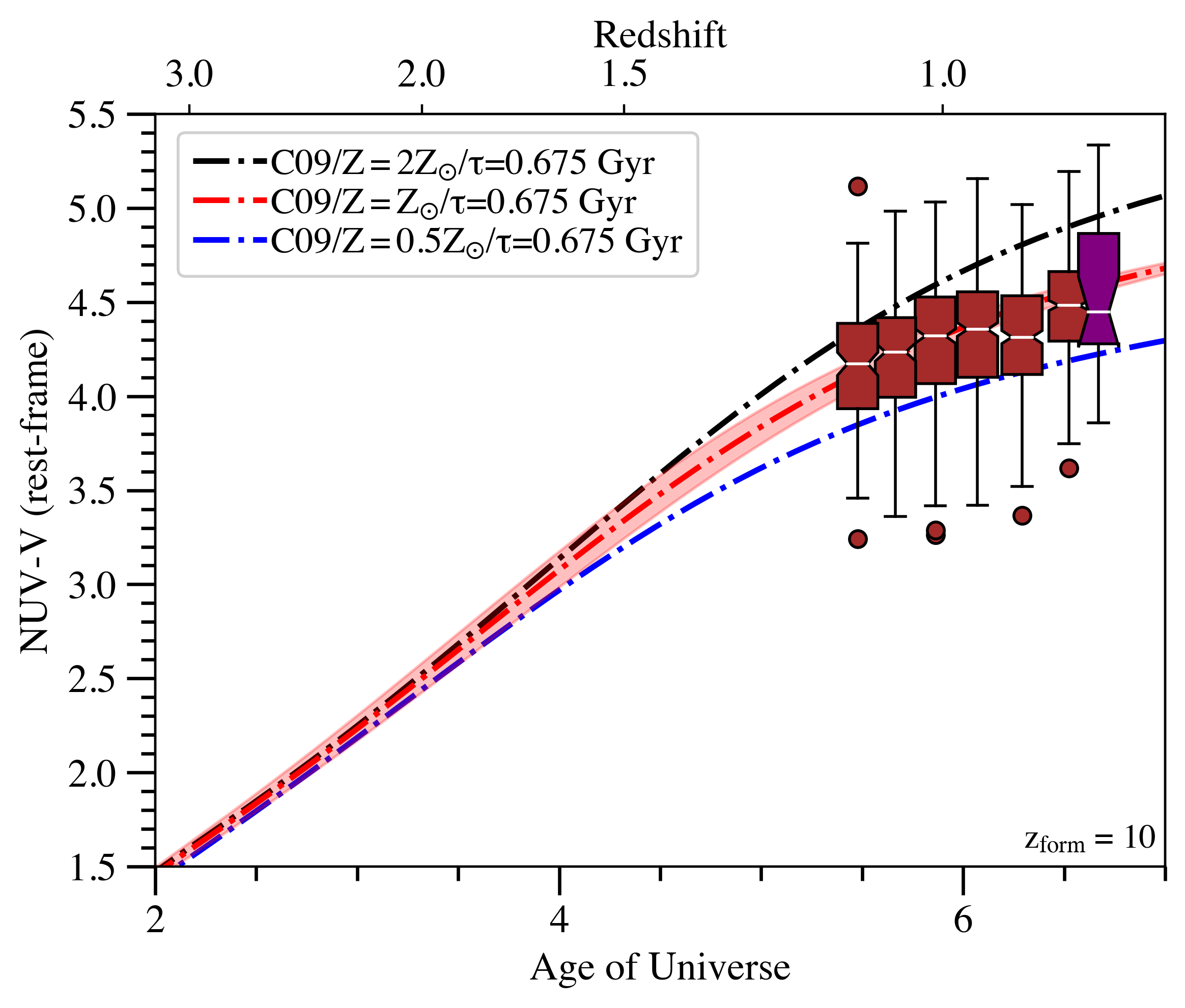}}
\caption{Evolution of the rest-frame $NUV-V$ (observed $g-y$ at $z=1$) color of cluster ETGs over redshift/lookback time as in Fig. \ref{fig:evo3}, but only for $z>0.8$ where the upturn is expected to no longer be present. Also plotted for comparison are C09 CSP models for $z_{form}=3, 5, 13$ (covering roughly 1.5 Gyrs in range) and varying e-folding timescales ($\tau$) that best fit the data for each formation redshift. The red dashed lines show the closest fit to the median at all redshift bins in each case (the red envelope indicates the uncertainty due to photometric error), with the black and blue lines showing the $Z$=2\(Z_\odot\) and $Z$=0.5\(Z_\odot\) models respectively. The results suggest that ETGs have $z_{form}=3-10$ and $\tau=0.35-0.7$ Gyrs respectively.}
\label{fig:evotau}
\end{figure*}

\begin{figure}
{\includegraphics[width=0.49\textwidth]{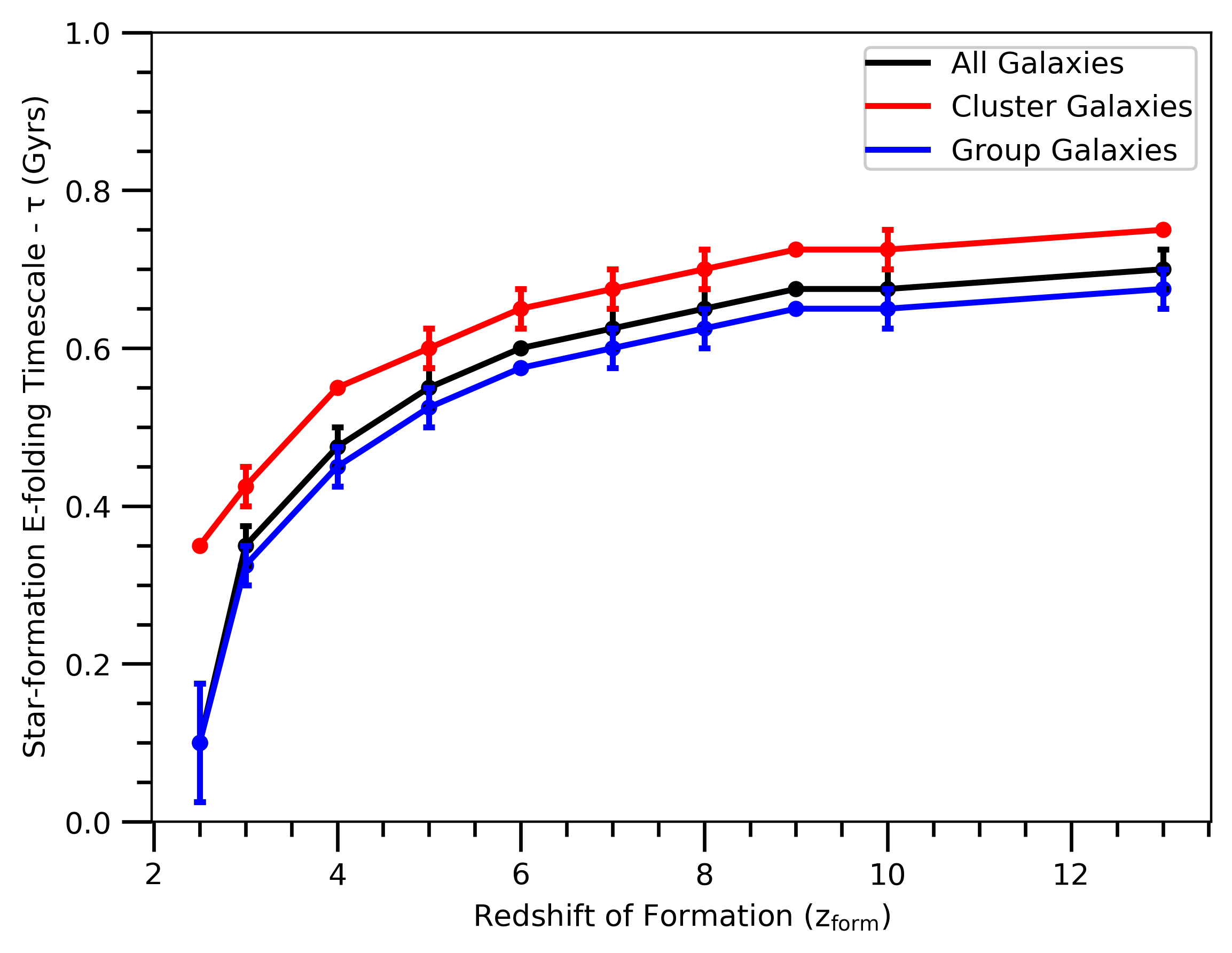}}
\caption{The star-formation e-folding timescale (for solar metallicity C09 CSP models) that would be required at each formation redshift to fit the rest-frame $NUV-V$ colors of cluster galaxies between $z=0.85-1.1$. Several examples of the model fits to the data are shown in Fig. \ref{fig:evotau}.}
\label{fig:zft}
\end{figure}

Given that the upturn has almost entirely faded by $z\sim0.8$ and the observed $NUV-V$ color matches that of a standard solar metallicity CSP model - it can be reasonably postulated that in this redshift regime the UV colors are mostly driven by the standard age/metallicity of the overall stellar population in these galaxies. The average stellar masses that we are probing can be estimated to have roughly solar metallicity, particularly for the median galaxy in our redshift bins, as this corresponds roughly to the $M^*$ characteristic magnitude in the luminosity function. We can thus use this knowledge to estimate the star-formation history of ETGs at $z>0.8$ using $NUV-V$, which is far more sensitive to age than optical/NIR colors, as it is likely being driven by the main sequence turnoff stars in the absence of young hot stars or old He-enhanced HB stars (\citealt{Lotz1999}). We use the data of all ETGs in redshift bins above $z>0.8$ (as in Fig. \ref{fig:evo3}) in order to fit models of varying $z_{form}$ (redshift of formation) and $\tau$ (e-folding timescale in Gyrs) and find the most likely combination of the two parameters which explains the star-formation history of cluster ETGs, attempting to break the degeneracy in $z_{form}$ and $\tau$.

For the purpose of this analysis we make use of the Flexible Stellar Population Synthesis ($FSPS$) models from \cite{conroy2009, conroy2010} - C09 hereafter. The results from a $Z$=\(Z_\odot\) C09 CSP with the power of the metallicity distribution function (pmetals) set to the default value of 2 match reasonably with the YEPS CSP infall models ($Y_{ini}=0.23$) used before, with the benefit of being able to adjust the early star-formation history of the CSPs. We generated a range of CSPs with $0<\tau<2$ Gyrs (in 0.025 Gyr intervals) and $2.5<z_{form}<13$ (with the highest redshift corresponding to the earliest observed galaxy - \citealt{robertson2023}). We minimize least squares to find the best fitting value of $\tau$ for each $z_{form}$ that produces UV colors matching the median observed color of galaxies in the $z>0.8$ redshift bins. Of particular importance is the color of the highest redshift bin ($z=1.1$) given that if $z_{form}$ is too late or $\tau$ is too long then the model will not have had enough time to evolve and become red enough to match the color of the highest redshift galaxies, which effectively places a strong limit on how long the e-folding timescale can be and how late the ETG can start forming stars based on the highly age sensitive UV data. Fig. \ref{fig:evotau} shows a sample of the best fitting models with $Z$=\(Z_\odot\) at $z_{form}=3, 5$ \& $10$ (roughly 1.5 Gyrs in range) as given by the dashed red line, with the $Z$=2\(Z_\odot\) and $Z$=0.5\(Z_\odot\) models given by the black and blue dashed lines for comparison. For $z_{form}=3, 5$ \& $10$, the best fit $\tau=0.35, 0.55$ \& $0.675$. Fig. \ref{fig:zft} (black line) shows the best fit $\tau$ with increasing $z_{form}$, where generally it is found that a higher formation redshift allows for and needs longer e-folding timescales to match the observed UV data. From our results we can reasonably infer that galaxies likely did not have $z_{form}<3$, as at $z_{form}=2.5$, $\tau$ needs to be $0.1$ Gyr, indicating that stars would have to have formed almost instantaneously in order for the model colors to match the observed colors. It is also not possible to push $z_{form}$ much higher than 10 as the value of $\tau$ appears to plateau at this redshift (Fig. \ref{fig:zft}) and we would start to push past the epoch of reionization. As such, our results suggest that the average $M^*$ cluster ETG realistically formed between $z_{form}=3-10$ (corresponding to about 1.5 Gyr range in time), with e-folding timescales of $\tau=0.35-0.7$ Gyr, meaning that the vast majority of stars in these galaxies formed rapidly at high redshift and the galaxies became passive thereafter - consistent with some of the high redshift quiescent galaxies now observed with JWST \citep{Nanayakkara2022,Antwi-Danso2023,Glazebrook2023,Looser2023,strait2023}. The results are also in general agreement with optical studies of massive quiescent galaxies at intermediate to high redshifts, where such galaxies are found to have started forming their stars at $z_{form}=5-10$ (\citealt{Tacchella2022}) with timescales shorter than 1 Gyr, and the peak of star-formation activity potentially occurring at $z\sim2.3-3.5$ (\citealt{Onodera2015,Eugenio2021}). However, our calculations of the star-formation history of ETGs can be further pinpointed with deep rest-frame UV data of cluster galaxies at $z=1.1-2$, in the redshift range where large galaxy clusters are still observed (such as in the MaDCoWs catalog; \citealt{gonzalez2019}). The HSC SSP survey is not deep enough to observe such galaxies, but surveys such as EUCLID may provide the requisite rest-frame UV data to further this analysis.

In addition, the presence of a He-rich population, if formed by the ejecta of stars in previous generations, and even assuming the maximum possible efficiency where 100\% of the material is recycled into stars (which is not unfeasible for galaxies that may be considered as closed boxes over the likely formation period of a few hundred Myr), implies that a very large fraction of the stellar mass must also have been present and {\it in situ} at $z>3$.

\subsection{Can Residual Star Formation Explain the UV Excess?}

Studies have found that the star-formation rate (SFR) in galaxies shows a steady increase in lookback time from present day out to the cosmic noon around $z\sim2$, when the SFR in the Universe peaked (e.g., see compilation in \citealt{Yuksel2008}). On average the cluster galaxies we observe follow the separate and opposite trend than would be expected from even residual star-formation (RSF) being the driver of the blue UV colors. Residual star formation \citep{Vazdekis2016,Rusinol2019} would need to increase with decreasing redshift to explain the increasing UV excess compared to the passively evolving main stellar population, opposite to what is observed for the evolution of SFRs in galaxies with redshift (e.g. \citealt{Finke2022}). \cite{Kaviraj2008} show that in their study that RSF increases with increasing redshift, from $\sim 7\%$ at $z= 0.5$ to $\sim 13\%$ at $z= 1$. The intensity of (recent) star formation in the most active early-types has therefore halved between $z \sim 0.7$ and present day. \cite{Ree2012} find that the number fraction of ETGs with RSF is negligible ($ < 5\%$) in their brightest ($M_r < -22$) subgroup, but gradually increases toward the fainter galaxies ($\sim 30\%$) at $z<0.12$. It is therefore unlikely that our findings can be explained by fine-tuning RSF in bright ETGs, especially in cluster environments where galaxies are likely to be more strongly quenched.

\begin{figure}
\centering
{\includegraphics[width=0.49\textwidth]{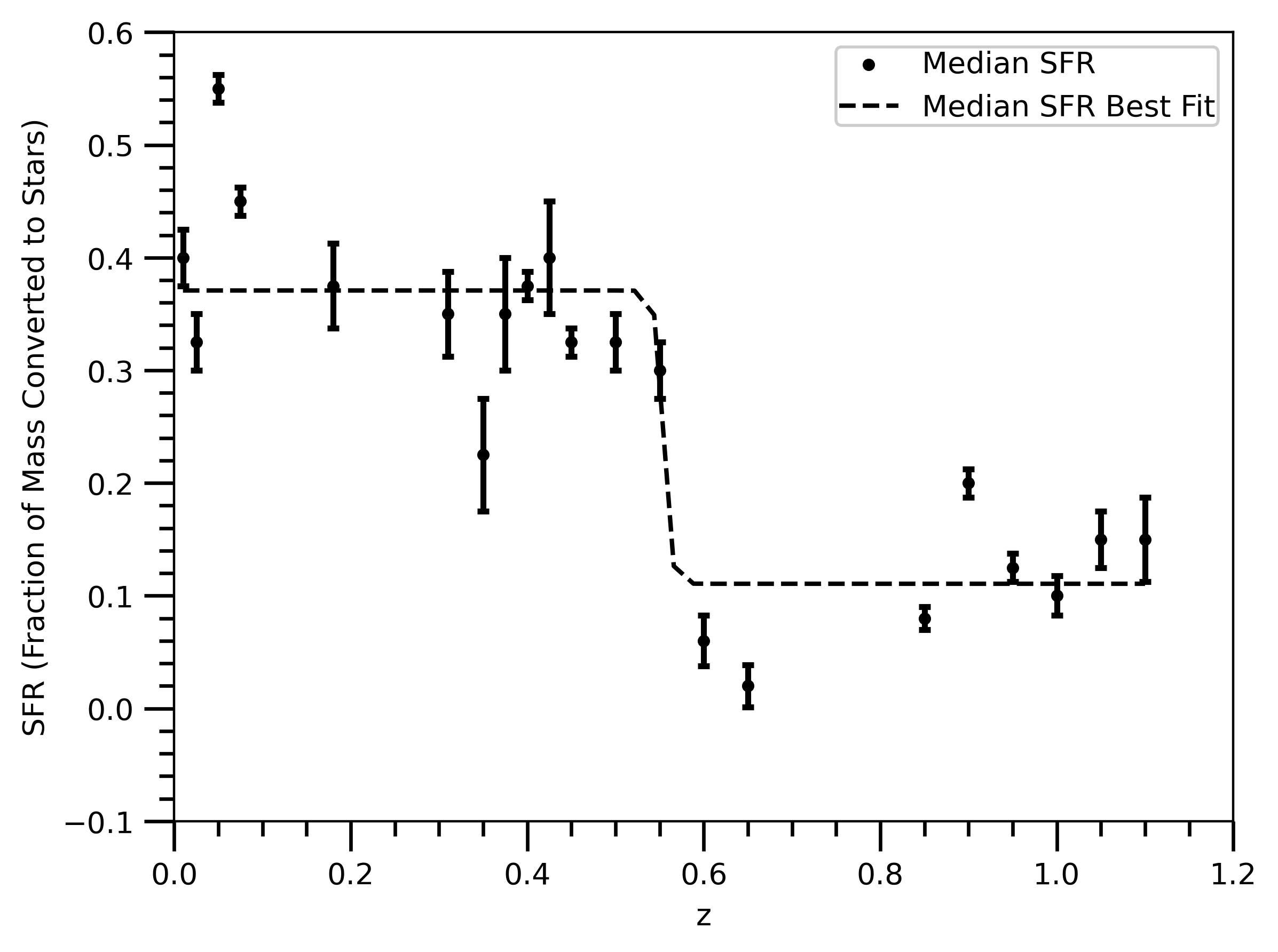}}
\caption{Plot showing the evolution of the median star-formation rate at each redshift in Fig. \ref{fig:evo3} calculated using the C09 CSP models ($Z$=\(Z_\odot\), $z_f=3$) with an initial burst of star-formation followed by a constant SFR as given by the fraction of total mass converted to stars in the models. This assumes that the entire UV emission observed in galaxies is attributed to constant star-formation and not the upturn. In order for star-formation to reasonably replicate the evolution of the UV output in ETGs across redshift, it would have to decrease around $z=0.6-0.8$, which would contradict the well established expectations of the SFR increasing at higher redshifts until the cosmic noon (\citealt{Yuksel2008}).}
\label{fig:sfr}
\end{figure}

As an experiment we use the C09 CSP models to calculate the average SFR in each redshift bin. The models have a fixed $Z$=\(Z_\odot\) and $z_f=3$, with varying levels of constant SFRs from which the best fit to the median UV color in each redshift bin is derived, hence giving us the median SFR across redshift for roughly $M^*$ galaxies. The SFR given by C09 is defined as the fraction of mass converted to stars in a constant mode of star-formation. In Fig. \ref{fig:sfr} we plot the best fit SFR to the median color of each redshift bin in Fig. \ref{fig:evo3}. As expected, the SFR in these galaxies is very low - less than $1\%$ given that most galaxies in our sample are selected to be passive from their optical colors. However, it can also be clearly seen that given the rise in the UV output at lower redshift in ETGs, the SFR increases below $z\sim0.6$ up to $0.4-0.5\%$, whereas at $z>0.6$ the SFR is nearly negligible at $0.1\%$. Altering the $Z$ or $z_{form}$ would change the SFR values slightly but the overall trend remains intact. Such an increase is opposite to the trend in SFR seen in key studies based on the more thoroughly analysed multi-wavelength data where the SFR is found to gradually increase with increasing redshift. It would also be difficult to physically explain why there would be a sudden jump in the SFR particularly around $z\sim0.6$ from the perspective of stellar population evolution, keeping in mind the results shown here are median values derived from hundreds to thousands of cluster galaxies at each bin and shows the general trend in quiescent ETGs at large. As such, an older stellar sub-population that evolves off the MS/RGB phase between $z=0.6-0.8$ and turns UV-bright in the HB phase can physically explain the results in a consistent and non-ad hoc manner. While the purpose of this study is not to derive highly accurate SFR values of cluster ETGs, this experiment does illustrate that while at any given redshift bin it may be possible to assume a small fixed SFR to account for the average UV color of ETGs, it is difficult to use star-formation to explain the overall evolution of the observed UV colors in ETGs over redshift.

\subsection{Effect of Environment on the UV emission \& SFH of ETGs}

\begin{figure*}
\centering
{\includegraphics[width=0.45\textwidth]{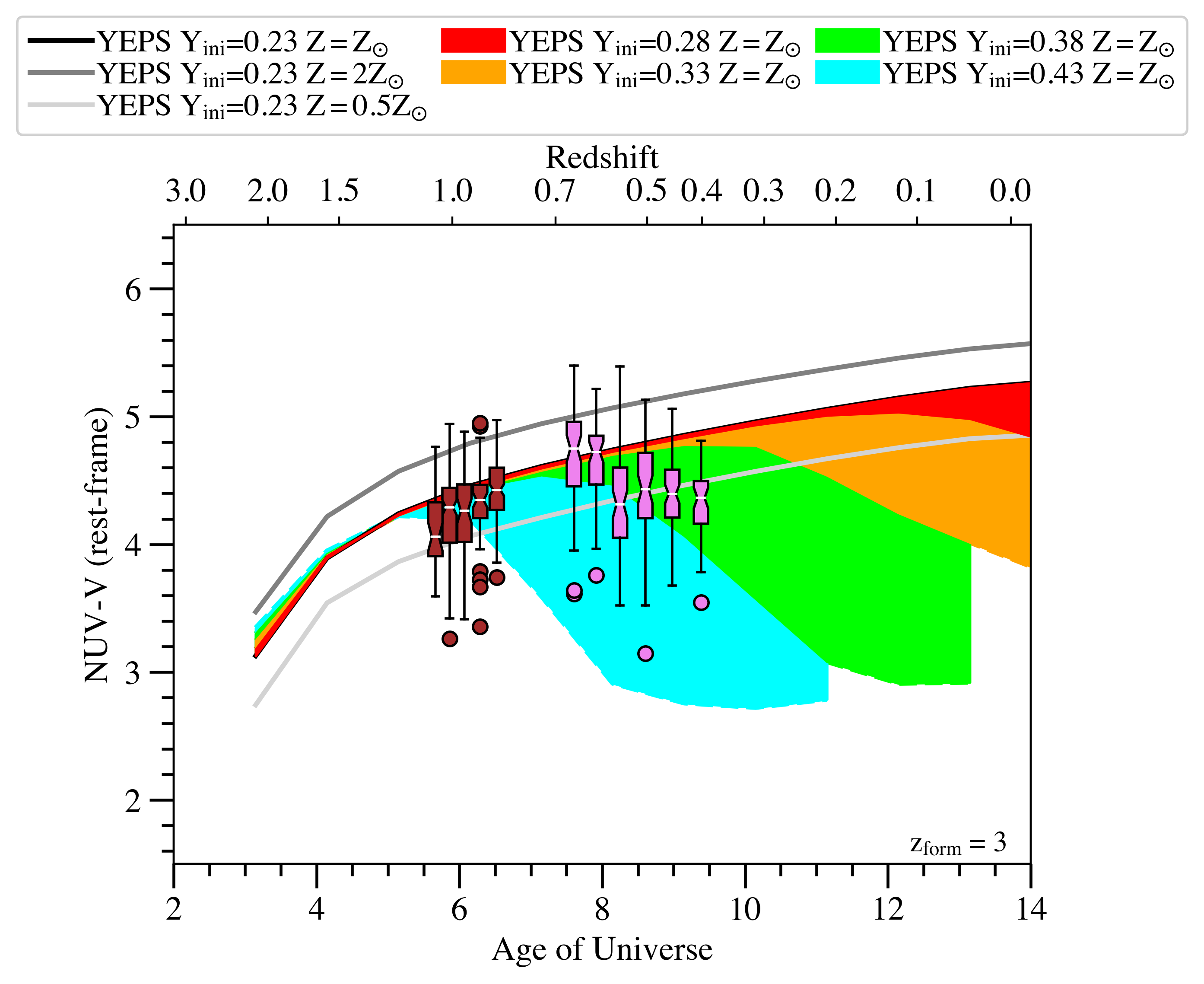}}
{\includegraphics[width=0.45\textwidth]{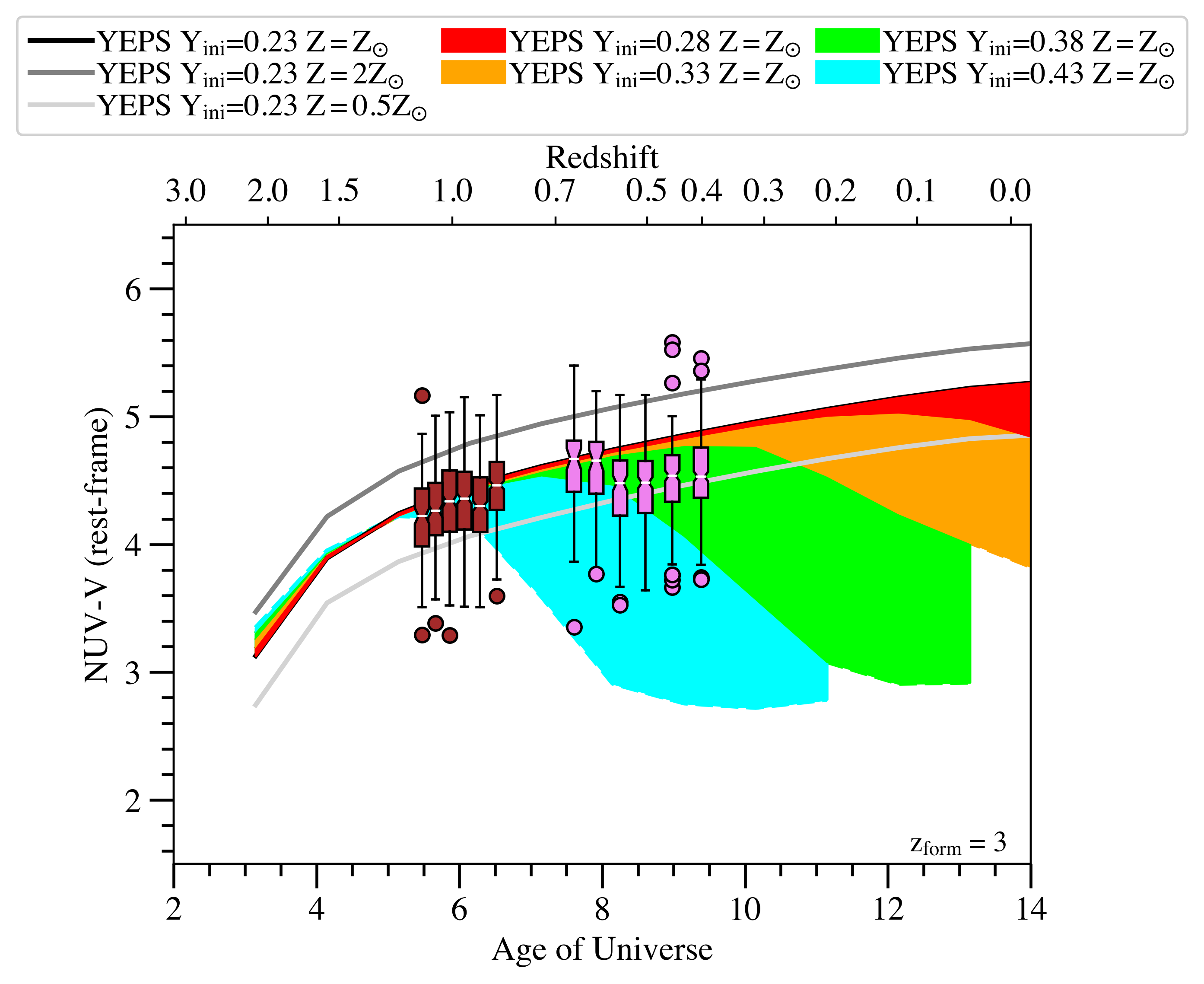}}
\caption{ Same as Fig. \ref{fig:evo3}. \textit{Left:} Shows only galaxies which belong to `clusters' with $N_{mem} \geq 20$. \textit{Right:} Shows only galaxies which belong to `groups' with $N_{mem} < 20$.}
\label{fig:evo3clgr}
\end{figure*}

\begin{figure*}
\centering
{\includegraphics[width=0.3\textwidth]{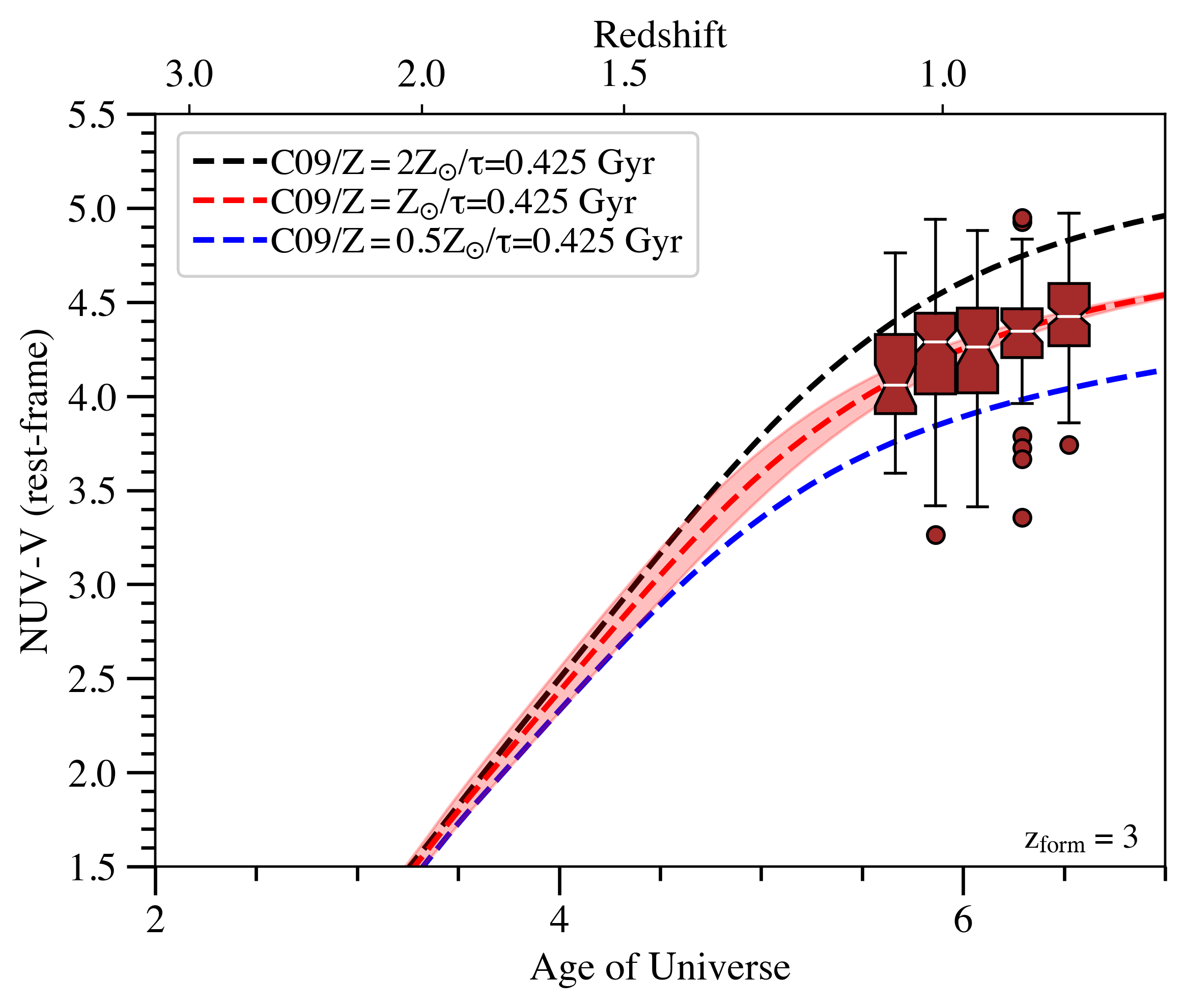}}
{\includegraphics[width=0.3\textwidth]{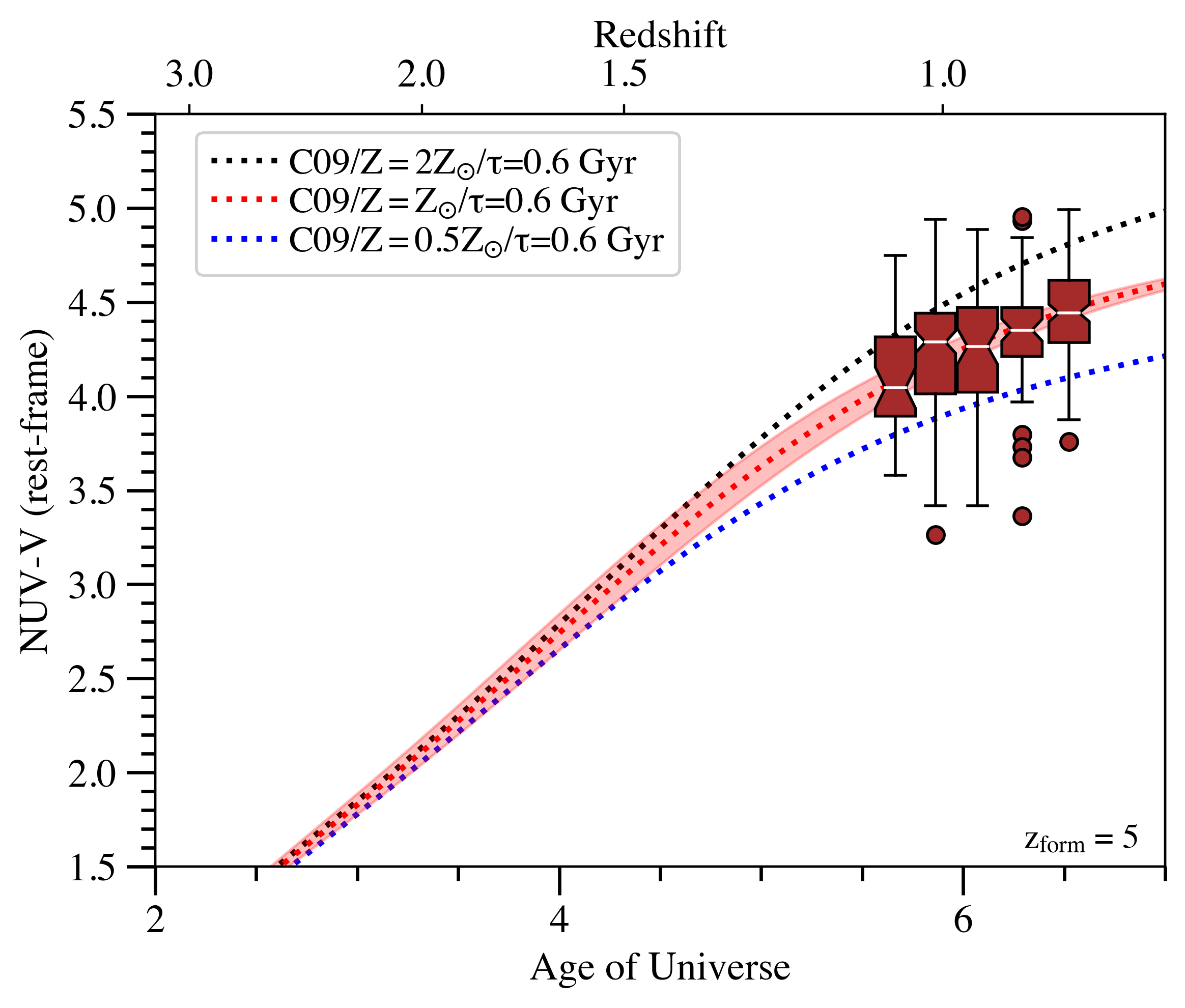}}
{\includegraphics[width=0.3\textwidth]{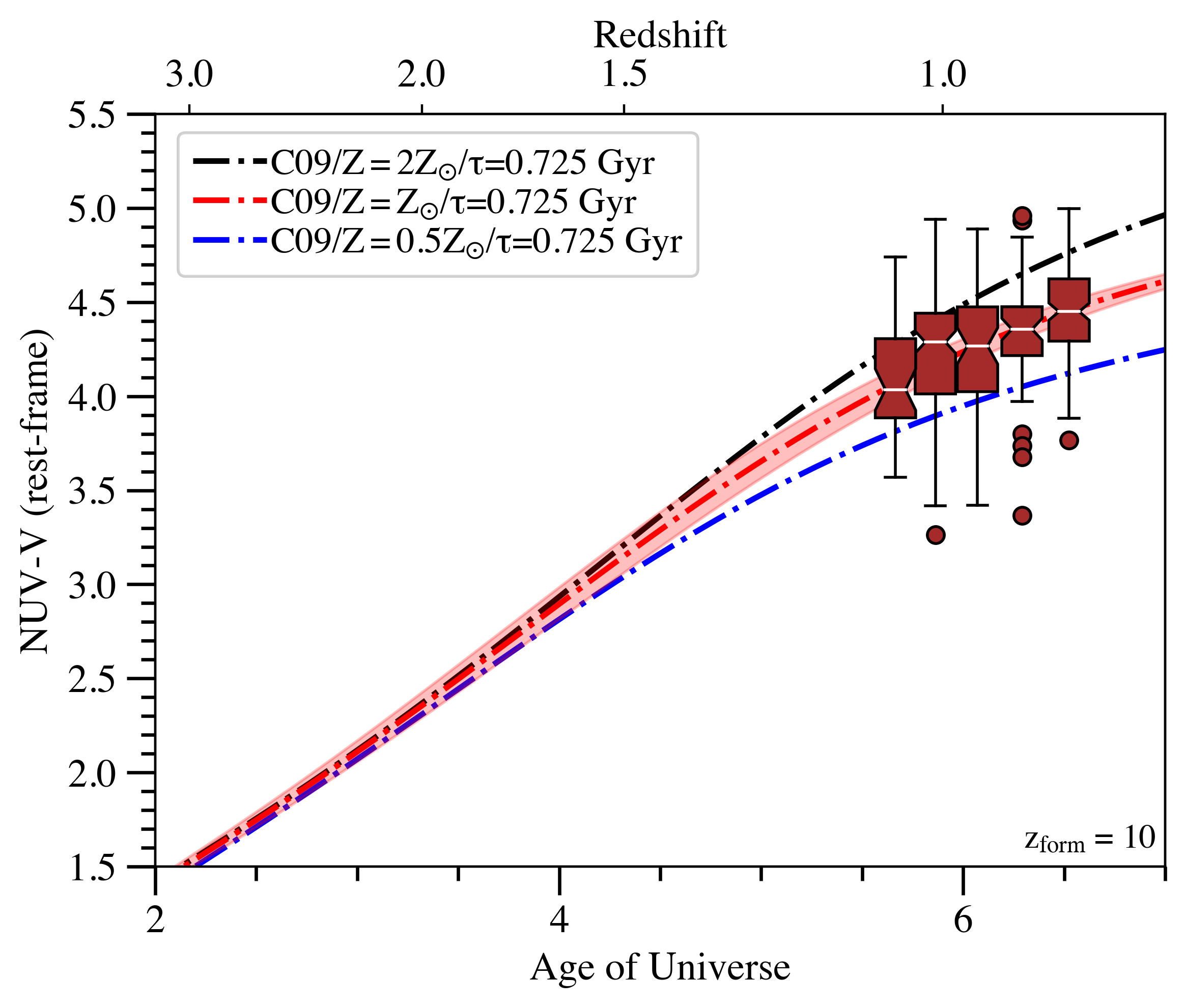}}
{\includegraphics[width=0.3\textwidth]{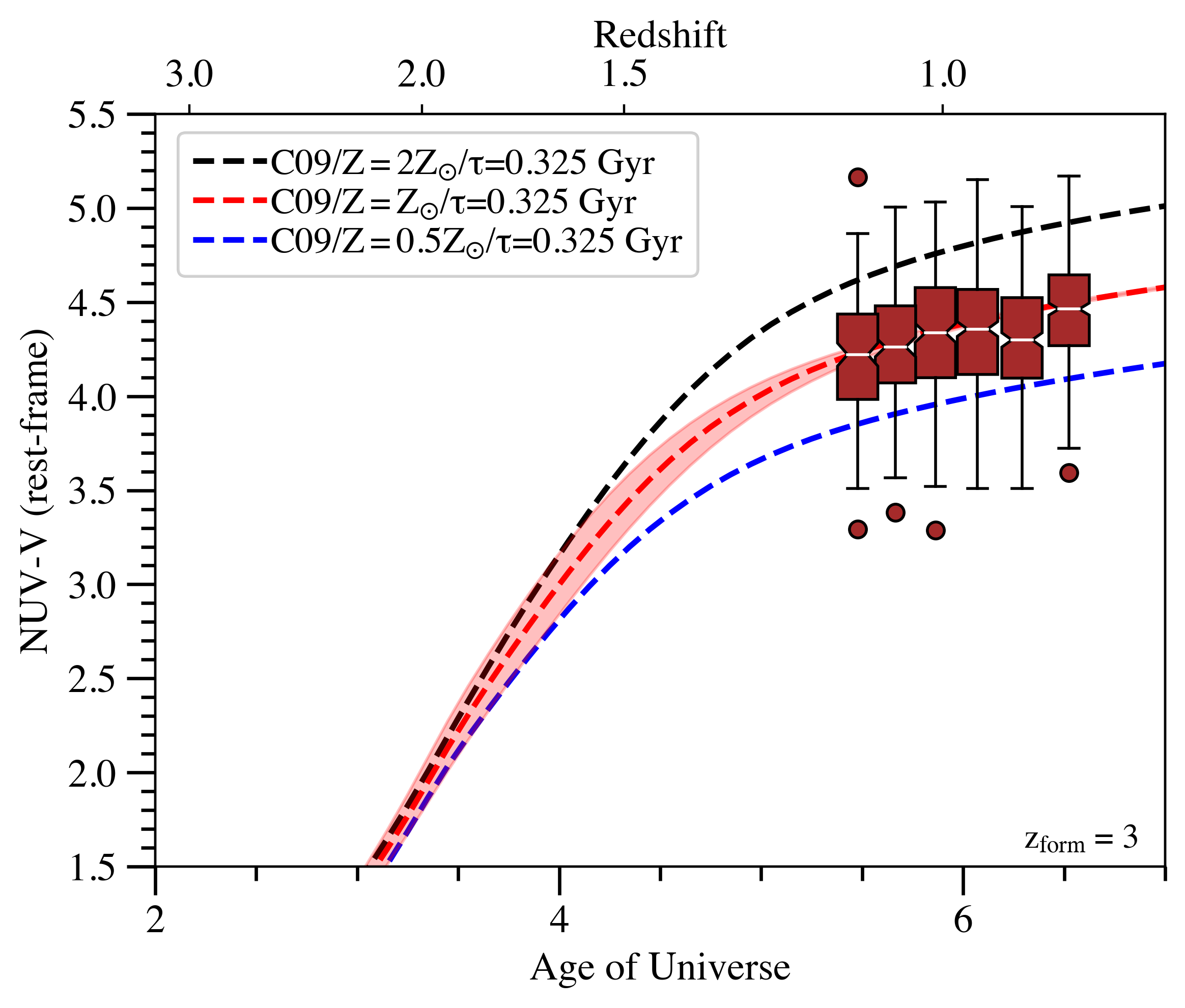}}
{\includegraphics[width=0.3\textwidth]{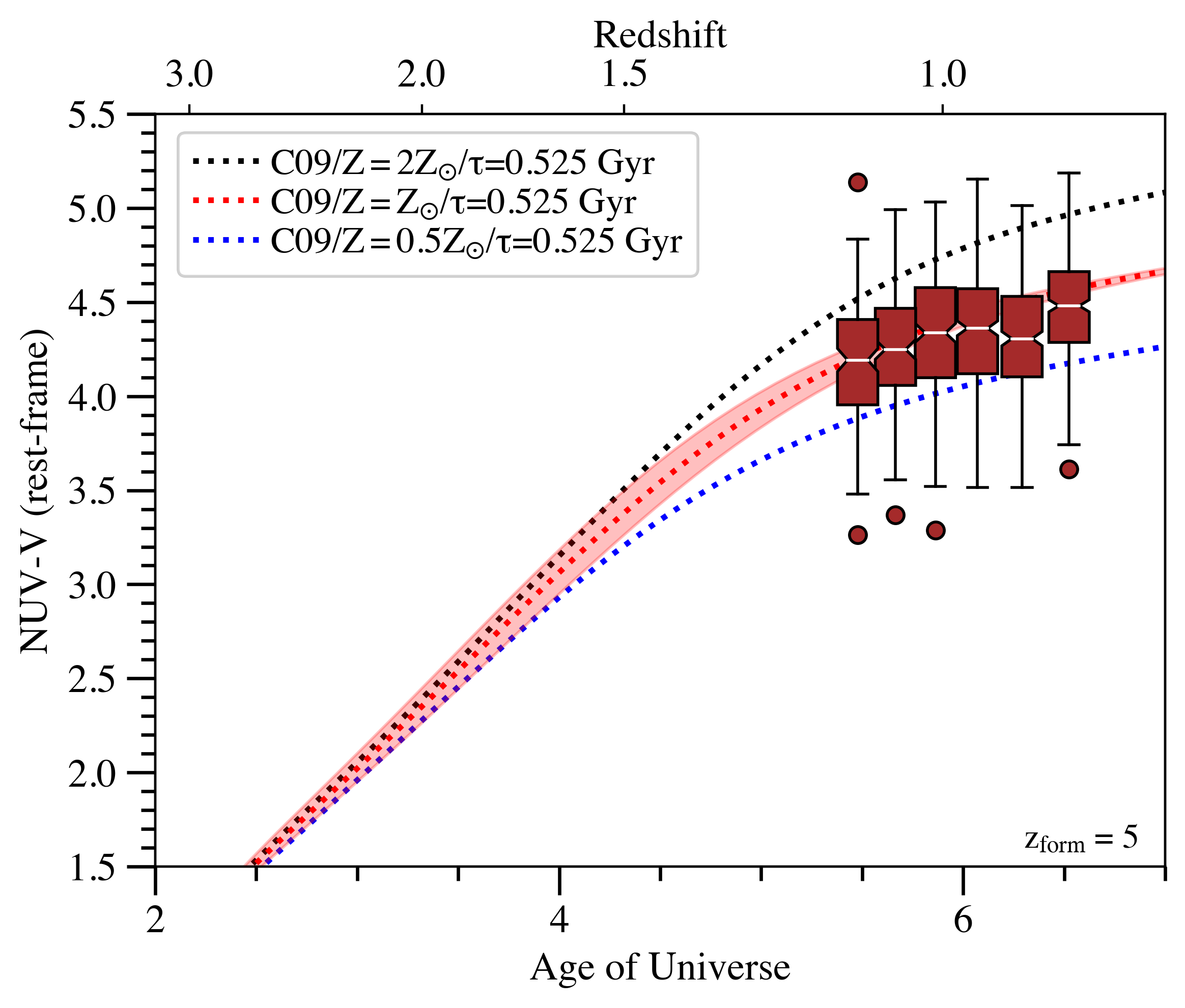}}
{\includegraphics[width=0.3\textwidth]{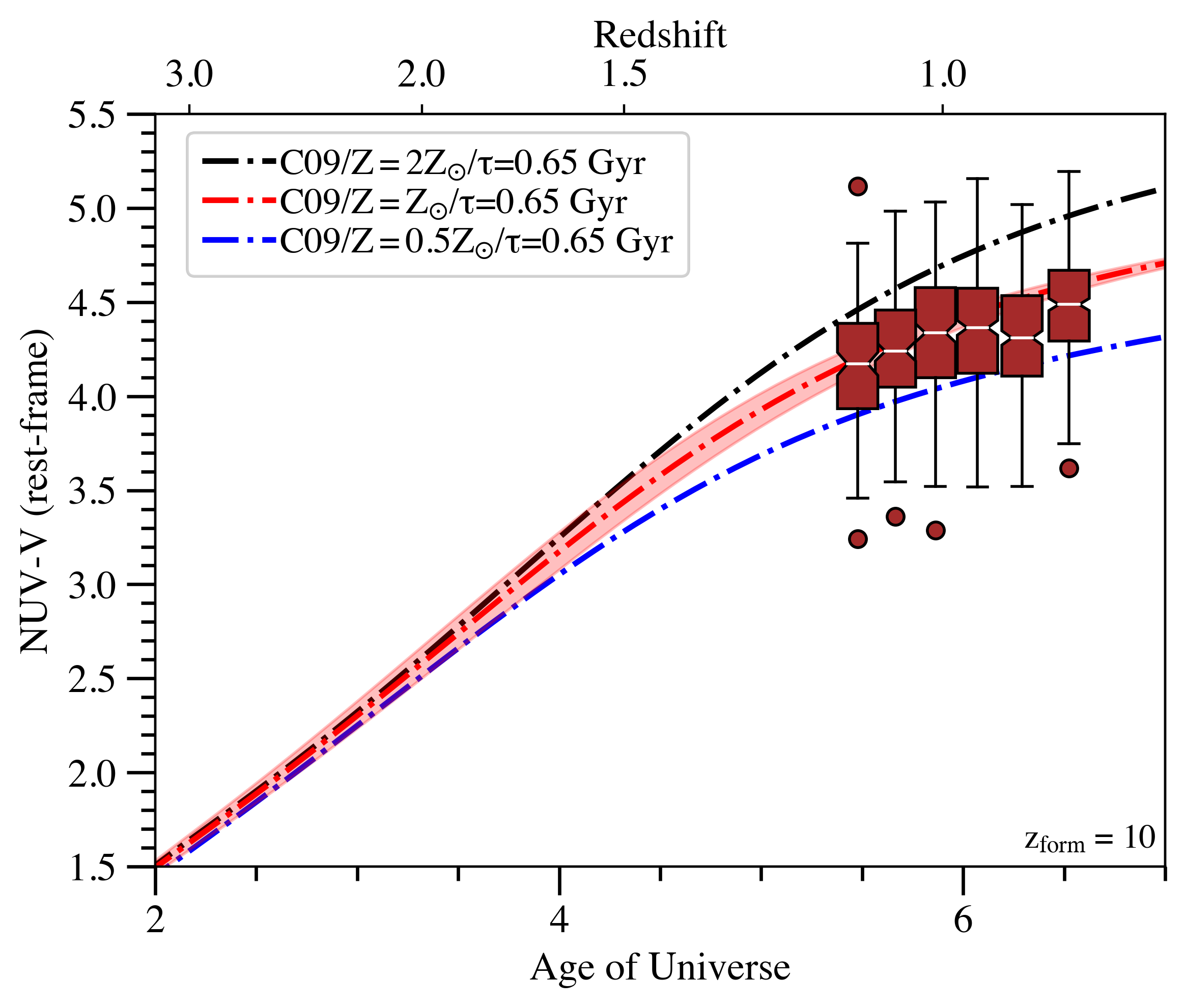}}
\caption{Same as Fig. \ref{fig:evotau}. \textit{Top:} Shows only galaxies which belong to `clusters' with $N_{mem} \geq 20$. \textit{Bottom:} Shows only galaxies which belong to `groups' with $N_{mem} < 20$.}
\label{fig:env}
\end{figure*}

It is well understood that the environment has a significant impact on the evolution of the morphology and stellar populations of galaxies, as is seen by the morphology-density relation (e.g. \citealt{dressler1997}) and the Butcher-Oemler effect (\citealt{butcher1978}). Clusters are found to host mostly an early-type galaxy population, particularly at the core where star-formation is quenched to a much stronger degree than galaxies in the outskirts, and even more so than in the field (see \citealt{boselli2016} and references therein). Given this, we were motivated to test whether the cluster size in which a galaxy resides has any significant impact on its UV color or its star-formation history derived from therein. For this analysis we make use of the cluster size information from the HSC DUD cluster catalog, in which the estimated richness of each cluster and its constituent member galaxies are provided. Thus we make a cut of all clusters with a richness above and below 20, dubbing for the purpose of this specific analysis those with $richness>20$ as `clusters' and $richness<20$ as `groups'. As would be expected, at higher redshift, particularly at $z>1$ the number of clusters declines significantly, with most galaxies being part of smaller groups that have 10-20 members. We re-create Fig. \ref{fig:evo3} - the evolution of the rest-frame $NUV-V$ color with redshift in Fig. \ref{fig:evo3clgr}, except for cluster and group galaxies separately in order to identify any difference in the color evolution between the two galaxy populations.

In general both the cluster and group galaxies show the same trend in regards to the evolution in their UV color, in that there is an excess in the UV emission below $z=0.6$, which largely disappears beyond $z=0.8$. This suggests, as with previous studies in local galaxy clusters (\citealt{ali2019}) and groups \citep{phillipps2020} that the UV upturn - driven by a sub-population of blue HB stars, is an intrinsic feature of all ETGs, or old stellar populations at large, which develops given sufficient time irrespective of the external environment. This would also explain the observation of blue HB stars in similarly old GCs in the Milky Way and elsewhere (e.g., the LMC and SMC, M31, M87) as discussed previously. If He-enhancement is the primary driver of blue colors, then such anomalous chemical enhancement likely arose in these extremely old stars through some universal mechanism in the early Universe.

Furthermore, using the evolution of the UV data at higher redshift, we wanted to test whether there is any clear difference in the star-formation history parameters, i.e. age and star-formation timescales between galaxies residing in larger clusters vs. smaller groups. As such we once again used the data of all ETGs above $z=0.8$ in order to fit models of varying $\tau$ for $z_{form}$ between $2.5-13$ and find the most likely combination of the two parameters which explains the star-formation history of ETGs in both cluster and group environments. The results are shown in Fig. \ref{fig:env} for a few select $z_{form}$ (similar to Fig. \ref{fig:evotau}). For clusters with $z_{form}=3-10$, $\tau$ is between $0.425-0.725$ Gyrs. For groups with $z_{form}=3-10$, $\tau$ is between $0.325-0.65$ Gyrs. Fig. \ref{fig:zft} shows the change in $\tau$ with $z_{form}$ between the two sets of galaxies for comparison. On average galaxies in both clusters and groups share similar formation histories, likely having finished most of their initial star-formation activity in very short timescales of $<0.8$ Gyrs and assembled at very high redshift - beyond $z=3$. The cluster galaxies do appear to on average have slightly longer e-folding timescales compared to group galaxies but this difference is minute ($\sim 0.1$ Gyrs). It should also be noted that at increasing redshift (particularly $z>1$), the number of galaxies within large cluster environments in the HSC SSP dataset is quite low - most ETGs at these redshifts are in smaller groups. In general there does not appear to be a significant divergence in star-formation history between group and cluster environments, though this finding may change with more data of large clusters at $z>1$, e.g. such as those identified in the MaDCoWS survey (\citealt{gonzalez2019}). It may then be of more significance to explore the UV emission and star-formation activity of field ETGs at $z>1$ to those of ETGs as part of larger structures, where the environmental differences may be more apparent.

\section{Conclusions}

From an analysis of rest-frame $NUV-V$ colors of numerous ETGs in hundreds of clusters at $0 < z < 1.1$ from multiple surveys/datasets, we conclude:

\begin{itemize}
    \item ETGs show an UV excess that has an onset epoch at about $z \sim 0.6-0.8$ and reaches colors as observed locally within less than 1 Gyr. This can best be explained by the presence of a sub-population of stars (about 10\% by mass but with a strong dependence on galaxy stellar mass) enriched in He at about the 44\% level or above, comparable to the more extreme blue HB stars observed in the globular clusters of the Milky Way.
    
    \item At $z > 0.8$ this component is not yet present and the colors of galaxies may be used to set stringent limits to their star formation timescales. Our data strongly constrain the majority of star formation to have occurred before $z=3$ and with e-folding timescales shorter than $0.75$ Gyrs in $z<1.2$ ETGs. Galaxies that are part of smaller `groups' and larger `clusters' appear to have similar star-formation redshifts and timescales, with only marginal differences.
    
    \item Based on the presence of He-rich stars at $z<0.6$, known evolutionary timescales and yields for chemical species, we also infer that the vast majority of stellar mass in ETGs must also have been present and assembled within galaxies by $z=3$. We note here that the stars may have also formed at an even earlier redshift within lower mass systems, which then merged hierarchically into the galaxies observed here, though this scenario may be less certain given the general metal-rich nature of the galaxies.

    \item The UV excess appears to be an intrinsic feature in ETGs and is unaffected by environment. Furthermore, residual star-formation cannot simply explain the evolutionary trend observed in the UV data, reinforcing the need of a highly evolved population of blue HB stars to be the primary driver of the UV emission in quiescent ETGs.
\end{itemize}

\acknowledgments

The Hyper Suprime-Cam (HSC) collaboration includes the astronomical communities of Japan and Taiwan, and Princeton University. The HSC instrumentation and software were developed by the National Astronomical Observatory of Japan (NAOJ), the Kavli Institute for the Physics and Mathematics of the Universe (Kavli IPMU), the University of Tokyo, the High Energy Accelerator Research Organization (KEK), the Academia Sinica Institute for Astronomy and Astrophysics in Taiwan (ASIAA), and Princeton University.  Funding was contributed by the FIRST program from the Japanese Cabinet Office, the Ministry of Education, Culture, Sports, Science and Technology (MEXT), the Japan Society for the Promotion of Science (JSPS), Japan Science and Technology Agency  (JST), the Toray Science  Foundation, NAOJ, Kavli IPMU, KEK, ASIAA, and Princeton University.

This paper is based [in part] on data collected at the Subaru Telescope and retrieved from the HSC data archive system, which is operated by Subaru Telescope and Astronomy Data Center (ADC) at NAOJ. Data analysis was in part carried out with the cooperation of Center for Computational Astrophysics (CfCA) at NAOJ.  We are honored and grateful for the opportunity of observing the Universe from Maunakea, which has the cultural, historical and natural significance in Hawaii.

These data were obtained and processed as part of the CFHT Large Area U-band Deep Survey (CLAUDS), which is a collaboration between astronomers from Canada, France, and China described in \cite{Sawicki2019}. CLAUDS is based on observations obtained with MegaPrime/MegaCam, a joint project of CFHT and CEA/DAPNIA, at the CFHT which is operated by the National Research Council (NRC) of Canada, the Institut National des Science de l’Univers of the Centre National de la Recherche Scientifique (CNRS) of France, and the University of Hawaii. CLAUDS uses data obtained in part through the Telescope Access Program (TAP), which has been funded by the National Astronomical Observatories, Chinese Academy of Sciences, and the Special Fund for Astronomy from the Ministry of Finance of China. CLAUDS uses data products from TERAPIX and the Canadian Astronomy Data Centre (CADC) and was carried out using resources from Compute Canada and Canadian Advanced Network For Astrophysical Research (CANFAR).

C.C. acknowledges support from the National Research Foundation of Korea to the Center for Galaxy Evolution Research (2022R1A6A1A03053472, 2022R1A2C3002992).

This work was supported by JSPS KAKENHI Grant Number JP21K03622.

This work made use of Astropy:\footnote{http://www.astropy.org} a community-developed core Python package and an ecosystem of tools and resources for astronomy \citep{astropy:2013, astropy:2018, astropy:2022}.

\bibliography{references}{}
\bibliographystyle{aasjournal}

\end{document}